\documentclass[12pt,twocolumn,tighten]{aastex63}
\usepackage{amsmath,amstext,amssymb}
\usepackage[T1]{fontenc}
\usepackage{apjfonts}
\usepackage[figure,figure*]{hypcap}
\usepackage{graphics,graphicx}
\usepackage{hyperref}
\usepackage{natbib}
\usepackage[caption=false]{subfig} 
\usepackage{enumitem} 
\usepackage{epigraph}

\newcommand{\tn}{TOI~837} 
\newcommand{\pn}{TOI~837b} 
\newcommand{\cn}{IC~2602} 

\newcommand{\kms}{\,km\,s$^{-1}$}

\newcommand{\stscilink}{\textsc{\url{archive.stsci.edu/hlsp/cdips}}}
\newcommand{\datasetlink}{\textsc{\dataset[doi.org/10.17909/t9-ayd0-k727]{https://doi.org/10.17909/t9-ayd0-k727}}}

\received{August 6, 2020}
\revised{September 6, 2020}
\accepted{September 16, 2020}
\submitjournal{AAS journals.}
\shorttitle{TOI~837 in IC~2602}

\begin{document}

\defcitealias{bouma_wasp4b_2019}{B19}

\title{
  Cluster Difference Imaging Photometric Survey. II.
  TOI~837: A Young Validated Planet in IC~2602
}

\correspondingauthor{L.\,G.\,Bouma}
\email{luke@astro.princeton.edu}

%
%
\author[0000-0002-0514-5538]{L. G. Bouma}
\affiliation{Department of Astrophysical Sciences, Princeton University, 4 Ivy Lane, Princeton, NJ 08540, USA}

\author[0000-0001-8732-6166]{J. D. Hartman}
\affiliation{Department of Astrophysical Sciences, Princeton University, 4 Ivy Lane, Princeton, NJ 08540, USA}

\author[0000-0002-9158-7315]{R. Brahm} 
\affiliation{Facultad de Ingenier\'{i}a y Ciencias, Universidad Adolfo Ib\'a\~nez, Av.\ Diagonal las Torres 2640, Pe\~nalol\'en, Santiago, Chile}
\affiliation{Millennium Institute for Astrophysics, Chile}

\author[0000-0002-5674-2404]{P. Evans}
\affiliation{El Sauce Observatory, Coquimbo Province, Chile}

\author[0000-0001-6588-9574]{K. A. Collins} 
\affiliation{Center for Astrophysics \textbar \ Harvard \& Smithsonian, 60 Garden St, Cambridge, MA 02138, USA}

\author[0000-0002-4891-3517]{G. Zhou}
\affiliation{Center for Astrophysics \textbar \ Harvard \& Smithsonian, 60 Garden St, Cambridge, MA 02138, USA}

\author[0000-0001-8128-3126]{P. Sarkis} 
\affiliation{Max-Planck-Institut f\"{u}r Astronomie, K\"onigstuhl 17, Heidelberg 69117, Germany }

\author[0000-0002-8964-8377]{S. N. Quinn} 
\affiliation{Center for Astrophysics \textbar \ Harvard \& Smithsonian, 60 Garden St, Cambridge, MA 02138, USA}

\author{J. de Leon}
\affiliation{Department of Astronomy, University of Tokyo, 7-3-1
Hongo, Bunkyo-ky, Tokyo 113-0033, Japan}

\author[0000-0002-4881-3620]{J. Livingston}
\affiliation{Department of Astronomy, University of Tokyo, 7-3-1
Hongo, Bunkyo-ky, Tokyo 113-0033, Japan}

\author[0000-0003-2989-7774]{C. Bergmann}
\affiliation{Exoplanetary Science at UNSW, School of Physics, UNSW
Sydney, NSW 2052, Australia}
\affiliation{Deutsches Zentrum f\"ur Luft- und Raumfahrt, M\"unchener
Str. 20, 82234 Wessling, Germany}

\author[0000-0002-3481-9052]{K. G. Stassun}
\affiliation{Vanderbilt University, Department of Physics \& Astronomy, 6301 Stevenson Center Lane, Nashville, TN 37235, USA}
\affiliation{Fisk University, Department of Physics, 1000 17th Avenue N., Nashville, TN 37208, USA}
%

%
%
\author[0000-0002-0628-0088]{W. Bhatti}
\affiliation{Department of Astrophysical Sciences, Princeton University, 4 Ivy Lane, Princeton, NJ 08540, USA}
\author[0000-0002-4265-047X]{J. N. Winn}
\affiliation{Department of Astrophysical Sciences, Princeton University, 4 Ivy Lane, Princeton, NJ 08540, USA}
\author[0000-0001-7204-6727]{G. \'A. Bakos}
\affiliation{Department of Astrophysical Sciences, Princeton University, 4 Ivy Lane, Princeton, NJ 08540, USA}

%
%

\author{L. Abe} 
\affiliation{Universit\'e C\^{o}te d'Azur, Observatoire de la C\^ote d'Azur, CNRS, Laboratoire Lagrange, Bd de l'Observatoire, CS 34229, 06304 Nice cedex 4, France}


\author[0000-0001-7866-8738]{N. Crouzet} 
\affiliation{European Space Agency, European Space Research and Technology Centre (ESA/ESTEC), Keplerlaan 1, 2201 AZ Noordwijk, The Netherlands}

\author[0000-0002-3937-630X]{G. Dransfield} 
\affiliation{School of Physics \& Astronomy, University of Birmingham, Edgbaston, Birmingham B15 2TT, United Kingdom}

\author[0000-0002-7188-8428]{T. Guillot} 
\affiliation{Universit\'e C\^{o}te d'Azur, Observatoire de la C\^ote d'Azur, CNRS, Laboratoire Lagrange, Bd de l'Observatoire, CS 34229, 06304 Nice cedex 4, France}

\author{W. Marie-Sainte} 
\affiliation{Institut Paul \'{E}mile Victor, Concordia Station, Antarctica}

\author{D. M\'ekarnia} 
\affiliation{Universit\'e C\^{o}te d'Azur, Observatoire de la C\^ote d'Azur, CNRS, Laboratoire Lagrange, Bd de l'Observatoire, CS 34229, 06304 Nice cedex 4, France}

\author[0000-0002-5510-8751]{A. H.M.J. Triaud} 
\affiliation{School of Physics \& Astronomy, University of Birmingham, Edgbaston, Birmingham B15 2TT, United Kingdom}

%
%

\author[0000-0002-7595-0970]{C.~G.~Tinney}
\affiliation{Exoplanetary Science at UNSW, School of Physics, UNSW Sydney, NSW 2052, Australia}

%
%

\author{T. Henning} 
\affiliation{Max-Planck-Institut f\"{u}r Astronomie, K\"onigstuhl 17, Heidelberg 69117, Germany }

\author[0000-0001-9513-1449]{N. Espinoza} 
\affiliation{Space Telescope Science Institute, 3700 San Martin Drive, Baltimore, MD 21218, USA}

\author[0000-0002-5389-3944]{A. Jord\'an} 
\affiliation{Facultad de Ingenier\'{i}a y Ciencias, Universidad Adolfo Ib\'a\~nez, Av.\ Diagonal las Torres 2640, Pe\~nalol\'en, Santiago, Chile}
\affiliation{Millennium Institute for Astrophysics, Chile}

\author{M. Barbieri} 
\affiliation{INCT, Universidad de Atacama, calle Copayapu 485, Copiap\'o, Atacama, Chile}

\author{S. Nandakumar} 
\affiliation{INCT, Universidad de Atacama, calle Copayapu 485, Copiap\'o, Atacama, Chile}

\author{T. Trifonov} 
\affiliation{Max-Planck-Institut f\"{u}r Astronomie, K\"onigstuhl 17, Heidelberg 69117, Germany }

\author[0000-0002-1896-2377]{J.~I.~Vines} 
\affiliation{Departamento de Astronom\'ia, Universidad de Chile, Camino El Observatorio 1515, Las Condes, Santiago, Chile}

\author{M. Vuckovic} 
\affiliation{Instituto de F\'isica y Astronom\'ia, Universidad de Vapara\'iso, Casilla 5030, Valpara\'iso, Chile}

%
%
\author[0000-0002-0619-7639]{C.~Ziegler} 
\affiliation{Dunlap Institute for Astronomy and Astrophysics, University of Toronto, 50 St. George Street, Toronto, Ontario M5S 3H4, Canada}


\author{N.~Law} 
\affiliation{Department of Physics and Astronomy, The University of North Carolina at Chapel Hill, Chapel Hill, NC 27599-3255, USA}

\author[0000-0003-3654-1602]{A.~W.~Mann} 
\affiliation{Department of Physics and Astronomy, The University of North Carolina at Chapel Hill, Chapel Hill, NC 27599-3255, USA}

%
%
\author{G. R. Ricker} 
\affiliation{Department of Physics and Kavli Institute for Astrophysics and Space Research, Massachusetts Institute of Technology, Cambridge, MA 02139, USA}
%
\author[0000-0001-6763-6562]{R. Vanderspek} 
\affiliation{Department of Physics and Kavli Institute for Astrophysics and Space Research, Massachusetts Institute of Technology, Cambridge, MA 02139, USA}
%
%
\author{S. Seager} 
\affiliation{Department of Earth, Atmospheric, and Planetary Sciences, Massachusetts Institute of Technology, Cambridge, MA 02139, USA}
%
\author[0000-0002-4715-9460]{J. M.~Jenkins} 
\affiliation{NASA Ames Research Center, Moffett Field, CA 94035, USA}

%

\author[0000-0002-7754-9486]{C.~J.~Burke} 
\affiliation{Department of Physics and Kavli Institute for Astrophysics and Space Research, Massachusetts Institute of Technology, Cambridge, MA 02139, USA}

\author[0000-0003-2313-467X]{D.~Dragomir}
\affiliation{Department of Physics and Astronomy, University of New Mexico, Albuquerque, NM, USA}
	


\author[0000-0001-8172-0453]{A.~M.~Levine} 
\affiliation{Department of Physics and Kavli Institute for Astrophysics and Space Research, Massachusetts Institute of Technology, Cambridge, MA 02139, USA}

\author{E.~V.~Quintana} 
\affiliation{NASA Goddard Space Flight Center, 8800 Greenbelt Road, Greenbelt, MD 20771, USA}

\author[0000-0001-8812-0565]{J.~E.~Rodriguez}
\affiliation{Center for Astrophysics \textbar \ Harvard \& Smithsonian, 60 Garden St, Cambridge, MA 02138, USA}

\author[0000-0002-6148-7903]{J. C. Smith} 
\affiliation{NASA Ames Research Center, Moffett Field, CA 94035, USA}
\affiliation{SETI Institute, Mountain View, CA 94043, USA}

\author[0000-0002-5402-9613]{B. Wohler} 
\affiliation{NASA Ames Research Center, Moffett Field, CA 94035, USA}
\affiliation{SETI Institute, Mountain View, CA 94043, USA}

\begin{abstract}
  We report the discovery of TOI 837b and its validation as a
  transiting planet.  We characterize the system using data from the
  NASA TESS mission, the ESA Gaia mission, ground-based photometry
  from El Sauce and ASTEP400, and spectroscopy from CHIRON, FEROS, and
  Veloce.  We find that TOI 837 is a $T=9.9$ mag G0/F9 dwarf in the
  southern open cluster IC 2602.  The star and planet are therefore
  $35^{+11}_{-5}$ million years old.  Combining the transit photometry
  with a prior on the stellar parameters derived from the cluster
  color-magnitude diagram, we find that the planet has an orbital
  period of $8.3\,{\rm d}$ and is slightly smaller than Jupiter
  ($R_{\rm p} = 0.77^{+0.09}_{-0.07} \,R_{\rm Jup}$).  From radial
  velocity monitoring, we limit $M_{\rm p}\sin i$ to less than 1.20
  $M_{\rm Jup}$ (3-$\sigma$).  The transits either graze or nearly
  graze the stellar limb.  Grazing transits are a cause for concern,
  as they are often indicative of astrophysical false positive
  scenarios.  Our follow-up data show that such scenarios are
  unlikely.  Our combined multi-color photometry, high-resolution
  imaging, and radial velocities rule out hierarchical eclipsing
  binary scenarios.  Background eclipsing binary scenarios, though
  limited by speckle imaging, remain a 0.2\% possibility.  TOI 837b is
  therefore a validated adolescent exoplanet.  The planetary nature of
  the system can be confirmed or refuted through observations of the
  stellar obliquity and the planetary mass.  Such observations may
  also improve our understanding of how the physical and orbital
  properties of exoplanets change in time.
\end{abstract}



\section{Introduction}

Over the first 100 million years of their lives, exoplanet systems are
expected to undergo major physical and dynamical changes.  For a
typical Sun-like star, the protoplanetary disk disperses within
roughly 1--10 million years
\citep{mamajek_initial_2009,fedele_timescale_2010,dullemond_inner_2010,williams_protoplanetary_2011}.
Gas giants presumably finish accreting before the end of disk
dispersal \citep{pollack_formation_1996}.  While rocky planets may
form within only a few million years \citep{dauphas_hf-w-th_2011},
they can also undergo significant growth over the next 10--100 million
years through giant impacts \citep[{\it
e.g.},][]{kleine_hf-w_2009,konig_earths_2011,morbidelli_building_2012,raymond_terrestrial_2014}.
The Moon, for instance, may have formed from debris ejected during a
collision between the proto-Earth and a planetesimal during Earth's
first 100 million years
\citep{cameron_origin_1976,canup_origin_2001,touboul_late_2007}.

A number of other processes are expected to shape young exoplanets.
After accreting, planets with gaseous envelopes are thought to cool
and contract, and their atmospheres are expected to undergo a mix of
photoevaporation and core-powered mass loss \citep[{\it
e.g.},][]{Fortney_et_al_2007,Owen_Wu_2013,Fulton_et_al_2017,gupta_sculpting_2019,gupta_signatures_2020}.
Predicted timescales for photoevaporation and core-powered mass loss
range from 10 million years to over 1 gigayear for typical transiting
sub-Neptunes
\citep{ginzburg_superearth_2016,owen_evaporation_2017,king_euv_2020}.
The relative importance of each process is set by the planetary
surface gravity and the radiation environment.  Both processes can be
directly observed \replaced{through observations of planetary winds
in}{in favorable cases using} the metastable 1083$\,$nm He line
\citep{spake_helium_2018,oklopcic_new_2018,mansfield_detection_2018}.

Beyond physical changes, dynamical changes are expected in the
semi-major axes, eccentricities, and stellar obliquities of young
planets.  When the gas disk is present, the planetary semi-major axis
is thought to change in step with the viscous evolution of the disk
\citep{lin_orbital_1996}.  High-eccentricity migration processes
including planet-planet scattering, secular chaos, and Kozai-Lidov
oscillations can also occur \citep[{\it
e.g.},][]{fabrycky_shrinking_2007,chatterjee_dynamical_2008,lithwick_secular_2014}.
The circularization timescale is thought to be such that for any giant
planets that \replaced{migrated early}{do migrate early}, their orbits
should circularize within 100 million years
\citep{zahn_tidal_1977,bonomo_gaps_2017}.

Finding and understanding systems undergoing these evolutionary
changes is a major goal in contemporary exoplanet research.  To
identify stars younger than say 1 \replaced{Gyr}{gigayear}, a number
of direct and indirect methods are available
\citep{soderblom_ages_2010}.  The traditional approach is to
isochronally age-date coeval groups of stars, hereafter ``clusters''
\citep[{\it
e.g.},][]{lada_embedded_2003,zuckerman_young_2004,krumholz_star_2019}.
Young field stars can also be identified isochronally, provided that
they are sufficiently massive \citep{berger_gaia-kepler_2020}.  Other
age indicators include stellar rotation periods, the abundance of
photospheric lithium, and chromospheric diagnostics such as calcium
emission and broadband UV emission.  Studies by, for instance,
\citet{sanchis-ojeda_kepler-63b_2013}, \citet{david_discovery_2018},
and G{.}~Zhou et al{.}~(2020, submitted) have combined these methods
to age-date individual field stars hosting transiting planets.  Many
of these latter methods were summarized by
\citet{mamajek_improved_2008}, and have since been calibrated by, {\it
e.g.},
\citet{irwin_rotational_2009,barnes_color-period_2015,meibom_spin-down_2015,angus_calibrating_2015}
and \citet{curtis_tess_2019} for stellar rotation,
\citet{zerjal_chromospherically_2017} for chromospheric activity, and
{\it e.g.}, \citet{berger_identifying_2018} and
\citet{zerjal_galah_2019} for lithium abundances.

\explain{NOTE: removed footnote to Damasso et al 2020 in prep, since
it has now been published, and added note to Donati et al 2020.}
\replaced{A}{To date, a} few dozen planets in clusters have been
detected, and fewer still have been closely characterized.  Despite
the challenges of starspot-induced radial velocity (RV) variations, RV
surveys found early success in the Hyades, NGC~\replaced{2523}{2423},
Praesepe, and M~67
\citep{Sato_et_al_2007,lovis_mayor_2007,Quinn_et_al_2012,Malavolta_et_al_2016,brucalassi_search_2017}.
RV surveys of highly active pre-main sequence stars in Taurus also led
to the youngest hot Jupiters yet reported orbiting V830~Tau, TAP~26,
and CI~Tau
\citep{donati_hj_2016,johns-krull_candidate_2016,yu_hot_2017,biddle_k2_2018,flagg_co_2019}.
\added{The planetary nature of at least two of these signals has been
debated \citep{donati_magnetic_2020,damasso_gaps_2020}.}

The transit method was comparatively slow to catch up.  Early deep
transit searches of open clusters by many groups did not yield
definitive planet detections
\citep{mochejska_planets_2005,mochejska_planets_2006,burke_survey_2006,aigrain_monitor_2007,irwin_monitordata_2007,miller_monitor_2008,pepper_photometric_2008,hartman_MMT_IV_2009}.
These searches were typically sensitive to planets larger than
Jupiter, on $\lesssim 3$ day orbital periods.  Hot Jupiter occurrence
rate limits were derived at the $\lesssim 5\%$ level \citep[{\it
e.g.},][]{burke_survey_2006,hartman_MMT_IV_2009}.  The modern 0.5-1\%
occurrence rate suggests that these early transit surveys would have
needed a greater data volume at higher precision for detection to be
possible
\citep{mayor_harps_2011,wright_frequency_2012,howard_planet_2012,petigura_metallicity_2018}.

Kepler observed a large enough number of stars with sufficient
baseline and precision to detect transiting planets in open clusters:
Kepler-66b and 67b, in the gigayear-old NGC~6811
\citep{borucki_kepler_2010,Meibom_et_al_2013}.  Though a broken
reaction wheel ended the prime Kepler mission, the repurposed K2
\citep{howell_k2_2014} switched between fields along the ecliptic
every quarter-year, and was able to observe far more clusters and
young stars.

The discoveries made by K2 through its surveys of Taurus, the Hyades,
Praesepe, and Upper Sco were a major inspiration for the present work
\citep[{\it
e.g.},][]{Mann_K2_25_2016,obermeier_k2_2016,Mann_et_al_2017,vanderburg_zeitVII_2018,ciardi_k2-136_2018,livingston_three_2018,mann_ZEITVI_2018,rizzuto_zeitVIII_2018,livingston_k2-264_2019}.
Observations with K2 convincingly showed that at least some close-in
planets must form within about 10 Myr
\citep{Mann_K2_33b_2016,David_et_al_2017}.  They also led to the first
hints that young planets in clusters may in fact be qualitatively
different from their field counterparts.  For instance, based on its
observed mass, radius, and UV environment, the 700$\,$Myr K2-100b is
probably actively losing its atmosphere, and should become a bare
rocky planet over the next few hundred Myr
\citep{Mann_et_al_2017,barragan_radial_2019}.  The four transiting
planets around V1298~Tau (23$\,$Myr) are also likely to be
photoevaporating, and could represent a precursor to Kepler's compact
multiple systems \citep{david_four_2019,david_warm_2019}.

To advance the young planet census, we have been using data from the
TESS spacecraft \citep{ricker_transiting_2015} to perform a Cluster
Difference Imaging Photometric Survey
\citep[CDIPS;][]{bouma_cluster_2019}.  Our targets in this survey are
candidate young stars that have been reported in the literature.  At
the time of writing, $\sim$$6\times10^5$ light curves from Year 1 of
TESS had been created and were available through
MAST\footnote{\stscilink}, and via \datasetlink.  Searching through a
subset of these light curves brought our attention to the candidate
transiting planet, \pn, that is the subject of this analysis.

The transits of \pn\ are grazing the stellar limb, which is a cause
for concern.  Particularly for a star near the galactic plane
($b=-5.8^\circ$), background eclipsing binaries are a major source of
astrophysical false positives \citep[{\it
e.g.},][Figure~30]{Sullivan_2015}.  Our follow-up data showed that
this and related scenarios were unlikely to the degree that we could
validate the planet, {\it i.e.}, determine that its probability of
being an astrophysical false positive was small.  We considered this
result worth reporting because of the planet's youth.

Section~\ref{sec:observations} describes the identification of the
candidate, and our follow-up observations.
Section~\ref{sec:validation} combines the available data to assess the
system's false positive probability, and validates \pn\ as a planet.
Section~\ref{sec:system} presents our knowledge of the cluster
(Section~\ref{subsec:cluster}), the star (Section~\ref{subsec:star})
and the planet (Section~\ref{subsec:planet}).  We conclude by
discussing avenues for confirmation and improved characterization in
Section~\ref{sec:discussion}.

\section{Identification and Follow-up Observations}
\label{sec:observations}

\subsection{TESS Photometry}
\label{subsec:tess}

\begin{figure*}[!t]
	\begin{center}
		\leavevmode
		\subfloat{
			\includegraphics[width=0.77\textwidth]{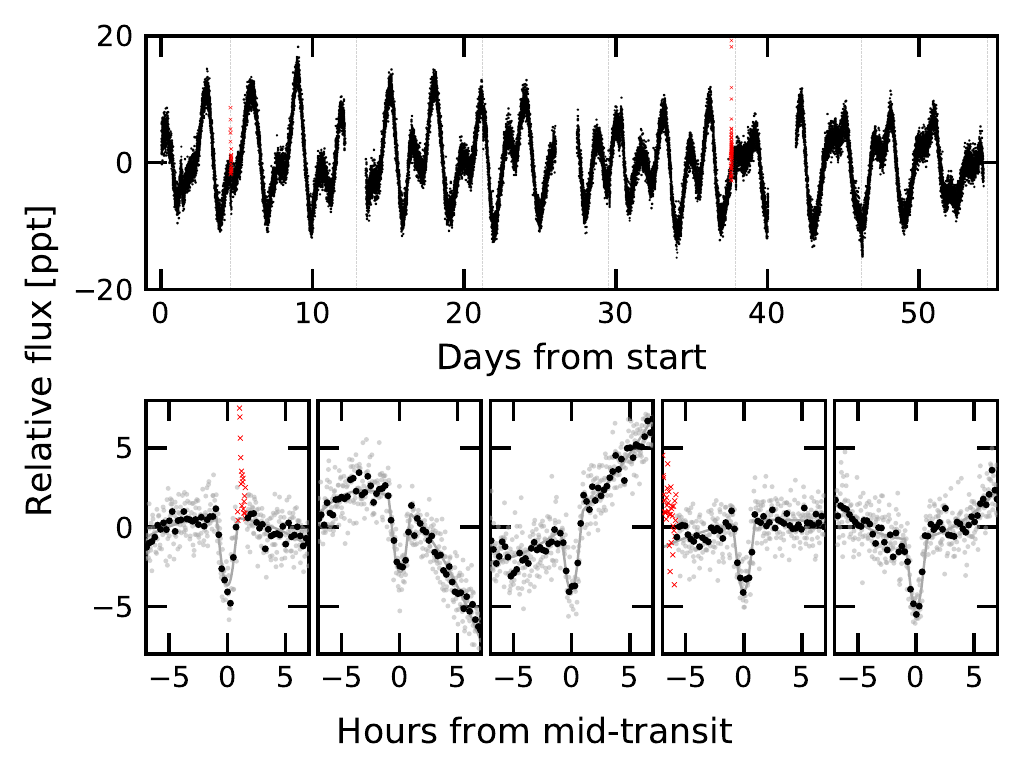}
		}
		
		\vspace{-0.5cm}
		\subfloat{
			\includegraphics[width=0.76\textwidth]{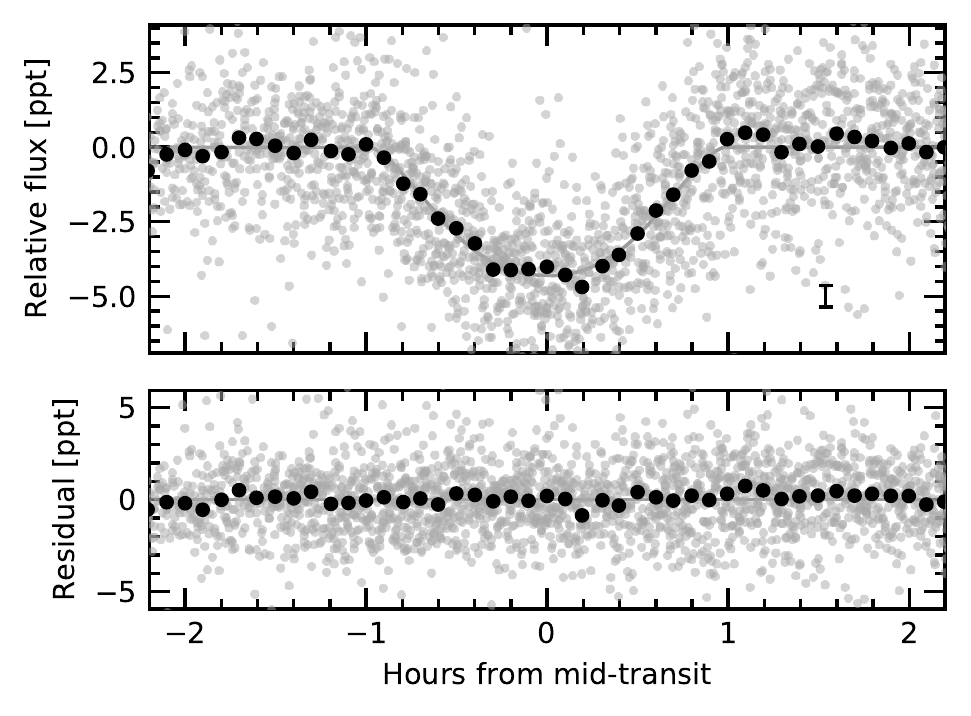}
		}
	\end{center}
	\vspace{-0.6cm}
	\caption{
    {\bf Light curves of \tn.}
    {\it Top}:
    TESS \texttt{PDCSAP} median-subtracted relative flux at 2-minute
    sampling in units of parts-per-thousand ($\times 10^{-3}$).
    Starspot-induced variability is the dominant signal; flares are
    shown with red crosses.  Dashed lines indicate the five transits
    observed by TESS over Sectors 10 and 11.  {\it Middle}: Individual
    TESS transits.  Gray lines are the best-fit model to the TESS
    \added{and ground-based data}, which includes a local quadratic
    trend for each transit. Gray points are 2-minute \texttt{PDCSAP}
    flux measurements, black points are binned to 15-minute intervals.
    {\it Bottom}: Phase-folded TESS and ground-based transits.
    \added{Section~\ref{subsec:groundphot} presents the ground-based
    data.} Gray points are flux measurements with the local
    spot-induced variation removed.  A weighted binning at 6-minute
    intervals yields the black points.  The black error-bar shows the
    median uncertainty for the black points.  The gray line is the
    best-fit model for the entire dataset.
		\label{fig:thephot}
	}
\end{figure*}

\tn\ was observed by TESS from 26 March 2019 until 20 May 2019, during
Sector 10 and Sector 11 of science operations
\citep{ricker_transiting_2015}.  The star was designated
TIC\,460205581 in the TESS Input Catalog
\citep{stassun_TIC_2018,stassun_TIC8_2019}.  Pixel data for an
$11\times11$ array surrounding the star were co-added and saved at
2-minute cadence.  The $2048\times2048$ image from the entire CCD was
also co-added into 30-minute stacks, and saved as a ``full frame
image'' (FFI).

The TESS Science Processing Operations Center
\citep{jenkins_tess_2016} processed the image data and identified the
transiting planet signature from two transits in Sector 10, again with
three transits in Sector 11, and then for a final time when Sectors
1--13 were searched at the end of the first year of the mission. The
transit signature was fitted with a limb-darkened transit model
\citep{li_kepler_2019} and passed all the diagnostic tests
\citep{twicken_kepler_2018}, including the odd/even depth test, the
weak secondary eclipse test, \deleted{the ghost diagnostic test,} and
the difference image centroiding test, which placed the transit source
within $\sim$2 arcsec of the location of TOI~837. No additional
transit-like features were identified in any of the SPOC searches. The
TESS Science Office alerted the community to this candidate transiting
planet on 17 June 2019.  Our subsequent blind search of the CDIPS FFI
light curves also showed the transits\added{, as did that of
\citet{nardiello_pathosII_2020}}. Given that the 2-minute data had
better sampling cadence, we opted to use the Presearch Data
Conditioning (PDC) light curve with the default aperture for our
analysis~\citep{smith_kepler_2012,stumpe_multiscale_2014,smith_finding_2016}.

\added{The top panel of} Figure~\ref{fig:thephot} shows the
\added{TESS} data.  The dominant modulation induced by starspots
coming into and out of view has a peak-to-peak amplitude of about
2.3\%, and a period of about 3 days.  The dips are suggestive of a
grazing transiting planet, recurring roughly every 8 days with a depth
of about 0.4\%.  A few flares are also visible.  A phase-folded view
of the TESS transits \deleted{with starspot variability
removed}\added{combined with ground-based follow-up photometry} is
shown in \added{the bottom panel of }Figure~\ref{fig:thephot}.
\added{The ground-based data }and our fitting procedure \deleted{in
these plots is described in }\added{are discussed in
Sections~\ref{subsec:groundphot} and~\ref{subsec:planet}
respectively}. \added{First though, some prerequisite context on the
stellar neighborhood of \tn\ is needed.}

\subsection{Gaia Astrometry and Imaging}
\label{subsec:gaia}

\begin{figure}[!t]
	\begin{center}
		\leavevmode
		\includegraphics[width=0.48\textwidth]{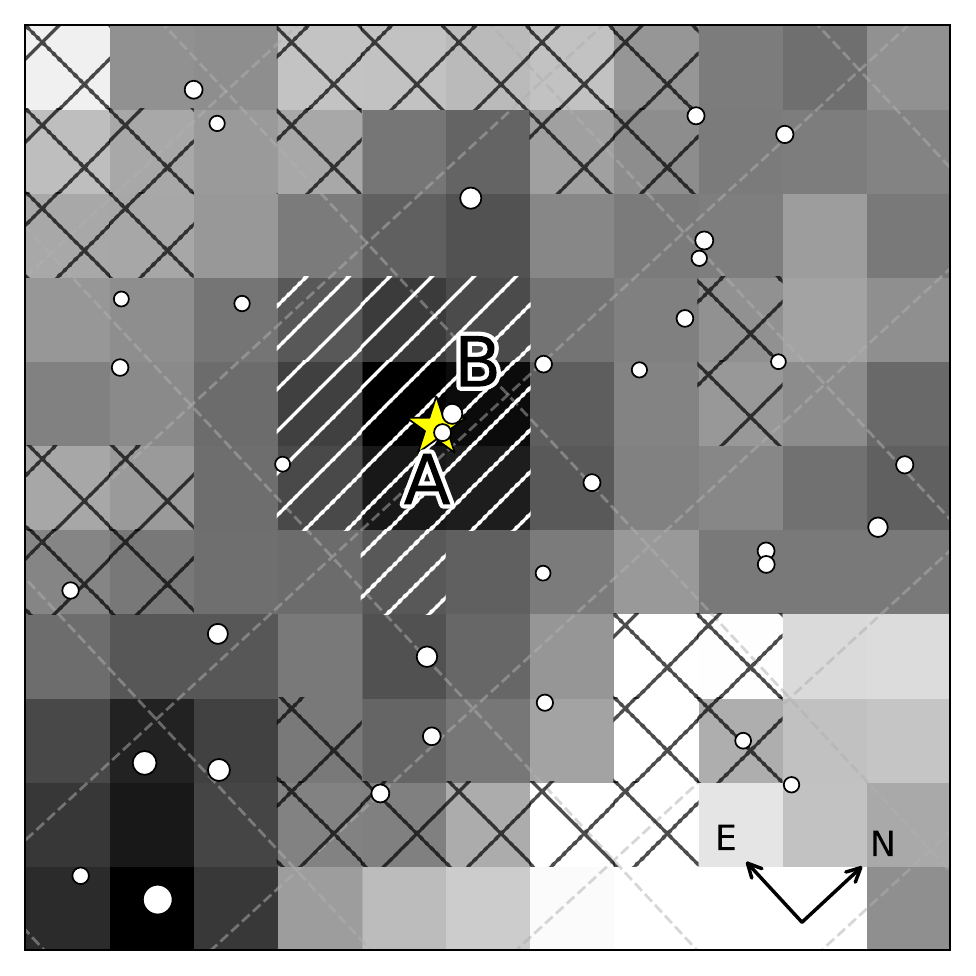}
	\end{center}
	\vspace{-0.6cm}
	\caption{ {\bf Scene of \tn.}
    Mean TESS image of \tn\ from Sector~10\added{ on an 11$\times$11
    pixel cutout}, with a logarithmic grayscale\added{ indicating the
    flux in each pixel}. The yellow star is the position of \tn.
    White circles are resolved Gaia sources with $T<16$\replaced{.
    Brighter stars are larger}{, with brighter stars being larger}.
    The black \texttt{X} and white \texttt{/} hatches show the
    apertures used to measure the background and target star flux,
    respectively. \replaced{Dashed lines of constant declination and
    right ascension are shown}{The compass shows cardinal directions
    in celestial coordinates. Dashed lines of constant declination are
    separated by 1$'$, while those of right ascension are separated by
    2$'$}.  Two stars of interest are ``Star A'' and ``Star B'', which
    were excluded as being possible sources of the transits.
		\label{fig:scene}
	}
\end{figure}

Between 25 July 2014 and 23 May 2016, the ESA Gaia satellite measured
about 300 billion centroid positions of 1{.}6 billion stars.  The
positions, proper motions, and parallaxes of the brightest 1{.}3
billion were calculated for the second data release (DR2)
\citep{gaia_collaboration_gaia_2016,lindegren_gaiasoln_2018,gaia_collaboration_gaia_2018}.
\tn\ was assigned the Gaia DR2 identifier 5251470948229949568, and had
276 ``good'' astrometric observations. Its brightness was measured in
the $G$, $Rp$, and $Bp$ bands of the Radial Velocity Spectrometer
\citep{cropper_gaia_2018,evans_gaia_2018}.  

The Gaia imaging, reduced to its point-source catalog, provides the
initial context for analyzing the TESS data.  Stars brighter than
$T=16$, as queried from the Gaia DR2 source catalog, are shown with
white circles in Figure~\ref{fig:scene}, overlaid on the TESS image.
Given its galactic latitude of $b=-6^\circ$, it is not surprising that
the field of \tn\ is crowded.  The resolved stars that were of
immediate concern for our false positive analysis were as follows.
\begin{itemize}
  \item \tn\ $\equiv$ TIC 460205581 ($T=9.9$). The target star.
  \item Star A $\equiv$ TIC 847769574 ($T=14.6$), $2.3''$ West. The
    proper motions and parallax of this star imply it is co-moving
    with \tn\ and that the two stars are separated by $6.6\pm
    0.1\,{\rm pc}$.  Star A is therefore likely an \cn\ member, but
    unlikely to be a bound binary companion.
  \item Star B $\equiv$ TIC 460205587 ($T=13.1$), $5.4''$ North.  The
    Gaia parallax implies this is a background giant star.
\end{itemize}
An additional source, TIC 847769581, is $4.9''$ from the target, but
too faint ($T=18.8$) to be the source of the observed transit signal.

The Gaia DR2 data for Star A seems poorly behaved.  While Star A has
$G=15.1$, and $Bp=14.9$, no $Rp$ magnitude is reported.
Correspondingly, no RUWE\footnote{ See the Gaia DPAC technical note
GAIA-C3-TN-LU-LL-124-01,
\url{http://www.rssd.esa.int/doc_fetch.php?id=3757412},
\texttt{2020-07-08}. } value is available.  We suspect that the
photometric failure to produce an $Rp$ magnitude as well as the poor
astrometric fit of this star are due to blending with \tn.

At the $\approx1'$ resolution of the TESS data, if either Star A or
Star B were eclipsing binaries, they could be the sources of the
transit signal.  A detailed analysis of ground-based seeing-limited
photometry was necessary to assess and rule out this possibility
(Section~\ref{subsec:groundphot} and Figure~\ref{fig:groundphot}).

\subsection{High-Resolution Imaging}
\label{subsec:speckle}

\begin{figure}[!t]
	\begin{center}
		\leavevmode
		\includegraphics[width=.44\textwidth]{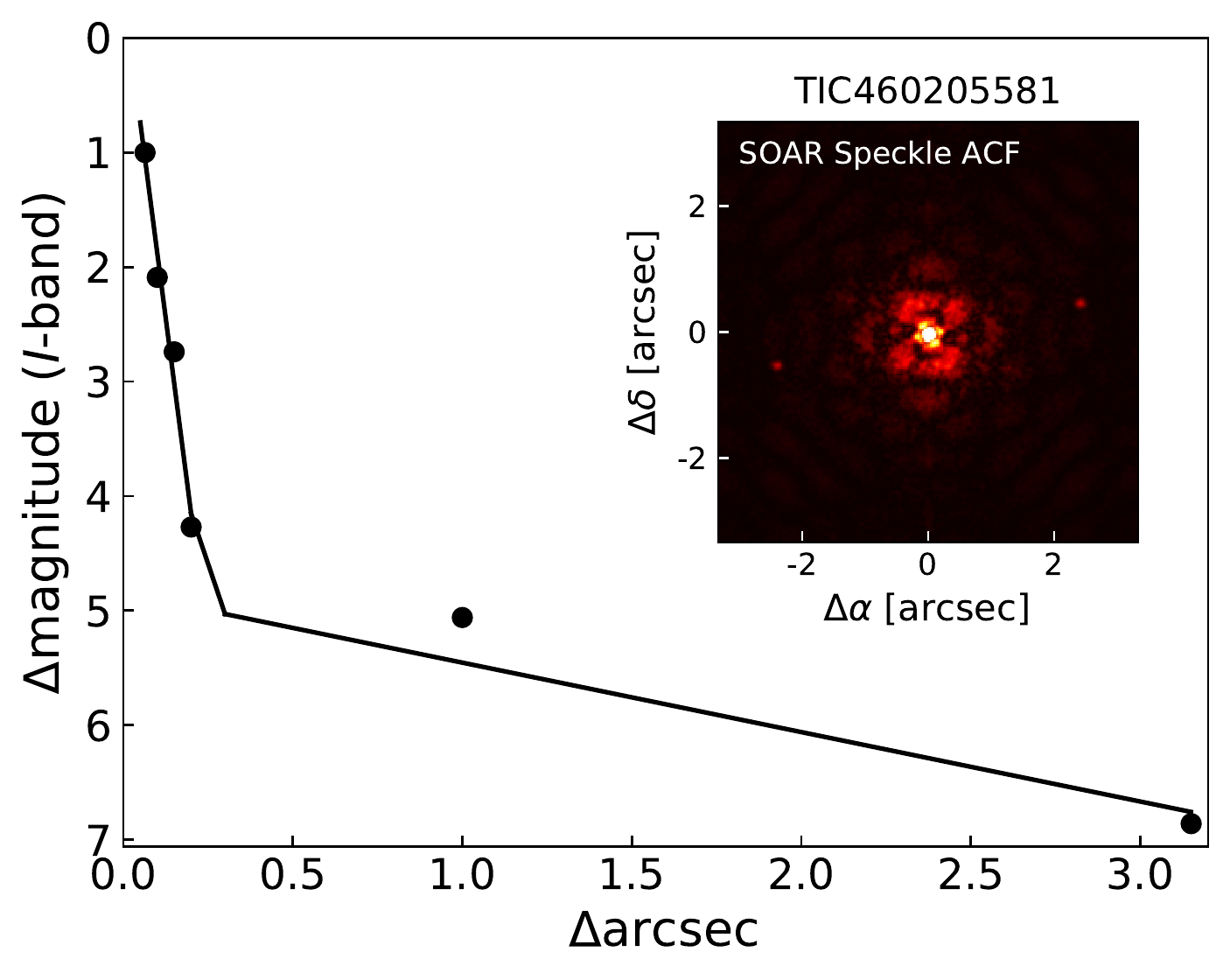}
	\end{center}
	\vspace{-0.7cm}
	\caption{
		{\bf Speckle-imaging of \tn.} Contrast limits from SOAR HRCam
		imaging were derived from point-source injection-recovery
		experiments. Star A ($\Delta T=4.7$, $2.3''$ West) is detected,
		and is also a resolved Gaia source.  It is co-moving with \tn, and
		its parallax and on-sky position imply that it is physically
		separated from \tn\ by $6.6\pm 0.1\,{\rm pc}$.
		\label{fig:speckle}
	}
\end{figure}

To determine if any fainter point sources existed closer to \tn\
inside of Gaia's point-source detection limits, we acquired
high-resolution speckle images. We then searched the autocorrelation
functions of these images for peaks indicative of nearby companions.

The observations of \tn\ were initially acquired by
\citet{ziegler_soar_2020} as part of the Southern Astrophysical
Research (SOAR) TESS Survey using the High Resolution Camera (HRCam;
\citealt{tokovinin_ten_2018}).  The HRCam $I$-band filter is described
by \citet{tokovinin_ten_2018}.  The points in Figure~\ref{fig:speckle}
show the resulting measured 5-$\sigma$ detectable contrasts.  The
lines are linear smoothing fits between the regimes of the diffraction
limit, the ``knee'' at $\approx 0.2''$, and the slow decrease until
$\approx 1.5''$, beyond which the speckle patterns become
de-correlated.  Star A (TIC 847769574) was detected at the expected
location and brightness contrast, and no additional companions were
found.  Star B was not detected; with a separation of $5.4''$ from
\tn, it fell outside the field of view.

\subsection{Ground-based Time-Series Photometric Follow-up}
\label{subsec:groundphot}

\begin{figure}[!t]
	\begin{center}
		\leavevmode
		\includegraphics[width=.48\textwidth]{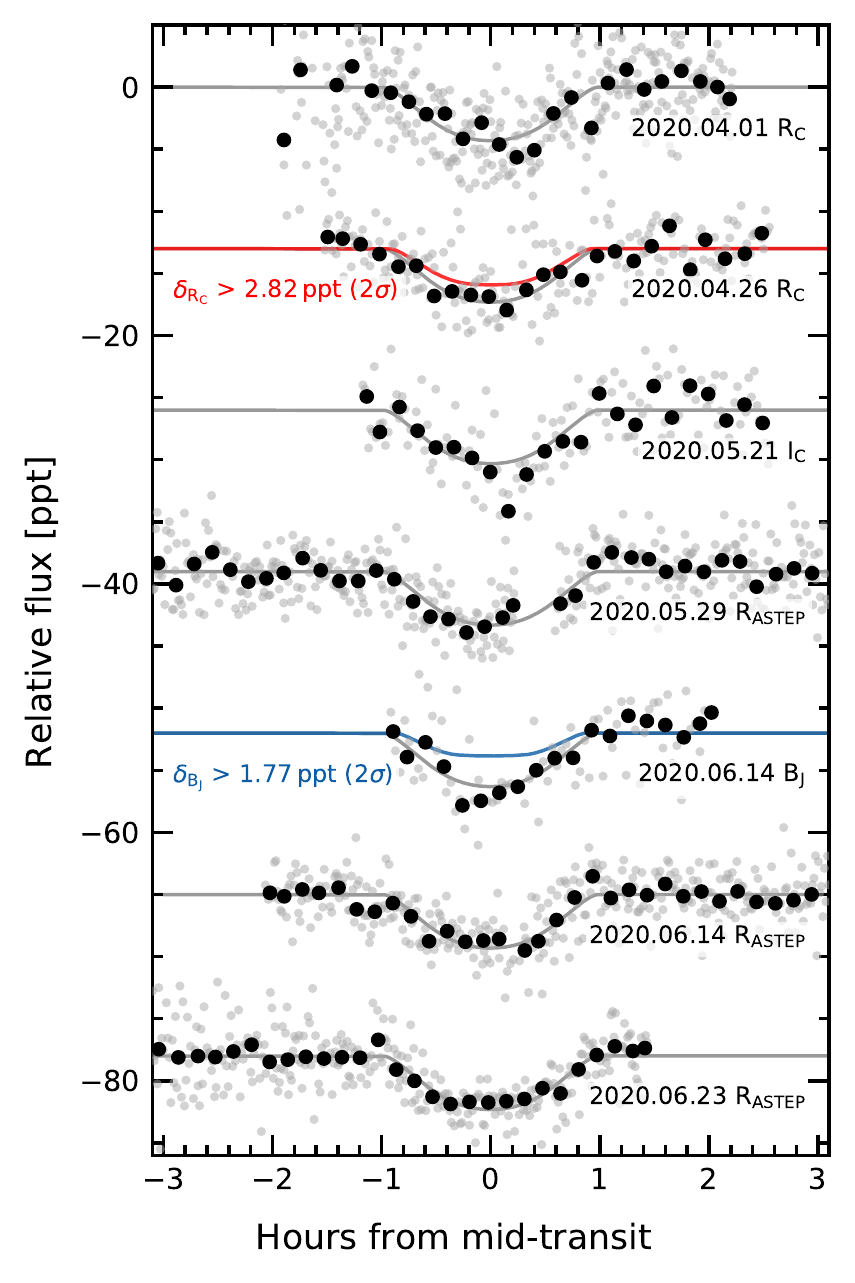}
	\end{center}
	\vspace{-0.7cm}
	\caption{
    {\bf Ground-based follow-up photometry.} The data were acquired
    using the 0.36$\,$m telescope at El Sauce and the 0.40$\,$m
    ASTEP400 telescope at Dome C.  Black points represent the
    measurements after binning at 10-minute intervals. The gray line
    is the model that best fits the combined TESS and ground-based
    data.  Red and blue lines show 2-$\sigma$ lower limits on the
    transit depths in the Cousins-R and Johnson-B bandpasses used to
    rule out specific false positive scenarios (see
    Section~\ref{subsec:colorphot}).
		\label{fig:groundphot}
	}
\end{figure}

We obtained ground-based seeing-limited time series photometric
observations of \tn\ bracketed around the times of transit.  These
observations confirmed that the transits occurred on-target to within
$\approx 2''$, and that they were achromatic. Both features are
essential for our ability to eliminate false-positive scenarios.

\subsubsection{El Sauce 0.36$\,$m}

\paragraph{Acquisition and reduction}
We observed four transits with the 0.36$\,$m telescope at Observatorio
El Sauce, located in the R\'io Hurtado Valley in Chile, and operated
by co-author P{.}~Evans.  The observations were obtained in Cousins-R
band on the nights of 1 April 2020 and 26 April 2020, Cousins-I band
on the night of 21 May 2020, and Johnson-B band on the night of 14
June 2020.  The final 14 June transit began shortly after twilight.

We scheduled our transit observations using the {\tt TESS Transit
Finder}, which is a customized version of the {\tt Tapir} software
package \citep{Jensen:2013}.  The photometric data were calibrated and
extracted using \texttt{AstroImageJ} \citep{collins_astroimagej_2017}.
Comparison stars of similar brightness were used to produce the final
light curves, each of which showed a roughly 4$\,$ppt dip near the
expected transit time.  The data are reported in Table~\ref{tab:phot}
and plotted in Figure~\ref{fig:groundphot}.

\paragraph{Custom aperture analysis}
Based solely on the TESS data, both Star A and Star B were possible
sources of blended eclipsing binary signals.  The typical FWHM for
stars in the El Sauce observations was $\approx 2.3''$, with a
variance of $\approx 0.2''$.  Star B is resolved in the 0.36$\,$m
images; Star A is not.

To rule out blend scenarios with the ground-based photometry, we
produced light curves centered on \tn\ with circular apertures of
radii ranging from 0.7$''$ to 5.1$''$.  We did not detect any
statistically significant variation in the depth of the transits with
aperture size.  \added{Beyond the difference image centroiding test
performed by the SPOC pipeline}, \replaced{Two lines of evidence}{two
additional lines of evidence} rule out Star B as the eclipsing source:
first, the transits were detected in the smallest apertures.  Second,
we made light curves with $2.1''$ apertures centered on Star B, and
they did not show the transit.

To assess the possibility that Star A is an eclipsing body, we created
light curves with a custom set of circular apertures with radii of
$2.1''$ and positions ranging from Star A (2.3$''$ West of \tn) to
2.3$''$ East of \tn.  We did not detect any variation of the transit
depth along this line of light curves. The apertures East of \tn\
exclude over 90\% of the flux from Star A.  The eclipse on Star A
would therefore need to have depth greater than unity to produce the
observed eclipse depth.  We therefore interpret the lack of asymmetry
between the Western-most (centered on Star A) and Eastern-most
(furthest from Star A) light curves as conclusive evidence that \tn\
is the source of the transit signal to within $\approx2.0''$.  To
verify self-consistency, we checked that the maximum dilution from
Star A ($\approx 1\%$) is less than the uncertainty of the transit
depth measurements ($\approx 15\%$), and so the lack of variation of
transit depth with aperture location is consistent with \tn\ being the
source of transits.

\added{An additional line of evidence for Star A not being the
transit host was also noted by the referee.  The $G$-band magnitude
and the parallax suggest that Star A is an M dwarf. As a probable
cluster member, it would have $Bp - Rp \approx 2.8$ (see
Section~\ref{subsec:hr}), which corresponds roughly to a mass in the
range of 0.15--0.45$\,M_\odot$, or densities roughly in the range of
2--3$\,{\rm g}\,{\rm cm}^{-3}$ based on the PARSEC isochrones
\citep{bressan_parsec_2012,chen_improving_2014,chen_parsec_2015,marigo_new_2017}.
These densities are inconsistent with those inferred from the transit
fits in Section~\ref{subsec:planet}.}

\subsubsection{ASTEP400}

We observed three transits with the 0.40$\,$m ASTEP telescope at the
Concordia base on the Antarctic Plateau \citep{daban_astep_2010}.
Concordia base is operated by the French and Italian polar institutes,
IPEV and PNRA.  Its position on the Antarctic Plateau allows it to
take advantage of the continuous night during Austral winter.  The
weather is of photometric quality for about two-thirds of each winter
\citep{crouzet_four_2018}.

ASTEP is equipped with a FLI Proline science camera with a KAF-16801E,
$4096\times4096$ front-illuminated CCD. The camera has an image scale
of 0.93\,arcsec pixel$^{-1}$ resulting in a $1^{\circ}\times1^{\circ}$
corrected field of view. The focal instrument dichroic plate splits
the beam into a blue wavelength channel for guiding, and a
non-filtered red science channel roughly matching a Cousins-R
transmission curve
\citep{abe_secondary_2013,mekarnia_transiting_2016}.  The images were
processed on-site using an automated aperture  photometry pipeline
based on the \texttt{daophot} package of the IDL astronomy user's
library \citep{landsman_1995}.

\tn\ was observed with ASTEP on 12 May 2020, 29 May 2020, 14 June
2020, and 23 June 2020 (UT).  Except for 12 May, our observations were
conducted under stable weather conditions, with clear skies,
temperatures of about $-70^\circ$C, and wind speeds less than
5\,m$\,$s$^{-1}$. Due to their poor quality, we exclude from the
analysis all data collected on 12 May. We found that the optimal
calibrated light curves of \tn\ correspond to an 11 pixel (10\,arcsec)
and 14 pixel (12\,arcsec) radius aperture for observations carried out
on June and May, respectively.  The data are reported in
Table~\ref{tab:phot}, and plotted in Figure~\ref{fig:groundphot}.

\startlongtable
\begin{deluxetable}{cccc}
    

\tabletypesize{\scriptsize}


\tablecaption{Ground-based \tn\ photometry.}
\label{tab:phot}


\tablehead{
  \colhead{Time [BTJD$_\mathrm{TDB}$]} &
  \colhead{Rel{.} Flux} &
  \colhead{Rel{.} Flux Err{.}} & 
  \colhead{Instrument}
}

\startdata
 1940.487018 & 0.999998 & 0.002430 & El Sauce \\
 1998.700107 & 0.998675 & 0.001621 & ASTEP \\
\enddata


\tablecomments{
Table~\ref{tab:phot} is published in its entirety in a
machine-readable format.  Two example entries are shown for guidance
regarding form and content.  To convert from BTJD to BJD, add
2{,}457{,}000.  See \citet{eastman_achieving_2010} for descriptions of
the barycentric and leap second corrections.
}
\vspace{-0.5cm}
\end{deluxetable}

\subsection{Spectroscopic Follow-up}
\label{subsec:spectra}

\begin{figure}[!t]
	\begin{center}
		\leavevmode
		\includegraphics[width=.48\textwidth]{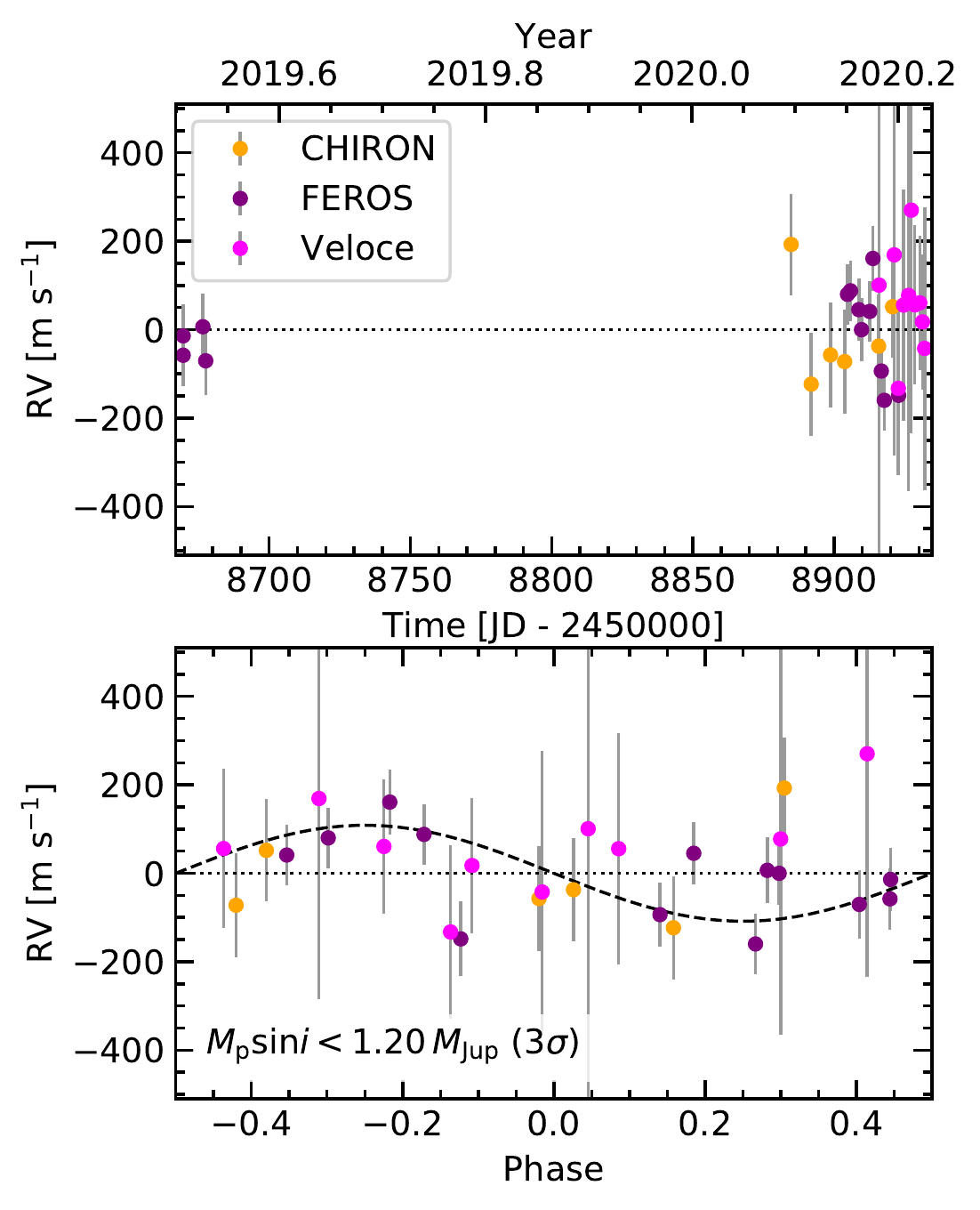}
	\end{center}
	\vspace{-0.7cm}
	\caption{
    {\bf Velocimetry of \tn.} {\it Top:} Radial velocity (RV)
    measurements, with best-fit instrument offsets and jitter terms
    included.  The expected scatter from starspots based on $v\sin i$
    and the photometric modulation amplitude is
    $\sim$300$\,$m$\,$s$^{-1}$.  {\it Bottom:} RV measurements phased
    to the orbital ephemeris of \pn.  The planet is not detected.  The
    dashed black line shows a circular Keplerian orbit representing
    the 3-$\sigma$ upper mass limit.
    \label{fig:rvs}
	}
\end{figure}

Reconnaissance spectroscopic follow-up is an essential step in vetting
planet candidates.  Medium to high-resolution spectra enable physical
characterization of the star and therefore planet.  Reducing multiple
spectra to radial velocities can enable planet mass measurements, and
can also lead to limits on the mass of nearby companions.  Finally, if
there are close or bright companions, reconnaissance spectra can also
reveal the presence of a secondary set of stellar lines.

\subsubsection{SMARTS 1.5$\,$m / CHIRON}
\label{subsec:chiron}

We acquired nine spectra using CHIRON at the SMARTS 1.5$\,$m telescope
at Cerro Tollo Inter-American Observatory, Chile
\citep{tokovinin_chironfiber_2013}.  Six met our signal-to-noise
requirements for radial velocity measurements and stellar parameter
extraction.  We used CHIRON in its image slicer configuration,
yielding a spectral resolution of $\approx 79{,}000$ across
415--880$\,$nm.

We derived radial velocities and spectroscopic line profiles from the
CHIRON observations using a least-squares deconvolution of the spectra
against non-rotating synthetic spectral templates
\citep{donati_1997_spectropolarimetric}. The spectral templates were
generated using ATLAS9 atmosphere models \citep{Castelli:2004} with
the \texttt{SPECTRUM} script \citep{gray_1994_spectrum}. These line
profiles were fitted with a broadening kernel that describes the
rotational, radial-tangential macroturbulent, and instrumental
broadening of the spectrum. The rotational and macroturbulent
broadening are computed per \citet{gray_2005_book}, following the
methods described in \citet{zhou_2018_hd106315obliq}. We fitted the
line profile from each observation independently, yielding the radial
velocities listed in Table~\ref{tab:rvs} and shown in
Figure~\ref{fig:rvs}. We found a mean rotational broadening velocity
of $v\sin I_\star = 16.2 \pm 1.1 \mathrm{km\,s}^{-1}$, and a
macroturbulent broadening of $v_\mathrm{mac} = 8.4 \pm 2.9
\mathrm{km\,s}^{-1}$.

To derive the stellar parameters, we matched the set of CHIRON spectra
against a library of observed spectra, previously obtained using the
Tillinghast Reflect Echelle Spectrograph
\citep[TRES,][]{furesz_tres_2008} on the 1.5\,m reflector at the Fred
Lawrence Whipple Observatory (FLWO), Arizona, USA, and classified
using the Stellar Parameter Classification pipeline
\citep{buchhave_hatp16b_class_2010}.  We found the best matching
stellar parameters to be $T_\mathrm{eff} = 5899 \pm 55$K, $\log g =
4.496 \pm 0.011$ dex, and $\mathrm{[Fe/H]} = -0.069 \pm 0.042$ dex.
We ultimately adopted a different set of stellar parameters for our
analysis (see Section~\ref{subsec:starparams}).

The spectroscopic line profiles were thoroughly examined for any signs
of secondary lines that might indicate the presence of another star,
either associated or in chance alignment with TOI 837. No such set of
lines was found. To set limits on the contributions of a close-by star
to the observed spectrum, we injected a \replaced{second set of lines
to the mean least-squares deconvolution profile}{secondary signal into
the mean least-squares deconvolution profile} derived from the CHIRON
observations.  The injection spanned 10,000 different combinations of
line broadening, velocity separation, and flux ratio $F_2/F_1$.  The
recovery results showed that for rotational broadenings of the
secondary of $5$, $15$, and $25\,\mathrm{km\,s}^{-1}$, we were able to
exclude sources with flux fractions $F_2/F_1$ brighter than roughly
$0.03$, $0.08$, and $0.20$, provided that the secondary was separated
from the primary by at least $\approx 15\,{\rm km\,s}^{-1}$.  At
smaller velocity separations, the injected lines begin to blend with
the target spectrum. \added{We verified these results by injecting
secondary lines directly into the spectrum and then deriving its LSD
broadening profile as we would for a normal observation.  The results
were nearly identical, save for greater computational cost.}

\subsubsection{FEROS}
TOI 837b was monitored with the FEROS echelle spectrograph
\citep{kaufer_commissioning_1999}, mounted on the MPG 2.2$\,$m
telescope at the ESO La Silla Observatory, in Chile. FEROS has a
resolution of $\approx$48,000 across a spectral range of 350–920 nm.
It has a high efficiency of $\approx$20\%. We obtained 13 spectra of
\tn\ between July 5 of 2019 and March 14 of 2020 in the context of the
Warm gIaNts with tEss (WINE) collaboration, which focuses on the
systematic characterization of TESS transiting giant planets with
moderately long orbital periods \citep[{\it
e.g.},][]{brahm:2019,jordan:2020}.  We adopted exposure times of 500
and 600 seconds and the observations were performed with the
simultaneous calibration mode for tracing the instrumental velocity
variations with a comparison fiber illuminated with a ThAr lamp. FEROS
data was processed with the \texttt{ceres} pipeline
\citep{brahm_2017_ceres}, which delivers precision radial velocities
and bisector span measurements through cross-correlation of the
extracted spectra with a binary mask resembling the properties of a
G2V star. The radial velocities are given in Table~\ref{tab:rvs}, and
shown in Figure~\ref{fig:rvs}.  To check for the presence of secondary
lines, we performed a similar injection-recovery exercise as with the
CHIRON data.  We achieved slightly worse limits, likely due to the
lower spectral resolution of FEROS, and therefore adopted the CHIRON
limits.

\subsubsection{Veloce}
We acquired 34 spectra over 10 visits of \tn\ using the Veloce
spectrograph, mounted on the 3.9$\,$m Anglo-Australian Telescope at
Siding Spring Observatory near Coonabarabran, Australia
\citep{gilbert_veloce_2018}.  The currently operational
``Veloce-Rosso'' channel provides coverage from 600--950$\,$nm at a
spectral resolution of $\approx 80{,}000$.  Many of the exposures were
taken in average or poor seeing conditions, when the SNR is lowest and
the fiber-to-fiber cross-contamination on the IFU-style fiber feed is
strongest.  To reduce the spectra to velocities, we cross-correlated
against a template of $\delta$ Pavonis, because with spectral type G8
IV it was the closest high-SNR \tn\ analog available in the Veloce
spectral database.  The velocity RMS seen across each visit was
hundreds of meters per second, likely due to uncorrected
fiber-to-fiber cross-contamination.  This cross-contamination severely
affects the wavelength solutions for the 19 individual science fibres,
ultimately leading to significantly increased RV scatter. For analysis
purposes, we averaged the single-shot RVs across each visit, and set
the velocity uncertainties to be the standard deviation of the
per-visit exposures.  The velocities are given in Table~\ref{tab:rvs},
and shown in Figure~\ref{fig:rvs}.

\startlongtable
\begin{deluxetable}{llll}
    

\tabletypesize{\scriptsize}


\tablecaption{\tn\ radial velocities.}
\label{tab:rvs}


\tablehead{
  \colhead{Time [BJD$_\mathrm{TDB}$]} &
  \colhead{RV [m$\,$s$^{-1}$]} &
  \colhead{$\sigma_{\rm RV}$ [m$\,$s$^{-1}$]} & 
  \colhead{Instrument}
}

\startdata
 8669.533150 &  -57.8 &    27.5 &   FEROS \\
 8669.540450 &  -13.9 &    29.4 &   FEROS \\
 8676.506930 &    6.7 &    37.8 &   FEROS \\
 8677.519150 &  -70.3 &    44.6 &   FEROS \\
 8884.787630 &  240.0 &    28.0 &  CHIRON \\
 8891.891180 &  -76.0 &    37.0 &  CHIRON \\
 8898.735330 &  -10.0 &    43.0 &  CHIRON \\
 8903.725760 &  -25.0 &    38.0 &  CHIRON \\
 8904.739930 &   80.1 &    24.5 &   FEROS \\
 8905.793630 &   88.0 &    21.7 &   FEROS \\
 8908.762520 &   45.3 &    28.3 &   FEROS \\
 8909.702140 &    0.0 &    31.8 &   FEROS \\
 8912.606750 &   41.3 &    24.1 &   FEROS \\
 8913.740580 &  161.1 &    37.3 &   FEROS \\
 8915.762170 &   10.0 &    33.0 &  CHIRON \\
 8916.714540 &  -93.5 &    33.6 &   FEROS \\
 8917.765720 & -159.7 &    24.8 &   FEROS \\
 8920.706100 &   99.0 &    32.0 &  CHIRON \\
 8922.845800 & -148.3 &    54.9 &   FEROS \\
 8915.924027 &   37.5 &   725.9 &  Veloce \\
 8921.284950 &  105.9 &   453.2 &  Veloce \\
 8922.733572 & -195.9 &   195.6 &  Veloce \\
 8924.583708 &   -7.6 &   262.3 &  Veloce \\
 8926.365810 &   14.3 &   442.6 &  Veloce \\
 8927.318146 &  207.0 &   505.2 &  Veloce \\
 8928.559780 &   -7.3 &   180.2 &  Veloce \\
 8930.324059 &   -2.6 &   152.0 &  Veloce \\
 8931.293091 &  -45.7 &   152.9 &  Veloce \\
 8932.065206 & -105.6 &   319.8 &  Veloce \\
\enddata


\tablecomments{
Times are in units of ${\rm BJD}_{\rm TDB} - 2{,}450{,}000$.
}
\vspace{-0.9cm}
\end{deluxetable}

\section{Assessment of False Positive Scenarios}
\label{sec:validation}

\begin{figure*}[!t]
	\begin{center}
		\leavevmode
		\includegraphics[width=0.75\textwidth]{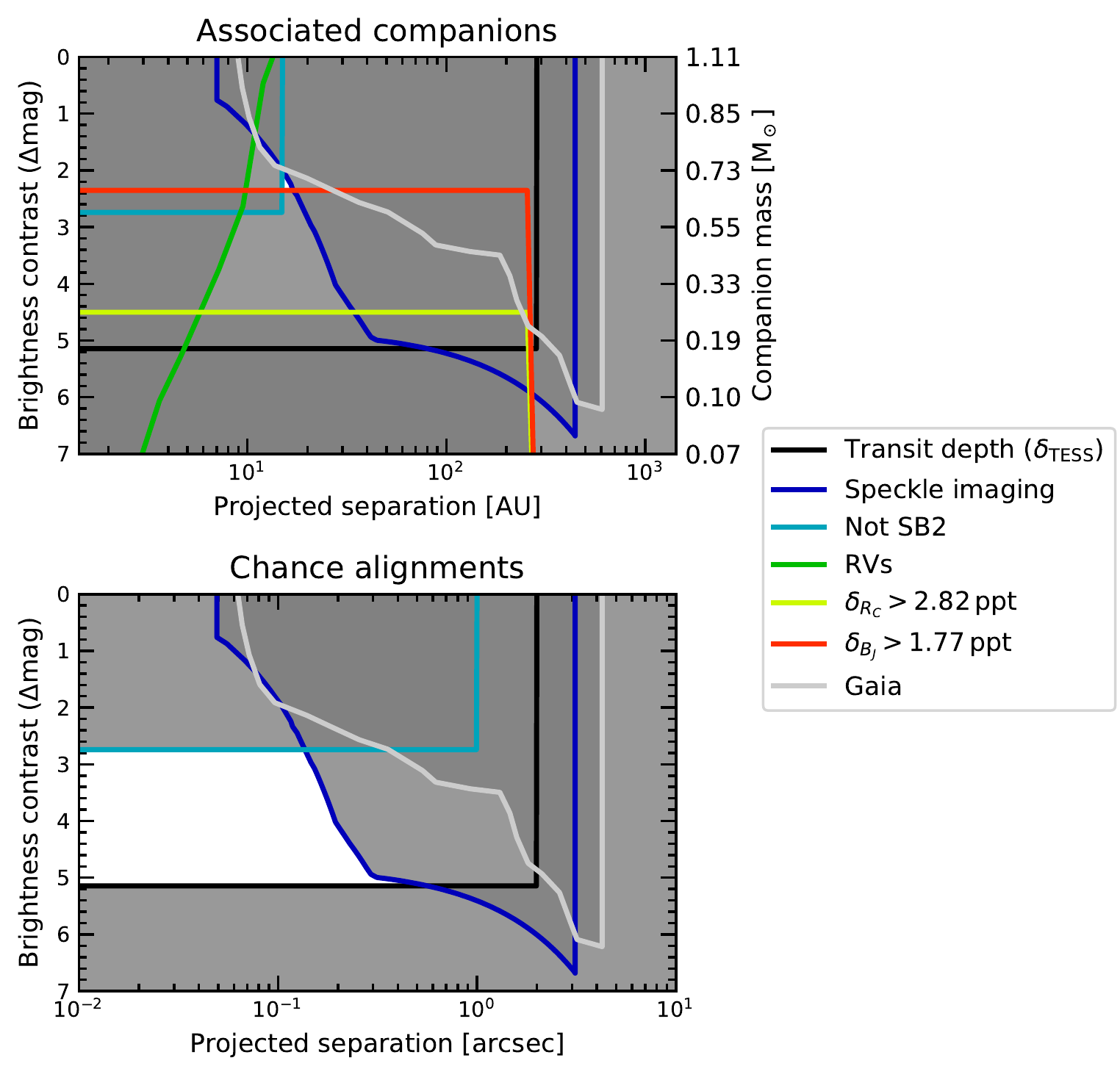}
	\end{center}
	\vspace{-0.6cm}
	\caption{
    {\bf Astrophysical false positive scenarios.} {\it Top}: for bound
    companions (EB and HEB scenarios), {\it Bottom}: for unassociated
    companions along the same line of sight (BEB scenarios).  Each
    constraint is described in Section~\ref{subsec:fp_constraints}.
    Gray regions are ruled out by at least one constraint.
		\label{fig:fpscenario}
	}
\end{figure*}

Validating a transiting planet means statistically arguing that the
data are much more likely to be explained by a planet than by an
astrophysical false positive. The concept of validation has been
developed and calibrated by {\it e.g.},
\citet{torres_modeling_2011,morton_efficient_2012,diaz_pastis_2014,santerne_pastis_2015,morton_false_2016}
and \citet{giacalone_triceratops_2020}.  ``Validation'' is different
from ``confirmation'', which means that there is overwhelming evidence
that the transits {\it must} be explained by a planet, through
elimination of all false positive scenarios and determination that the
planet's mass is in the substellar regime.

Assuming an eclipse has been localized to the target star, potential
false positive scenarios include eclipses of a\added{n unresolved}
background binary (BEB), eclipses of a hierarchical system bound to
the primary star (HEB), and the possibility that the eclipses are
caused by a stellar companion, rather than a planetary one (EB).

Figure~\ref{fig:fpscenario} provides a visual summary of the possible
astrophysical false positive scenarios, as well as our ability to rule
them out based on our combined photometry, velocimetry, and imaging.
In this Section we describe each constraint in turn, and then present
a calculation using \texttt{VESPA} \citep{morton_efficient_2012} to
demonstrate that the probability of \tn\ being an astrophysical false
positive is small enough to validate it as a planet.

\subsection{Constraints on False Positive Scenarios}
\label{subsec:fp_constraints}

\subsubsection{Transit Depth}
In HEB and BEB scenarios, the flux from \tn\ and the true eclipsing
binary host blend together, and reduce the ``true'' TESS-band eclipse
depth $\delta_{\rm true}$ to the observed depth $\delta_{\rm obs}$:
\begin{equation}
  \delta_{\rm obs}
  = 
  \delta_{\rm true} \frac{F_{\rm comp}}{F_{\rm total}},
\end{equation}
where the total system flux and the flux from only the companion
(``comp'') binary are labeled as such.  The requirement that the
eclipse is produced by fusion-powered stars and that $\delta_{\rm true} < 0.5$
translates to a bound on the faintest possible blended companion
system:
\begin{equation}
  \Delta m < -\frac{5}{2} \log_{10}
             \left( \frac{0.5}{\delta_{\rm obs}} \right).
\end{equation}
For \tn\ ($T=9.93$), this implies that any stellar companion invoked
to explain the transit depth must be brighter than $T=15.07$.  In
Figure~\ref{fig:fpscenario}, we set the spatial limit to 2$''$ based
on the precision at which we have localized the transits using
seeing-limited ground-based photometry.

If the transit were box-shaped, this argument could be extended to
even more restrictive depths \citep[{\it
e.g.},][]{seager_unique_2003,vanderburg_hr858_2019,rizzuto_tess_2020}.
Since the transits of \tn\ could be grazing, the second and third
contact points do not necessarily occur, and the shape of the transit
is not particularly restrictive.

\subsubsection{Speckle Imaging}
\label{subsec:speckleconstraint}
The contrast limits obtained through the SOAR $I$-band speckle imaging
(Section~\ref{subsec:speckle}) are shown in
Figure~\ref{fig:fpscenario}.  While ``Star A'' was detected in the
SOAR images, our ground-based photometry rules it out as a possible
source of the eclipse signal (Section~\ref{subsec:groundphot}).  To
convert the remaining contrast constraints to limits on the masses of
bound companions, we used the \citet{baraffe_evolutionary_2003} models
for sub-stellar mass objects and the MIST models for stellar mass
objects
\citep{paxton_modules_2011,paxton_modules_2013,paxton_modules_2015,dotter_mesa_2016,choi_mesa_2016}.
We assumed that the system age was 35 Myr, so that companions would be
at a plausible state of contraction.

To convert from theoretical effective temperatures and bolometric
luminosities to expected magnitudes in instrumental bandpasses, we
made the simplifying assumption that all sources had blackbody
spectra.  Using the theoretical stellar parameters and the measured
transmission functions \citep{tokovinin_ten_2018}, we then calculated
the apparent magnitudes of stellar companions of different masses, and
interpolated to produce the scale shown on the upper-right in
Figure~\ref{fig:fpscenario}.

\subsubsection{Not SB2}
We derived limits on blended spectroscopic companions using the
stacked CHIRON spectra (see Section~\ref{subsec:chiron}).  For a
slowly rotating stellar companion well-separated in velocity, the
spectra would have revealed companions with flux fractions $F_2/F_1
\gtrsim 3\%$.  For a companion with rotational broadening of $15\,{\rm
km\,s}^{-1}$, roughly equivalent to that of \tn, we were able to
exclude companions with flux fractions exceeding $\approx$8\%.  For
plotting purposes, in Figure~\ref{fig:fpscenario} we have assumed the
latter flux-fraction limit of 8\% ($\Delta {\rm mag} \approx 2.7$).
\added{The outer limit in projected separation for associated
companions is the distance at which the Keplerian orbital velocity is
well below the rotational broadening. This condition translates to a
projected separation of $10$--$20\,{\rm AU}$, depending on the
companion mass.  For chance alignments, the same restrictions on
velocity separation apply, but out to a projected separation equal to
the CHIRON slit width of $\approx 1''$.}

\subsubsection{RVs}
The radial velocities from FEROS, CHIRON, and Veloce can be used to
detect massive bound companions orbiting \tn.  We searched for
\deleted{the presence of }planetary and stellar-mass companions
\replaced{using \texttt{radvel} \citep{fulton_radvel_2018}.  We
assumed circular orbits, and performed two sets of fits.}{in two
different regimes: first, at the orbital period of the transiting
object, and second, at longer orbital periods to constrain the
presence of a massive bound companion.}

For the first fit we set a prior on the period and time of conjunction
using the known ephemeris from the transit.  We then fitted for the
semi-amplitude, instrument offsets, and jitter parameters \added{using
\texttt{radvel} \citep{fulton_radvel_2018}, and assuming circular
orbits}.  This yielded a non-detection of the planet's orbit.  The
corresponding 3-$\sigma$ (99.7$^{\rm th}$ percentile) upper limit on
$M_{\rm p} \sin i$ is $1.20 M_{\rm Jup}$.  The data and corresponding
model are shown in Figure~\ref{fig:rvs}.

The above exercise ruled out the possibility that the observed
eclipses are caused by a stellar-mass object orbiting \tn.  The lack
of a linear radial velocity trend, particularly in the FEROS data,
further constrains the presence of a hierarchical binary system.
Fitting a line to the FEROS velocities yielded a 3-$\sigma$ limit on
linear radial velocity trends of $|\dot{\gamma}| < 0.82\,{\rm
m\,s}^{-1}\,{\rm day}^{-1}$, over the 253 day FEROS baseline.  The
agreement between the mean Gaia DR2 velocity ($17.44 \pm 0.64\,$\kms)
and that from FEROS ($18.0 \pm 0.1$\kms) in theory places an
additional limit on linear trends, since the two observation epochs
are separated by roughly five years.

To place limits on the properties of a possible bound hierarchical
companion, we \added{performed the following injection-recovery
exercise.  We simulated $10^6$ two-body systems with random orbital
phases and inclinations, and drew their semi-amplitudes and periods
from logarithmic distributions: $K\ [{\rm m\,s^{-1}}] \sim
\log\mathcal{U}(1,10^7)$, and $P\ [{\rm days}] \sim \log
\mathcal{U}(1, 10^{15})$. Again assuming circular orbits, we then
analytically evaluated what the radial velocities would have been at
the observed FEROS times if the system had the assumed parameters.  We
then calculated what the linear slope would have been for each
simulated system.  If the absolute value of the slope exceeded our
3-$\sigma$ limit of $|\dot{\gamma}| < 0.82\,{\rm m\,s}^{-1}\,{\rm
day}^{-1}$, we assumed that we would have detected such a system.
Figure~\ref{fig:fpscenario} shows the resulting limits; weakened
sensitivity at harmonics of the baseline occur at lower masses and
smaller projected separations than shown on the plot. The
interpolation from mass to brightness contrast was performed using the
same isochrone models and assumptions as in
Section~\ref{subsec:speckleconstraint}.}\deleted{fitted the radial
velocity data for a Keplerian orbit assuming wide logarithmic priors
on the semi-amplitude and period: $K\ [{\rm m\,s^{-1}}] \sim
\log\mathcal{U}(1,10^5)$, and $P\ [{\rm days}] \sim \log
\mathcal{U}(0.1, 10^{15})$.  We then fitted for the semi-amplitude,
period, time of conjunction, instrument offsets, and jitter
parameters.  We converted the resulting posterior in period and
semi-amplitude to minimum mass and semi-major axis assuming Kepler's
third law.  The resulting 3-$\sigma$ limits are shown in
Figure~\ref{fig:fpscenario}.}

\subsubsection{Multicolor Photometry}
\label{subsec:colorphot}

\paragraph{Multicolor photometry and HEB scenarios}
The most plausible HEB scenarios for \tn\ involve pairs of eclipsing M
dwarfs (Figure~\ref{fig:fpscenario}).  Eclipses of such stars are much
redder than eclipses of the G-dwarf \tn.  Limits on whether the
transit depth decreases in bluer bandpasses can therefore rule out
certain HEB scenarios.

We fitted for the observed depths in different bandpasses using a
machinery similar to that described below in
Section~\ref{subsec:planet}.  We fitted each ground-based transit
individually for the planet-to-star size ratio, the impact parameter,
and a local quadratic trend (the ephemeris was assumed from an initial
fit of only the TESS data).  The corresponding 2-$\sigma$ lower limits
on the transit depths in Cousins-R and Johnson-B band light curves
were 2.82 and 1.77$\,$ppt, respectively, and are shown in
Figure~\ref{fig:groundphot}.  Particularly in our Johnson-B light
curve, the transit depth is correlated with the mean and linear slope
of the light curve: a smaller depth is allowed if the data are fitted
with a larger linear slope and a larger mean.  Our quoted limits
marginalize over these correlations, and the depth measurement itself
is nearly Gaussian.

To determine what classes of HEB are eliminated by these limits, we
performed the following calculation.  We assumed that each system was
composed of the primary (\tn), plus a tertiary companion eclipsing a
secondary companion every 8.3 days.  For secondary masses ranging from
0.07 to 1.10 $M_\odot$, and mass ratios ($M_3/M_2$) from 0.1 to 1, we
then calculated the observed maximal eclipse depth caused by Star 3
eclipsing Star 2 in each observed bandpass.  As before, we
interpolated between mass, effective temperature, and radius assuming
the MIST isochrones for a 35$\,$Myr old system, and also assumed that
each source had a blackbody spectrum.  We used the transmission
functions from the SVO filter profile
service\footnote{\url{http://svo2.cab.inta-csic.es/theory/fps/}}.  For
a typical HEB system ({\it e.g.}, $M_2=M_3=0.2M_\odot$), the bluest
optical bandpasses produced eclipses roughly 10 times shallower than
in TESS-band, because the M-dwarf blackbody function turns over at
much redder wavelengths than the G-dwarf blackbody (Wien's law).

For a fixed secondary mass, we then asked whether any tertiary
companions existed for which the maximal expected eclipse depth could
have been larger than the observed depth.  We could not rule out
hierarchical eclipsing binary systems in cases for which the answer
was yes.  Conversely, we ruled out systems for which at fixed
secondary mass no tertiary mass could enable eclipses of the necessary
depth (in ${\rm R_C}$-band, or in ${\rm B_J}$-band).  The ${\rm
R_C}$-band limit corresponded to a secondary mass limit of $M_2 > 0.27
M_\odot$, and the ${\rm B_J}$-band corresponded to a stronger limit of
$M_2 > 0.70 M_\odot$.

\paragraph{Multicolor photometry and BEB scenarios}

While the above constraints rule out HEBs, certain configurations of
BEB systems ({\it e.g.}, a background G0V+K3V binary) can produce blue
eclipses while remaining undetected along the line of sight.  Such
scenarios are constrained by the lack of an observed secondary
eclipse, and therefore require either eccentric orbits to avoid
secondary eclipses, or else a background twin binary system at double
the orbital period.  The only way to definitively rule out such
scenarios is to prove that the loss of light is from the target star,
for instance by detecting the Rossiter-McLaughlin effect during a
transit, and confirming that the spectroscopic transit is consistent
with the photometric transit.

\subsubsection{Gaia}

The ``Gaia'' curve in Figure~\ref{fig:fpscenario} combines both
point-source detections from imaging and sources showing an
astrometric noise excess relative to the single-source astrometric
model.  The curve was interpolated from Figure~4 of
\citet{rizzuto_zeitVIII_2018}.  \tn\ has a RUWE statistic of 1.022,
indicative that there are no obviously present astrometric companions.
The UWE statistic (square-root of the reduced astrometric $\chi^2$) is
1.38, which is consistent with stars of similar brightness and color
\citep[][Appendix A]{lindegren_gaiasoln_2018}.

\subsubsection{Patient Imaging}

Archival SERC-J and AAO-SES plates are available for the \tn\
field\footnote{\url{https://archive.stsci.edu/cgi-bin/dss_form}}.
These plates were acquired in 1982 and 1992, respectively.  For high
proper motion stars archival imagery can be used to detect slowly
moving background stars that might be an astrophysical false-positive
source \citep[{\it
e.g.},][]{bakos_stellar_2006,huang_pimen_2018,vanderburg_hr858_2019}.
However \tn\ has only moved $\approx0.7''$ between 1982 and present,
in comparison to the $\approx2.0''$ FWHM of the target on the plates.
We therefore cannot resolve it from background sources not already
resolved through more modern imaging.

\subsection{False positive probability}
\label{subsec:fpp}

The constraints on false-positive scenarios summarized in
Figure~\ref{fig:fpscenario} rule out the possibilities that {\it i)}
the eclipses are caused by a star orbiting \tn, {\it ii)} the eclipses
are caused by hierarchical blends\added{\footnote{There is a small gap
in the upper panel of Figure~\ref{fig:fpscenario}, corresponding to a
$\approx 0.7 M_\odot$ companion HEB at a projected separation of
$\approx$15$\,$AU. This region of parameter space is small, and we
ignore it in the remaining analysis.}}, and {\it iii)} the eclipses
are caused by neighboring stars outside $\approx 2''$.  The only
scenario not formally ruled out is a background eclipsing binary.  A
simple, and fallacious, argument against background blends follows
from counting statistics.  The local density of $T<15.1$ stars around
\tn, found by counting from TIC8, is $3.7\times10^{-4} \,{\rm
arcsec}^{-2}$.  Therefore within the relevant $\approx0.3''$ radius
not excluded by the SOAR HRCam contrast curve, for a randomly selected
star we would expect $1.0\times10^{-4}$ potential $T<15.1$
contaminants, which appears small.

The reason the above statement is an insufficient argument against
BEBs is that \tn\ is not a randomly selected star---it was selected
because it shows eclipses.  Given a foreground star that shows
eclipses, the probability of a background star being present is much
greater than for an arbitrary foreground star.  \replaced{The relevant
populations need to be modeled at the Monte Carlo level}{A
probabilistic framework is required to calculate the chance that a
background eclipsing binary causes the eclipses}.  We \replaced{opt to
use \texttt{VESPA} to model the populations}{adopt the Bayesian
framework implemented in \texttt{VESPA}}
\citep{morton_efficient_2012,vespa_2015}.

\texttt{VESPA} calculates the false positive probability for a transit
signal as
\begin{equation}
  {\rm FPP} = 1 - P_{\rm pl},
\end{equation}
where in our case the probability that the signal comes from a planet,
$P_{\rm pl}$, is given by
\begin{align}
  P_{\rm pl} = 
  \frac{
    \mathcal{L}_{\rm pl}\pi_{\rm pl}
  }{
    \mathcal{L}_{\rm pl}\pi_{\rm pl} + \mathcal{L}_{\rm BEB}\pi_{\rm BEB}
  },
\end{align}
where $\mathcal{L}_i$ is the model likelihood for the planet and BEB
scenarios, and $\pi_i$ is the model prior.  The terms labeled as
``BEB'' usually include other false positive scenarios (HEBs and EBs),
but our followup data have excluded these possibilities.  The priors
are evaluated using a combination of galactic population synthesis
\citep{girardi_star_2005}, binary star statistics
\citep{raghavan_survey_2010}, and specific planet occurrence rates
\citep[][Section~3.4]{morton_efficient_2012}.  The likelihoods are
evaluated by forward-modeling a representative population of eclipsing
bodies for each model class, in which each population member has a
particular trapezoidal eclipse depth, total duration, and ingress
duration.  The likelihood is then calculated by multiplying the
probability distribution function of the simulated population's shape
parameters with the posterior probability of the actual observed
eclipse shape.

We ran \texttt{VESPA}\footnote{We used \texttt{VESPA-0.6} and
\texttt{isochrones-1.2.2}.}, and directly incorporated our constraints
of the SOAR $I$-band contrast curve and a non-detection of secondary
eclipses with a depth set at roughly twice the limits from the SPOC
vetting report (0.1\%).  \added{This limit applies across all phases.}
We verified that changing the secondary eclipse depth limit did not
significantly affect the results.  We set the maximum aperture radius
at $2''$, based on our ground-based photometry.  Incorporating the
constraints from Figure~\ref{fig:fpscenario}, our nominal false
positive probability analysis excluded EB and HEB scenarios.  This
yielded an FPP of 0.21\% for \pn, sufficient for formal validation as
a planet \citep{morton_efficient_2012}.  We did not incorporate our
constraint that \tn\ is not double-lined, which rules out an
additional portion of BEB parameter space.  Had we not acquired
multicolor ground-based photometry, and been unable to exclude HEB
scenarios, the FPP would have risen to 8\%.  Since the transits are
achromatic (Figure~\ref{fig:groundphot}), particularly in Johnson-B
band, we can rule out HEB scenarios.

One potential caveat in our approach is that \texttt{VESPA} uses a
galactic population synthesis to model the sight-line. Since \tn\ is
in the foreground of \cn\ (see Section~\ref{subsec:star}), for roughly
$25\,{\rm pc}$ behind the sightline to the star, the number of
background stars is higher than \texttt{VESPA} would predict due to
the presence of the cluster.  To quantify the importance of this
effect, we assessed the sky-plane density of potential contaminants by
counting stars brighter than $T=15.07$ within 0.5 degrees of \tn\
\citep{stassun_TIC8_2019}. We then compared this density against
sightlines rotated in galactic longitude towards and away from the
galactic center. Within $\pm10^\circ$ in galactic longitude, the
sky-plane density of stars fluctuated at the level of $\approx 15\%$,
with a local maximum a few degrees away from \tn, towards the center
of \cn.  The overall density also slowly increased towards the
galactic center.  We therefore do not expect this consideration to
significantly alter our FPP calculation.

\section{System Modeling}
\label{sec:system}

\begin{figure}[!t]
	\begin{center}
		\leavevmode
		\includegraphics[width=0.48\textwidth]{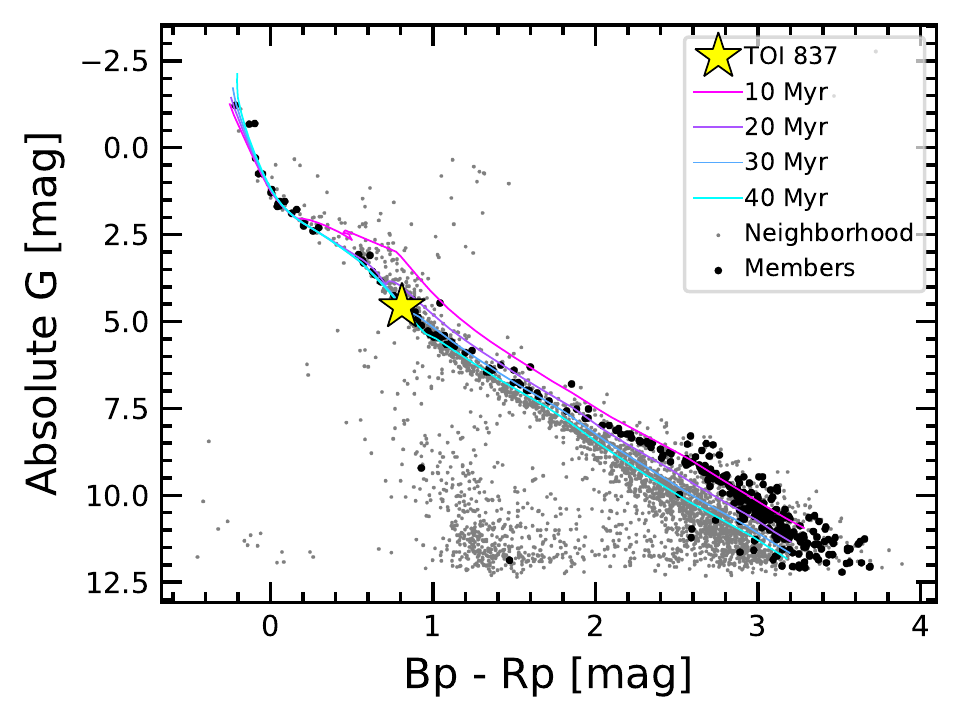}
	\end{center}
	\vspace{-0.7cm}
	\caption{ 
  {\bf Hertzsprung-Russell diagram of \tn\ and members of \cn.}
  Members (black circles) were identified by
  \citet{cantatgaudin_gaia_2018}.  Gray circles are non-member stars
  with right ascension, declination, and parallax similar to \cn.
  They are selected by drawing from a $\{\alpha, \delta, \pi\}$ cube
  centered on the cluster with boundaries set at $5\times$ the
  standard deviation in the cluster parameters.  $G$ denotes Gaia
  broadband magnitudes, $Bp$ Gaia blue, and $Rp$ Gaia red.  MIST
  isochrones (colored lines) fit the upper main sequence well, but
  diverge from the data for $M_\star \lesssim 0.7 M_\odot$. This is a
  known issue with the M dwarf models (see Section~\ref{subsec:hr}).
  \label{fig:hr}
	}
\end{figure}

\begin{figure*}[!t]
	\begin{center}
		\leavevmode
		\includegraphics[width=0.65\textwidth]{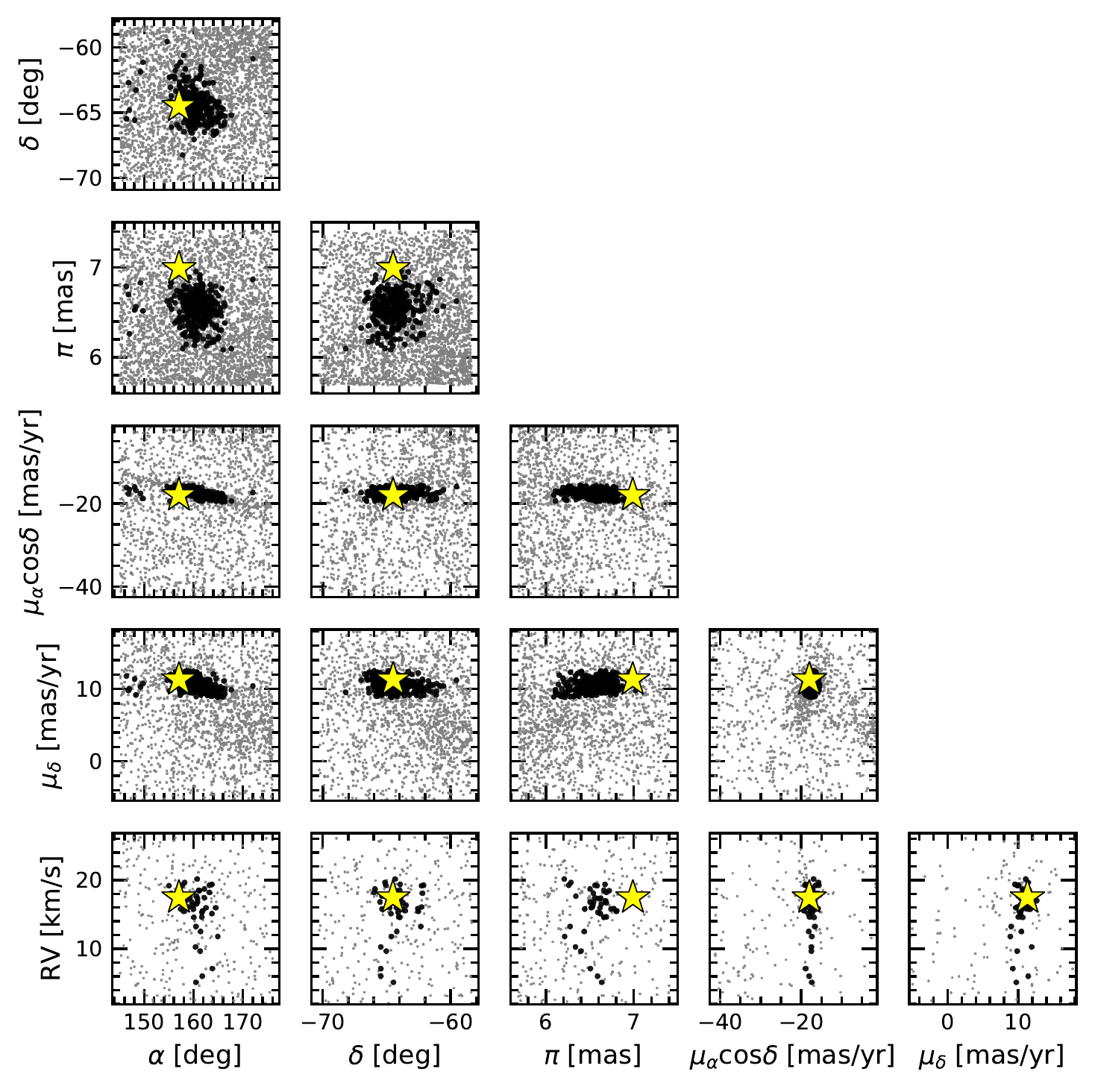}
	\end{center}
	\vspace{-0.5cm}
	\caption{ 
  {\bf Positions and kinematics of \tn\ (star), \cn\ members (black
  circles), and stars in the neighborhood (gray circles).} Members
  were identified by \citet{cantatgaudin_gaia_2018}.  Neighbors are as
  in Figure~\ref{fig:hr}. $\alpha$ denotes right ascension, $\delta$
  declination, $\pi$ parallax, $\mu_{\rm \delta}$ and $\mu_{\rm
  \alpha}$ proper motion in each equatorial direction, and ${\rm RV}$
  radial velocity reported by Gaia DR2.  The RVs are for unblended
  spectra of bright stars ($G\lesssim 12$).  The proper motion
  projection ($\mu_{\delta}$ vs{.} $\mu_{\rm \alpha}\cos\delta$)
  highlights incompleteness in the membership selection function.
  \label{fig:full_kinematics}
	}
\end{figure*}

\begin{figure}[!t]
	\begin{center}
		\leavevmode
		\subfloat{
			\includegraphics[width=0.45\textwidth]{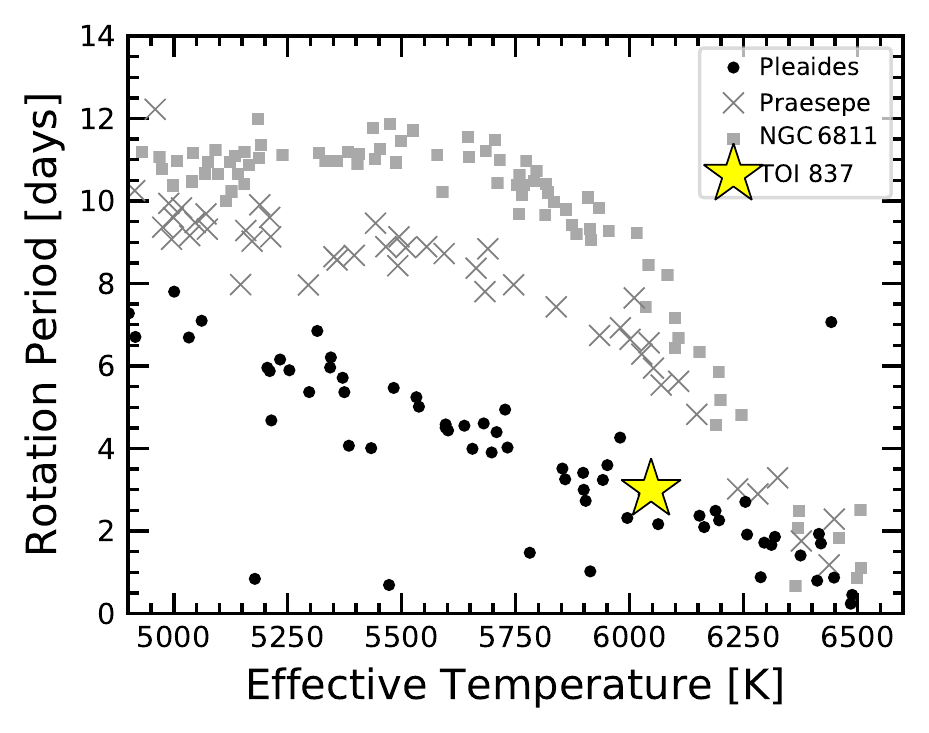}
		}
		
		\vspace{-0.5cm}
		\subfloat{
			\includegraphics[width=0.45\textwidth]{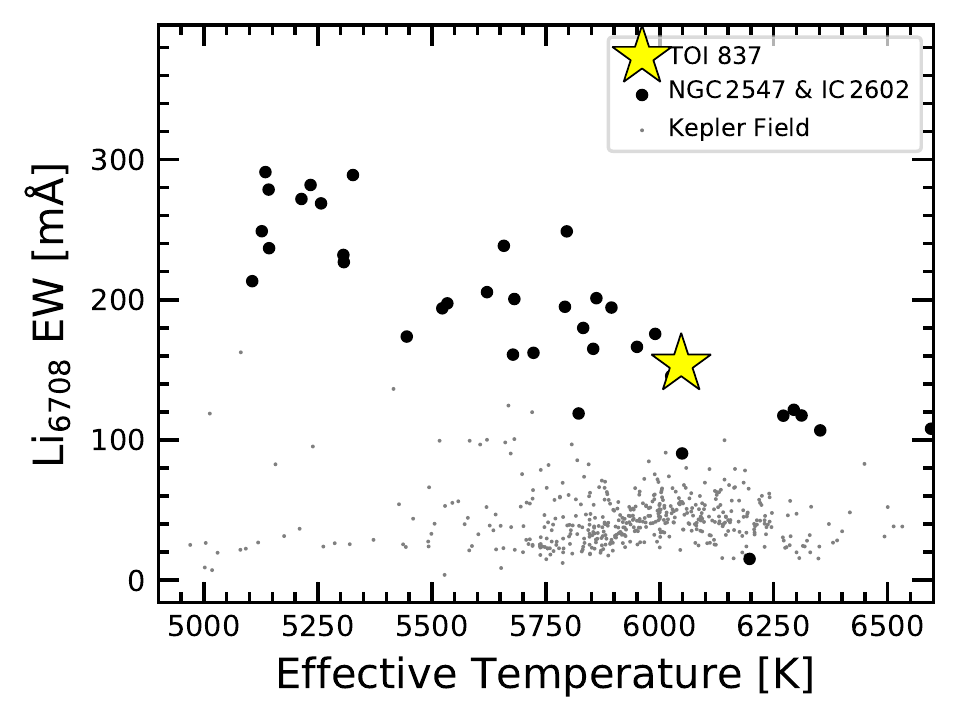}
		}
	\end{center}
	\vspace{-0.7cm}
	\caption{ {\bf Youth diagnostics.}
  {\it Top:} Rotation periods for \tn\ and selected open clusters.
  The Pleiades (120$\,$Myr), Praesepe (670$\,$Myr), and NGC$\,$6811
  (1000$\,$Myr) are shown.  Their rotation periods were measured by
  \citet{rebull_rotation_2016a,douglas_poking_2017,douglas_k2_2019},
  and \citet{curtis_temporary_2019}, respectively.  {\it Bottom:}
  Lithium 6708\AA\ equivalent widths for \tn, field stars, and young
  open clusters.  The field star sample is drawn from Kepler planet
  hosts, and was measured by \citet{berger_identifying_2018} using
  Keck-HIRES.  The young open cluster members were surveyed by
  \citet{randich_gaiaeso_2018} using the UVES and GIRAFFE
  spectrographs at the ESO VLT.  \citet{randich_gaiaeso_2018} found
  lithium depletion boundary ages for these clusters of
  $37.7^{+5.7}_{-4.8}\,{\rm Myr}$ (NGC$\,$2547) and
  $43.7^{+4.3}_{-3.9}\,{\rm Myr}$ (\cn).  
    \label{fig:lithium_rotation}
	}
\end{figure}

\subsection{The Cluster}
\label{subsec:cluster}

\begin{deluxetable}{ccc}
    

\tabletypesize{\scriptsize}


\caption{Previously reported ages for the open cluster IC~2602.}
\label{tab:ages}


\tablehead{\colhead{Method} & \colhead{Age [Myr]} & \colhead{Reference}} 

\startdata
MSTO isochrone & $36.3$ & \citet{mermilliod_comparative_1981} \\
PMS+MSTO isochrone & $30 \pm 5$ & \citet{stauffer_rotational_1997} \\
Isochrone (a) & $67.6$ & \citet{kharchenko_astrophysical_2005} \\
Isochrone (b) & $221$ & \citet{Kharchenko_et_al_2013} \\
Isochrone  & $67.6$ & \citet{van_leeuwen_parallaxes_2009} \\
LDB (c) & $46^{+6}_{-5}$ & \citet{dobbie_ic_2010} \\
MSTO isochrone (d) & $41-46$ & \citet{david_ages_2015} \\
MSTO isochrone (e) & $37-43$ & \citet{david_ages_2015} \\
Li selection + isochrone & $43.7^{+4.3}_{-3.9}$ & \citet{bravi_gaia-eso_2018} \\
Isochrone (f) & $30^{+9}_{-7}$ & \citet{randich_gaiaeso_2018} \\
LDB & $43.7^{+4.3}_{-3.9}$ & \citet{randich_gaiaeso_2018} \\
Isochrone & $35.5^{+0.8}_{-1.6}$ & \citet{bossini_age_2019} \\
Isochrone & $35.5^{+14.6}_{-10.4}$ & \citet{kounkel_untangling_2019} \\
\enddata


\tablecomments{
  MSTO $\equiv$ main sequence turn-off.
  PMS $\equiv$ pre-main-sequence.
	LDB $\equiv$ lithium depletion boundary.
 (a) Based on location in HR diagram of just two stars.
 (b) Notes major age change since \citet{kharchenko_astrophysical_2005}.
 (c)
  \citet{dobbie_ic_2010} performed a dedicated study of the LDB in IC~2602.  Comparing to early isochronal ages, they write their age is ``consistent with the general trend delineated by the Pleiades, $\alpha$-Per, IC$\,$2391, and NGC$\,$2457, whereby the LDB age is ~120-160 per cent of the estimates derived using more traditional techniques'' such as isochrone fitting.
 (d)
  Using \citet{ekstrom_grids_2012} evolutionary models.
 (e)
  Using PARSEC evolutionary models \citep{bressan_parsec_2012}.
 (f)
  Averaged across PROSECCO, PARSEC, MIST models in $(J, H, K_{\rm s})$ and $(J, H, K_{\rm s}, V)$ planes.
} 
\vspace{-1cm}
\end{deluxetable}

\subsubsection{Physical Characteristics}
\label{subsec:clusterchar}

The \cn\ cluster is about 150$\,$pc from the Earth, and is near the
galactic plane with $(l,b)\approx(289.6^\circ, -5.0^\circ)$
\citep{cantatgaudin_gaia_2018}.  It is also sometimes called the
$\theta$ Carinae cluster, after its brightest member, or also the
``Southern Pleiades''.  While \cn\ is close to the Lower Centaurus
Crux subgroup of the Scorpio-Centaurus OB2 association in both
position and proper motion space, its older age and clear kinematic
separation indicate that it is a distinct stellar population
\citep{de_zeeuw_hipparcos_1999,damiani_stellar_2019}.

Reliable ages reported for \cn\ range from 30 to 46$\,$Myr.  We have
collected ages reported over the years in Table~\ref{tab:ages}.  The
Li depletion boundary technique yields slightly older absolute ages
than isochrone fitting \citep{dobbie_ic_2010,randich_gaiaeso_2018}.
Rather than redetermine the age of the cluster and add another line to
the table, we simply adopt an absolute age range for \tn\ of
30--46$\,$Myr.

Reported mean metallicity values ${\rm [Fe/H]}$ for the cluster range
between slightly super-solar ($0.04\pm0.01$,
\citealt{baratella_gaia-eso_2020}) and slightly sub-solar ($-0.02 \pm
0.02$, \citealt{netopil_metallicity_2016}).  The extinction $E(B-V)$
is rather low, with reported values ranging from 0.03 to 0.07
\citep[{\it e.g.},][]{randich_gaiaeso_2018}.

Kinematically, IC~2602 seems to be supervirial, in the sense that the
observed stellar velocity dispersion is larger than the value expected
if it were in virial equilibrium by about a factor of two
\citep{bravi_gaia-eso_2018}.  \citet{damiani_stellar_2019} also
reported evidence for the ongoing evaporation of \cn, in the form of a
diffuse $\approx10^\circ$ halo of young stars around the central
density cusps.  A gyrochronological study of these stars could confirm
that these stars are truly coeval with the cluster.

\subsubsection{HR Diagram}
\label{subsec:hr}

Figure~\ref{fig:hr} shows a Hertzsprung-Russell diagram of \tn, the
\cn\ cluster, and the neighborhood of spatially nearby stars.  Stars
labeled as cluster members are those reported by
\citet{cantatgaudin_gaia_2018} based on Gaia DR2 positions, proper
motions, and parallaxes.  We included candidate members with formal
membership probability exceeding 10\%.  Most members appear to be
young and coeval.  \replaced{\tn\ is in its expected position relative
to the other members along the cluster isochrone.  This
photometrically limits the presence of binary companions in the \tn\
system to be less than half the brightness ($\approx0.75$ magnitudes)
of the target star.}{\tn\ lies on the single-star sequence.  Any
hypothetical companions to \tn\ must therefore be $\lesssim 50\%$ of
its brightness; brighter companions would have made the total system
$\gtrsim 0.44$ magnitudes brighter than the single-star sequence,
which can be ruled out based on the photometric uncertainties and the
intrinsic scatter in the HR diagram.}

Figure~\ref{fig:hr} suggests that the membership census of \cn\ is
incomplete.  We defined the reference neighborhood as the group of at
most $10^4$ randomly selected non-member stars within 5 standard
deviations of the mean \cn\ right ascension, declination, and
parallax.  In other words, the neighborhood members are chosen based
on the observed spread in the cluster's parameters.  We queried Gaia
DR2 for these stars using \texttt{astroquery} \citep{astroquery_2018}.
Many low-mass stars appear above the main sequence, even though they
were not identified as 5-dimensional kinematic members through the
unsupervised \citet{cantatgaudin_gaia_2018} membership assignment
process.

Figure~\ref{fig:hr} also compares the data to the MIST isochrones
\citep{choi_mesa_2016}.  We used the web
interface\footnote{\url{http://waps.cfa.harvard.edu/MIST/interp_isos.html},
\texttt{2020-07-08}} to interpolate isochrones at 10, 20, 30, and 40
million years. We assumed solar metallicity, and a fixed extinction
value of $A_V = 0.217$ \citep{randich_gaiaeso_2018}.  The 30 and 40
Myr models align well with the data for stars with masses ranging from
roughly $0.7$--$7\,M_\odot$.  The pre-main-sequence (PMS) K and M
dwarf models are bluer than observed in the Gaia photometry.  This
discrepancy was noted and discussed at length by
\citet{choi_mesa_2016}.  One suggested explanation was that strong
magnetic fields in low-mass pre-main-sequence stars inhibit convection
and produce a high filling factor of starspots \citep[{\it
e.g.},][]{stauffer_why_2003,feiden_magnetic_2013}.  This explanation
however fails to explain poor isochrone fits in both old open clusters
({\it e.g.}, M$\,$67) and the field, particularly in blue bandpasses.
An alternative explanation is that the molecular line lists for M
dwarf atmospheres are incomplete in these wavelength ranges
\citep{rajpurohit_effective_2013,mann_spectrothermometry_2013}.

\subsection{The Star}
\label{subsec:star}

\subsubsection{Membership of \tn\ in \cn}
\label{subsec:member}

\tn\ has been reported as a member of \cn\ by many independent
investigators \citep[{\it
e.g.},][]{Kharchenko_et_al_2013,oh_comoving_2017,cantatgaudin_gaia_2018,damiani_stellar_2019,kounkel_untangling_2019}.
The simplest way to verify the membership is through inspection of the
Gaia DR2 position and kinematics.  Figure~\ref{fig:full_kinematics}
shows the six-dimensional positions and kinematics of \tn, \cn\
members, and nearby stars.  The ``neighborhood'' is defined as in
Figure~\ref{fig:hr}.  The axes limits for the right ascension,
declination, and parallax dimensions are set by being within 5
standard deviations of the mean \cn\ right ascension, declination, and
parallax.  The axes limits for the proper motion and radial velocity
dimensions are set at the 25$^{\rm th}$ and 75$^{\rm th}$ percentiles,
in order to give a sense of the population's distribution, while
excluding outliers.  The radial velocities suffer the greatest
incompleteness due to the current $G\approx12$ magnitude limit of the
Gaia DR2 data processing.

Figure~\ref{fig:full_kinematics} provides strong evidence that \tn\ is
a member of \cn.  The only dimension that could lead to some worry is
the parallax, as \tn\ is one of the closest \cn\ members reported by
\citet{cantatgaudin_gaia_2018}.  Fortunately, there are independent
means of verifying the star's youth.

\subsubsection{Rotation}

As stars get older, their rotation rates incrementally slow due to
magnetic braking \cite{weber_angular_1967,skumanich_time_1972}.  One
way to verify the youth of \tn\ is by comparing its rotation period to
other stars with known ages.

We measured the rotation period from the TESS \texttt{PDCSAP} light
curve using the Lomb-Scargle periodogram implemented in
\texttt{astropy}
\citep{lomb_1976,scargle_studies_1982,vanderplas_periodograms_2015}.
We fitted the light curve without masking out the transits or flares,
as these represent a small fraction of the overall time series.  To
derive the uncertainty on the best period, we fitted a Gaussian to the
dominant peak, after first ensuring that we had oversampled the
initial frequency grid.  This gave a rotation period of $P_{\rm rot} =
2.987 \pm 0.056\,{\rm d}$ when allowing for a single Fourier term in
the periodogram model, and $P_{\rm rot} = 3.004 \pm 0.053\,{\rm d}$
when allowing for two Fourier terms.  As the latter model provides a
better fit to the data, we adopt it as the rotation period.

As we will discuss in Section~\ref{subsec:starparams}, we measured the
star's radius by combining the spectroscopic effective temperature
with a broadband photometry SED fit.  We would expect, combining our
$R_\star$ and $P_{\rm rot}$ measurements, that the equatorial velocity
$v$ of the star would be $17.67 \pm 0.32 \,{\rm km\,s}^{-1}$.  Our
spectroscopically measured $v\sin i$ from CHIRON, $16.2 \pm 1.1 \,{\rm
km\,s}^{-1}$ agrees reasonably well with this expectation.

The star is clearly a rapid rotator.
Figure~\ref{fig:lithium_rotation} compares its rotation period with
rotation periods that have been measured in a number of well-studied
open clusters.  \tn\ seems to be gyrochronologically coeval with the
Pleiades sequence.  This is not to say that \tn\ is ``Pleaides-aged'',
because the observed scatter in the rotation-period diagram for the
first 10--100$\,$Myr is quite high \citep[see Figure~9
of][]{rebull_rotation_2020}.  Instead, we interpret the rotation
period as evidence to support the claim that \tn\ is younger than
$\sim$ $500\,{\rm Myr}$.

\subsubsection{Lithium}

Lithium depletion for early G-dwarfs like \tn\ requires hundreds of
megayears \citep{soderblom_ages_2014}. This is because their
convective envelopes are shallow, and so transport of photospheric
lithium to the hot core takes place over diffusive timescales, rather
than convective timescales.  Nonetheless, comparison of early G-dwarfs
in the field to {\it e.g.}, 600 Myr old Hyads has shown that the
depletion does indeed happen over many gigayears
\citep{berger_identifying_2018}.

The spectra of \tn\ all show the 6708$\,$\AA\ lithium doublet in
absorption. Opting to use our FEROS spectra because of their high S/N,
we measured the line's equivalent width (EW) to be $154 \pm 9 \,
$m\AA.  Figure~\ref{fig:lithium_rotation} compares this EW to stars in
the field, and other young open cluster members.  The field star
measurements were collected by \citet{berger_identifying_2018}; we
show their reported lithium detections with ${\rm S/N}>3$.  The young
open cluster members were selected for the presence of lithium, as
described by \citet{randich_gaiaeso_2018}.  The measured \tn\ Li EW is
much larger than observed for field stars, and is consistent with
lithium absorption seen in stars with similar colors in sub-100$\,{\rm
Myr}$ moving groups.

\subsubsection{Stellar Parameters}
\label{subsec:starparams}

Select properties of \tn\ from the literature and our analysis are
presented in Table~\ref{tab:starparams}.  We calculated the stellar
parameters using two different approaches.

\explain{NOTE: the description of why we avoided Method 1 has been moved
from a paragraph at the end of this section, to the paragraph below.}

In ``Method 1'', we measured spectroscopic parameters from each of the
CHIRON spectra (Section~\ref{subsec:chiron}).  We then calculated the
stellar radius and reddening following \citet{stassun_accurate_2017}.
We first derived the bolometric flux by combining available broadband
magnitudes from Gaia, Tycho-2, APASS, 2MASS, and WISE.  We then fitted
the SED with the \citet{kurucz_atlas12_2013} stellar atmopshere
models, and summed to find $F_{\rm bol}$.  When fitting the atmosphere
model, we varied the extinction ($A_{\rm V}$) and the overall
normalization.  This procedure yielded $A_{\rm V} = 0.20\pm0.03$,
which agrees with the average from the \cn\ isochrone fits of
\citet{randich_gaiaeso_2018}.  Combining the spectroscopic effective
temperature, bolometric flux, and Gaia distance, we determined the
stellar radius using the Stefan-Boltzmann law.  Combining this radius
with the spectroscopic $\log g$ also yields a stellar mass.
\added{The stellar mass however seemed to be high relative to the
observed CHIRON effective temperature ($1.21 M_\odot$ to $5946\,{\rm
K}$, with relative uncertainties of a few percent on each). We
therefore explored a second method, and ultimately adopted
it because its systematic uncertainties were easier to quantify.}

In ``Method 2'', we \deleted{simply} used the observed location of
\tn\ in the HR diagram and interpolated against the 40$\,$Myr
\added{MIST} isochrone.  This method leverages the relative location
of \tn\ within the \deleted{narrow} \cn\ isochrone to derive precise,
theoretically self-consistent constraints on all of the stellar
parameters.  Although this approach would fail for a low-mass star,
\tn\ is above the stellar masses where the Gaia photometry and
isochrone models begin to diverge. \added{The statistical
uncertainties yielded by this approach are of order 1\% for the
stellar mass and radius.  To quantify the systematic uncertainties, we
compared the parameters derived from the MIST isochrones with those
from the PARSEC\footnote{\url{http://stev.oapd.inaf.it/cmd}}
isochrones
\citep{bressan_parsec_2012,chen_improving_2014,chen_parsec_2015,marigo_new_2017}.
The PARSEC isochrones gave a stellar mass 5\% lower, effective
temperature 3\% lower, logarithmic surface gravity 1\% lower, and
radius 8\% lower than the MIST isochrones.  For the sake of
self-consistency, in Table~\ref{tab:starparams} and the ensuing
analysis we adopted the stellar parameter values from MIST.  We took
the uncertainties to be the quadrature sum of the statistical and
systematic components.}

\deleted{
Method 1 yielded a stellar mass that seemed to be high relative to the
observed CHIRON effective temperature ($1.21 M_\odot$ to $5946\,{\rm
K}$, with relative uncertainties of a few percent on each). To avoid
poorly understood systematics,  we adopted the stellar parameters from
Method 2, and report them in Table~\ref{tab:starparams}.
}

\subsection{The Planet}
\label{subsec:planet}

\begin{figure}[!t]
	\begin{center}
		\leavevmode
		\includegraphics[width=0.48\textwidth]{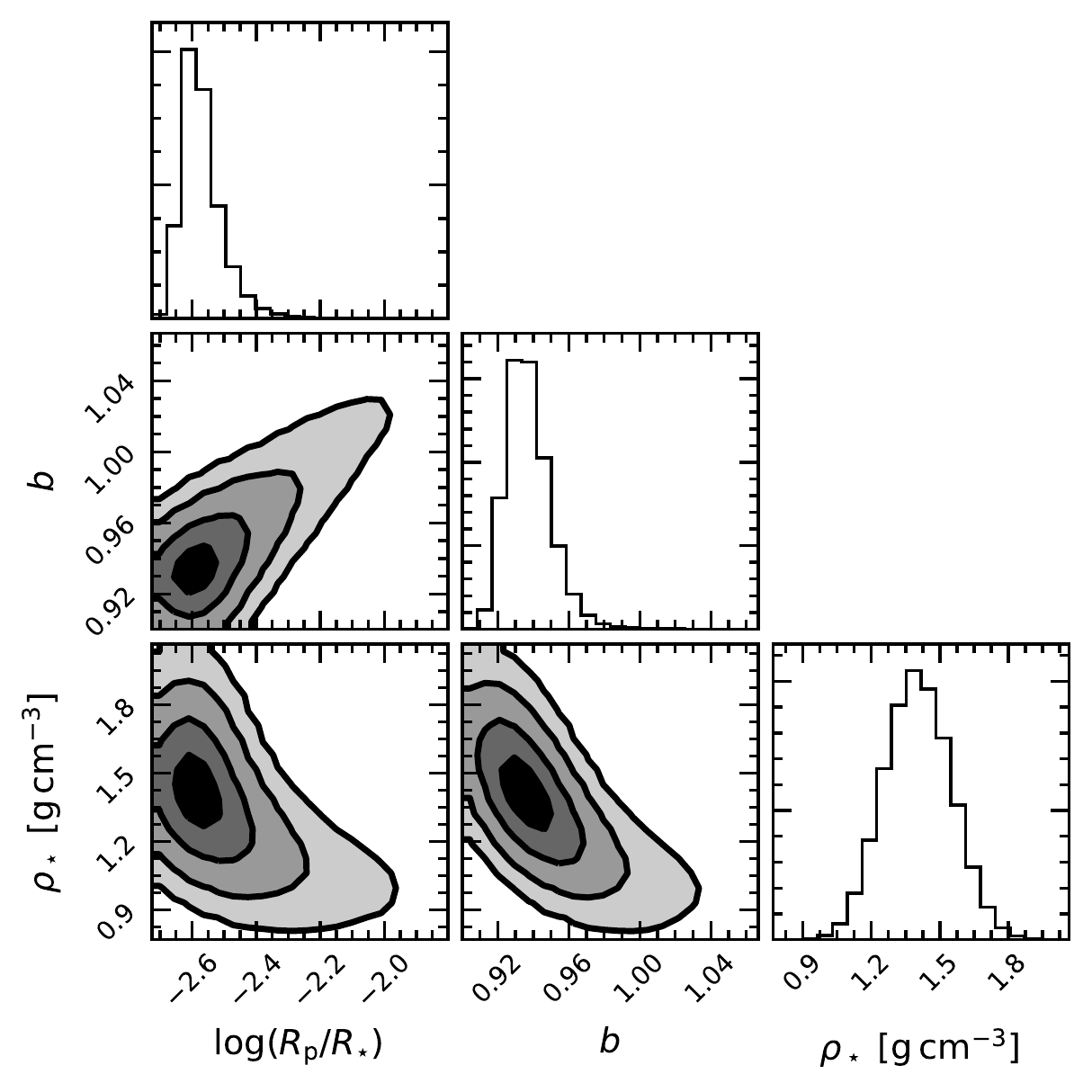}
	\end{center}
	\vspace{-0.6cm}
	\caption{ 
    {\bf Posterior probabilities of impact parameter, planet-to-star
    size ratio, and stellar density.} Contours are shown at 1, 2, 3,
    and 4-$\sigma$ confidence.  The planet-to-star size ratio
    corresponds to a planet size between \replaced{$0.68\,R_{\rm Jup}$
    and $0.97\,R_{\rm Jup}$}{$0.62\,R_{\rm Jup}$ and $0.95\,R_{\rm
    Jup}$} (3$^{\rm rd}$--97$^{\rm th}$ percentile).  This plot was
    made using \texttt{corner} \citep{corner_2016}.
		\label{fig:subsetcorner}
	}
\end{figure}

We also considered two different approaches for fitting the available
time-series photometry of \pn.  \replaced{To derive the most precise
possible ephemeris,}{In the first approach,} we fitted the ground and
space-based transits simultaneously.  \replaced{To derive the physical
parameters of the planet}{In the second}, we fitted the TESS data
alone\deleted{, due to our better understanding of the underlying
systematic trends and the higher precision}.

To clean the TESS \texttt{PDCSAP} light curve, we first eliminated
points that had quality flags corresponding to any of bits $\{3, 4, 6,
8, 11, 12\}$.  This excluded cadences affected by coarse spacecraft
pointing, reaction wheel desaturation events, manual flags, cosmic ray
hits, and straylight from the Earth or Moon being present.  Inspecting
the data, we also manually excluded the two flares shown in
Figure~\ref{fig:thephot}.  We then trimmed the TESS data to windows of
$\pm 7\,{\rm hr}$ centered on each transit.

Our model for time-series photometry data was an
\citet{exoplanet:agol20} transit with physical and orbital parameters
shared across all transit windows, plus a local quadratic trend
allowed within each window.  Select parameters and priors are listed
in Table~\ref{tab:posterior}, for the \replaced{TESS-only model}{joint
model of the TESS and ground-based data}.  In brief, we fitted for the
shared stellar parameters $\{\log g, R_\star, u_0, u_1\}$, and the
shared planetary parameters $\{t_0, P, b, \log(R_{\rm p}/R_\star)\}$.
There were also three free trend parameters for each transit window to
account for the local rotational variability.  In the TESS-only model
this yielded 23 free parameters, of which 8 were physically relevant
and 15 were nuisance parameters.  In the combined TESS and
ground-based model, there were an additional 7 transits, and therefore
an additional 21 nuisance parameters for a total of 44 free
parameters.

We fitted the model\added{s} using \texttt{PyMC3}
\citep{salvatier_2016_PyMC3,exoplanet:theano}.  For the exoplanet
transit, we used the \texttt{exoplanet} code
\citep{exoplanet:exoplanet}.  After initializing each model with the
parameters of the maximum {\it a posteriori} model, we assumed a
Gaussian likelihood, and sampled using \texttt{PyMC3}'s gradient-based
No-U-Turn Sampler \citep{hoffman_no-u-turn_2014}. We used $\hat{R}$ as
our convergence diagnostic \citep{gelman_inference_1992}.

We opted for this approach rather than a joint fit of the photometry
and radial velocities because the RVs on their own did not show
evidence for a planetary signal.  \deleted{Our preference for using
only the TESS data to derive the transit parameters was in our view
justified by the systematic uncertainties inherent to ground-based
light curve production, particularly in comparison star selection.}
Our assumption of a constant radius across all bandpasses was tested
by independently fitting each ground-based transit while letting the
planetary radius float (Section~\ref{subsec:colorphot}). The transit
depths did not significantly change between different bandpasses.  Our
assumption in the false-positive probability calculation
(Section~\ref{subsec:fpp}) of no odd-even variations was tested by
independently fitting all odd and all even transits separately.  The
resulting best-fit depths were consistent within 1-$\sigma$.

\explain{The paragraphs below were heavily modified and/or newly added
after submission, given the updated systematic uncertainties on the
stellar parameters, and our revision in favor of the TESS+ground
model. }

The posteriors from fitting the TESS \added{and ground-based }data
\deleted{alone }are given in Table~\ref{tab:posterior}.  The condition
for a grazing transit is whether the impact parameter $b$ is
\replaced{below}{above} $1 - R_{\rm p}/R_\star$.  The relevant
posterior probabilities are shown in Figure~\ref{fig:subsetcorner}.
The transit is either grazing, or nearly grazing.
\added{The planet radius and impact parameter based on the
TESS and ground-based data are as follows.
\begin{align}
  R_{\rm p} &= 0.768^{+0.091}_{-0.072} R_{\rm Jup}, \\
  b &= 0.936^{+0.013}_{-0.010},
\end{align}
where we quote the median, 86$^{\rm th}$, and 14$^{\rm th}$
percentiles of the marginalized one-dimensional posteriors.
}
\deleted{From experimenting with the priors, we found that in the
absence of a strong prior on the stellar density, the inherent
degeneracy between the impact parameter and planet-to-star size ratio
would have been much stronger.  In the absence of precise information
about the star, a larger fraction of the posterior would therefore
have been grazing.  However, our priors on the stellar parameters from
the cluster isochrone fit break this degeneracy, enabling us to report
two-sided limits on the planet-to-star size ratio.}

\added{The second model, which used just the TESS data and the
cluster-isochrone stellar parameter priors,
formally yielded only a one-sided limit on the planet radius.  The
reason is that the $b$--$R_{\rm p}/R_\star$ degeneracy was not broken:
the combination of uncertain stellar parameters and
the grazing geometry allowed very
large planet-to-star radius ratios for very large impact parameters.
Based on our mass upper limit of $1.20\,M_{\rm Jup}$, we might argue
in favor of discarding the large-radius solution, since no sub-Jovian
mass objects larger than $\sim$$3R_{\rm Jup}$ are known to exist.  Had
we imposed this additional prior, then the TESS-only model would have
yielded
\begin{align}
  R_{\rm p} &= 0.836^{+0.208}_{-0.121} R_{\rm Jup} \\
  b &= 0.957^{+0.027}_{-0.017}.
\end{align}
Although these parameters are in 1-$\sigma$ agreement with our adopted
joint model of the TESS and ground-based data, we preferred the first
model both because it included all available data, and because it
succeeded in breaking the $b$--$R_{\rm p}/R_\star$ degeneracy without
requiring the adoption of informed priors.
}

\section{Discussion}
\label{sec:discussion}

\begin{figure}[!t]
	\begin{center}
		\leavevmode
		\subfloat{
			\includegraphics[width=0.48\textwidth]{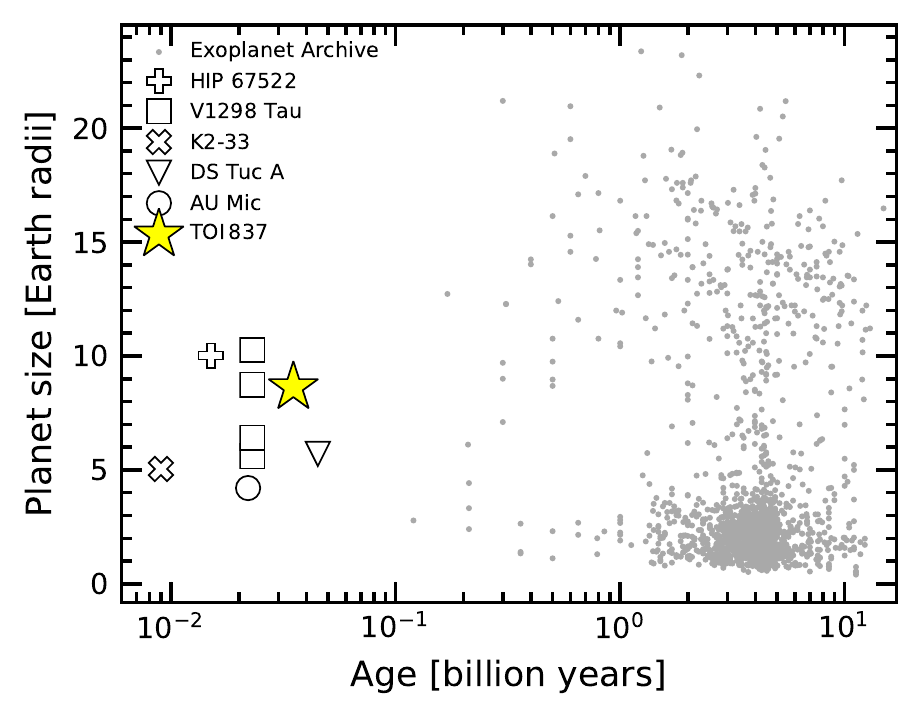}
		}
		
		\vspace{-0.5cm}
		\subfloat{
			\includegraphics[width=0.48\textwidth]{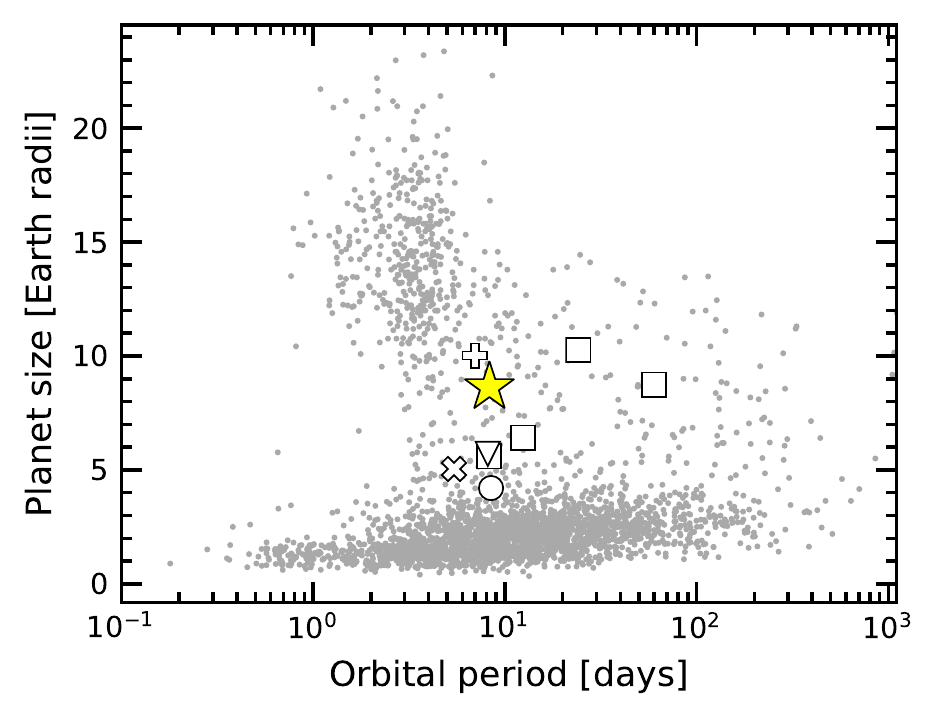}
		}
	\end{center}
	\vspace{-0.6cm}
	\caption{
    {\bf \tn\ compared to known transiting planets}.
    {\it Top}: Planet radii versus ages.  Systems younger than
    100 Myr are emphasized.  Ages and radii are from the NASA
    Exoplanet Archive on 27 Aug 2020.  Precise ages are known for only
    a small fraction of the gray points.
    {\it Bottom}: Planet radii versus orbital periods. The youngest
    known transiting planets do not obviously overlap with the
    populations of known hot Jupiters or sub-Neptunes.
		\label{fig:population}
	}
	\vspace{-0.2cm}
\end{figure}

\tn\ joins a number of other young planetary systems reported from
TESS, including DS~Tuc~Ab, HIP~67522b, TOI~1726, and AU~Mic~b
\citep{newton_tess_2019,zhou_well_2020,montet_young_2020,rizzuto_tess_2020,mann_tess_2020,plavchan_planet_2020,palle_transmission_2020,addison_youngest_2020,martioli_magnetism_2020,hirano_limits_2020}.
\replaced{ Figure~\ref{fig:rpvsage} shows \tn\ in the space of planet
sizes and ages.  \tn\ is among the youngest transiting planets known.
}{ In the space of planet sizes and ages, the top panel of
Figure~\ref{fig:population} shows that \tn\ is among the youngest
transiting planets known.  }

\added{In the space of planet sizes and orbital periods, the bottom
panel of Figure~\ref{fig:population} highlights a peculiar feature of
the known sub-100$\,$Myr transiting planets: they do not overlap with
the known populations of either hot Jupiters or sub-Neptune sized
planets.  The young planets instead have sizes ranging from
4.2$\,R_\oplus$ (AU Mic b) to slightly smaller than Jupiter.  The lack
of sub-Neptune sized planets could be a selection effect, because
larger planets are easier to detect around highly variable stars.
Another (speculative) explanation is that the known
sub-100$\,$Myr planets are currently enveloped by primordial H/He
atmospheres, and that they will become sub-Neptune sized planets after
undergoing atmospheric escape \citep[{\it
e.g.},][]{Fortney_et_al_2007,Owen_Wu_2013,gupta_sculpting_2019,gupta_signatures_2020}.
}

While we have statistically validated that \tn\ is a planet, the
possibility that it could be a background eclipsing binary has not
been excluded with sufficient confidence to call the planet
``confirmed''.  The distinction is methodological.  Our calculations
have shown that at a population level we expect negligibly few BEBs
within $\approx0.3''$ of \tn\ to produce eclipses of the appropriate
shape across all bandpasses, with no observed secondary eclipse or
odd-even variations.  This statement is
tautologically ``validation'', but it is weaker than having data on
hand that conclusively rules in favor of the planetary interpretation.

The easiest way to confirm the planetary nature of \tn\
will be a Rossiter-McLaughlin (RM) measurement.  Detection of an RM
signal consistent with the photometric transit would rule out BEB and
HEB scenarios, as it would imply that the eclipsing object is bound to
the target star. Combined with our non-detection of the planet's mass
from radial velocity monitoring, this would confirm that \pn\ is a
planet.

The maximum amplitude of the Rossiter-McLaughlin anomaly is
\citep{gaudi_prospects_2007}
\begin{align}
  \Delta V_{\rm RM} &\approx f_{\rm LD} \cdot \delta \cdot v\sin i \cdot \sqrt{1-b^2}
  \approx 14\,{\rm m\,s}^{-1},
\end{align}
for
\begin{align}
  f_{\rm LD} &= 1 - u_1 ( 1 - \mu ) - u_2 (1 - \mu )^2,
\end{align}
where $\mu \approx (1 - b^2)^{1/2}$, $u_i$ are the limb-darkening
parameters, and for calculation purposes we
assumed $b=0.95$ and used stellar and transit parameters from
Tables~\ref{tab:starparams} and~\ref{tab:posterior}.  Although
challenging, for a 1.9$\,$hr transit of a $V=10.6$ star, a detection
could be achieved with modern spectrographs.  The next viable total
transit windows from Chile occur in January and February of 2021;
there are also a few visible per season from other southern locations.
The most precise available ephemeris, found from our joint fit of the
TESS and ground-based photometry, is as follows.
\begin{align}
  t_0\ [{\rm BJD_{TDB}}] &= 2458574.272527 \pm 0.000593 \nonumber \\
  P\ [{\rm d}] &= 8.3248762 \pm 0.0000157 \nonumber \\
  T_{14}\ [{\rm hr}] &= 1.96 \pm 0.04.
  \label{eq:ephem}
\end{align}

The Rossiter-McLaughlin approach is more likely to yield short-term
success than a direct mass measurement because of the RV noise
expected to be induced by stellar rotation.  The photometric amplitude
induced by starspots on \tn\ is $\approx 2\%$.  The spot-induced RV
variation expected over the course of the $\approx 3\,{\rm d}$
rotation period can be estimated by multiplying the photometric
amplitude and spectroscopic equatorial velocity.  This gives
$\sigma_{\rm RV,rot} \approx 300\,{\rm m\,s}^{-1}$, and is consistent
with the scatter we observe in our radial velocities from FEROS.
Detecting a planet's Keplerian motion in this regime is challenging,
and requires a significant amount of data and care in signal
extraction \citep{barragan_radial_2019,stefansson_k2-25_2020}.  The
Rossiter-McLaughlin measurement avoids the majority of this issue
because the transit occurs over a much shorter duration than a single
stellar rotation period.  

If the RM measurements prove that the validated planet is real,
measuring its mass may be worth the effort, because it would improve
understanding of the planet's \replaced{contraction and
photoevaporation}{composition and future atmospheric evolution}.  If
an RV campaign were timed to coincide with TESS Sectors 36 and 37 (3
March 2021 through 28 April 2021), it would significantly ease
extraction of the Keplerian signal.  The reason is that the RVs,
activity indicators, and photometry could be modeled simultaneously
\citep[{\it e.g.},][]{aigrain_simple_2012,rajpaul_gaussian_2015}.
Combining photometric and radial velocity data from non-overlapping
epochs would also constrain the models, but perhaps not quite as
convincingly.

While we hope that RV observations will be pursued, data acquired
during the TESS mission extension may also help in understanding the
system \citep{bouma_extend_2017,huang_expected_2018}.\deleted{ Any
misalignment between the star's spin axis and the planet's orbit
should induce nodal precession, which could yield large changes in the
transit duration given the system's high impact parameter.  If
observed, this could be a spectroscopy-free method for confirming the
planet.}\added{ In particular, additional photometry will likely
enable more detailed exploration of whether the orbit of \tn\ is
eccentric, and also whether the system could host additional
transiting planets. }


\acknowledgements
\raggedbottom

\added{The authors thank K.~Anderson for fruitful discussions, and are
also grateful to the anonymous referee for their constructive comments
and suggestions.}
L.G.B. and J.H. acknowledge support by the TESS GI Program, program
G011103, through NASA grant 80NSSC19K0386.
This study was based in part on observations at Cerro Tololo
Inter-American Observatory at NSF's NOIRLab (NOIRLab Prop. ID
2020A-0146; PI: L{.}~Bouma), which is managed by the Association of
Universities for Research in Astronomy (AURA) under a cooperative
agreement with the National Science Foundation.
This paper includes data collected by the TESS mission, which are
publicly available from the Mikulski Archive for Space Telescopes
(MAST).
Funding for the TESS mission is provided by NASA's Science Mission
directorate.  \tn\ was included on the TESS 2-minute target list in
part thanks to the Guest Investigator program of G.\ Sacco (G011265).

The ASTEP project acknowledges support from the French and Italian
Polar Agencies, IPEV and PNRA, and from Universit\'e C\^ote d'Azur
under Idex UCAJEDI (ANR-15-IDEX-01). We thank the dedicated staff at
Concordia for their continuous presence and support throughout the
Austral winter.
This research received funding from the European Research Council
(ERC) under the European Union's Horizon 2020 research and innovation
programme (grant n$^\circ$ 803193/BEBOP), and from the
Science and Technology Facilities Council (STFC; grant n$^\circ$
ST/S00193X/1).
%
%

This research was based in part on observations obtained at the
Southern Astrophysical Research (SOAR) telescope, which is a joint
project of the Minist\'{e}rio da Ci\^{e}ncia, Tecnologia e
Inova\c{c}\~{o}es (MCTI/LNA) do Brasil, the US National Science
Foundation's NOIRLab, the University of North Carolina at Chapel Hill
(UNC), and Michigan State University (MSU).

This research made use of the Exoplanet Follow-up Observation
Program website, which is operated by the California Institute of
Technology, under contract with the National Aeronautics and Space
Administration under the Exoplanet Exploration Program.

This research made use of the SVO Filter Profile Service
(\url{http://svo2.cab.inta-csic.es/theory/fps/}) supported from the Spanish
MINECO through grant AYA2017-84089.

Resources supporting this work were provided by the NASA High-End
Computing (HEC) Program through the NASA Advanced Supercomputing (NAS)
Division at Ames Research Center for the production of the SPOC data
products.

A.J.\ and R.B.\ acknowledge support from project IC120009 ``Millennium
Institute of Astrophysics (MAS)'' of the Millenium Science Initiative,
Chilean Ministry of Economy. A.J.\ acknowledges additional support
from FONDECYT project 1171208.  J.I.V\ acknowledges support from
CONICYT-PFCHA/Doctorado Nacional-21191829.  R.B.\ acknowledges support
from FONDECYT Post-doctoral Fellowship Project 3180246.
C.T.\ and C.B\ acknowledge support from Australian Research Council
grants LE150100087, LE160100014, LE180100165, DP170103491 and
DP190103688.
C.Z.\ is supported by a Dunlap Fellowship at the Dunlap Institute for
Astronomy \& Astrophysics, funded through an endowment established by
the Dunlap family and the University of Toronto.
D.D.\ acknowledges support through the TESS Guest Investigator Program
Grant 80NSSC19K1727.

\software{
  \texttt{arviz} \citep{arviz_2019},
  \texttt{astrobase} \citep{bhatti_astrobase_2018},
	\texttt{AstroImageJ} \citep{collins_astroimagej_2017},
  \texttt{astropy} \citep{astropy_2018},
  \texttt{astroquery} \citep{astroquery_2018},
  \texttt{ceres} \citep{brahm_2017_ceres},
  \texttt{cdips-pipeline} \citep{bhatti_cdips-pipeline_2019},
  \texttt{corner} \citep{corner_2016},
  \texttt{exoplanet} \citep{exoplanet:exoplanet}, and its
  dependencies \citep{exoplanet:agol20, exoplanet:kipping13, exoplanet:luger18,
  	exoplanet:theano},
	\texttt{IDL Astronomy User's Library} \citep{landsman_1995},
  \texttt{IPython} \citep{perez_2007},
	\texttt{isochrones} \citep{morton_2015_isochrones},
	\texttt{lightkurve} \citep{lightkurve_2018},
  \texttt{matplotlib} \citep{hunter_matplotlib_2007}, 
  \texttt{MESA} \citep{paxton_modules_2011,paxton_modules_2013,paxton_modules_2015}
  \texttt{numpy} \citep{walt_numpy_2011}, 
  \texttt{pandas} \citep{mckinney-proc-scipy-2010},
  \texttt{pyGAM} \citep{serven_pygam_2018_1476122},
  \texttt{PyMC3} \citep{salvatier_2016_PyMC3},
  \texttt{radvel} \citep{fulton_radvel_2018},
  \texttt{scipy} \citep{jones_scipy_2001},
  \texttt{tesscut} \citep{brasseur_astrocut_2019},
	\texttt{VESPA} \citep{morton_efficient_2012,vespa_2015},
  \texttt{webplotdigitzer} \citep{rohatgi_2019},
  \texttt{wotan} \citep{hippke_wotan_2019}.
}
\ 

\facilities{
 	{\it Astrometry}:
 	Gaia \citep{gaia_collaboration_gaia_2016,gaia_collaboration_gaia_2018}.
 	{\it Imaging}:
    Second Generation Digitized Sky Survey,
    SOAR~(HRCam; \citealt{tokovinin_ten_2018}).
 	{\it Spectroscopy}:
	CTIO1.5$\,$m~(CHIRON; \citealt{tokovinin_chironfiber_2013}),
    MPG2.2$\,$m~(FEROS; \citealt{kaufer_commissioning_1999}),
	AAT~(Veloce; \citealt{gilbert_veloce_2018}).
 	{\it Photometry}:
 	  ASTEP:0.40$\,$m (ASTEP400),
	  El Sauce:0.356$\,$m,
 	TESS \citep{ricker_transiting_2015}.
}

\begin{table*}
\scriptsize
\setlength{\tabcolsep}{2pt}
\centering
\caption{Literature and Measured Properties for TOI$\,$837}
\label{tab:starparams}
\begin{tabular}{llcc}
  \hline
  \hline
Other identifiers\dotfill & \\
\multicolumn{3}{c}{TIC 460205581} \\
\multicolumn{3}{c}{GAIADR2 5251470948229949568} \\
\hline
\hline
Parameter & Description & Value & Source\\
\hline 
$\alpha_{J2015.5}$\dotfill	&Right Ascension (hh:mm:ss)\dotfill & 10:28:08.95 & 1	\\
$\delta_{J2015.5}$\dotfill	&Declination (dd:mm:ss)\dotfill & -64:30:18.76 & 1	\\
$l_{J2015.5}$\dotfill	&Galactic Longitude (deg)\dotfill & 288.2644 & 1	\\
$b_{J2015.5}$\dotfill	&Galactic Latitude (deg)\dotfill & -5.7950 & 1	\\
\\
B\dotfill			&Johnson B mag.\dotfill & 11.119 $\pm$ 0.107		& 2	\\
V\dotfill			&Johnson V mag.\dotfill & 10.635 $\pm$ 0.020		& 2	\\
${\rm G}$\dotfill     & Gaia $G$ mag.\dotfill     & 10.356$\pm$0.020 & 1\\
${\rm Bp}$\dotfill     & Gaia $Bp$ mag.\dotfill     & 10.695 $\pm$0.020 & 1\\
${\rm Rp}$\dotfill     & Gaia $Rp$ mag.\dotfill     & 9.887$\pm$0.020 & 1\\
${\rm T}$\dotfill     & TESS mag.\dotfill     & 9.9322$\pm$0.006 & 2\\
J\dotfill			& 2MASS J mag.\dotfill & 9.392  $\pm$ 0.030	& 3	\\
H\dotfill			& 2MASS H mag.\dotfill & 9.108 $\pm$ 0.038	    &  3	\\
K$_{\rm S}$\dotfill			& 2MASS ${\rm K_S}$ mag.\dotfill & 8.933 $\pm$ 0.026 &  3	\\
W1\dotfill		& WISE1 mag.\dotfill & 8.901 $\pm$ 0.023 & 4	\\
W2\dotfill		& WISE2 mag.\dotfill & 8.875 $\pm$ 0.021 &  4 \\
W3\dotfill		& WISE3 mag.\dotfill &  8.875 $\pm$ 0.020& 4	\\
W4\dotfill		& WISE4 mag.\dotfill & 8.936 $\pm$ N/A &  4	\\
\\
$\pi$\dotfill & Gaia DR2 parallax (mas) \dotfill & 6.989 $\pm$ 0.022 &  1 \\
$d$\dotfill & Distance (pc)\dotfill & $143.1 \pm 0.5$ & 1 \\
$\mu_{\alpha}$\dotfill		& Gaia DR2 proper motion\dotfill & -18.017 $\pm$ 0.039 & 1 \\
                    & \hspace{3pt} in RA (mas yr$^{-1}$)	&  \\
$\mu_{\delta}$\dotfill		& Gaia DR2 proper motion\dotfill 	&  11.307 $\pm$ 0.037 &  1 \\
                    & \hspace{3pt} in DEC (mas yr$^{-1}$) &  \\
RV\dotfill & Systemic radial \hspace{9pt}\dotfill  & $17.44 \pm 0.64$$^{\dagger}$ & 1 \\
                    & \hspace{3pt} velocity (\kms)  & \\
\\
$v\sin{i_\star}$\dotfill &  Rotational velocity (\kms) \hspace{9pt}\dotfill &  16.2 $\pm$ 1.1 & 5 \\
$v_{\rm mac}$\dotfill &  Macroturbulence velocity (\kms) \hspace{9pt}\dotfill &  8.4 $\pm$ 2.9 & 5 \\
${\rm [Fe/H]}$\dotfill &   Metallicity \hspace{9pt}\dotfill & -0.069 $\pm$ 0.042 & 5 \\
$T_{\rm eff}$\dotfill &  Effective Temperature (K) \hspace{9pt}\dotfill & 6047 $\pm$ 162 &  6  \\
$\log{g_{\star}}$\dotfill &  Surface Gravity (cgs)\hspace{9pt}\dotfill &  4.467 $\pm$ 0.049  &  6 \\
Li EW\dotfill & 6708\AA\ Equiv{.} Width (m\AA) \dotfill & $154 \pm 9$  & 7 \\
$P_{\rm rot}$\dotfill & Rotation period (d)\dotfill & $3.004\pm 0.053$  & 8 \\
Age & Adopted stellar age (Myr)\dotfill & $30$--$46$  &  9 \\
%
Spec. Type\dotfill & Spectral Type\dotfill & 	G0/F9 V & 5 \\
$R_\star$\dotfill & Stellar radius ($R_\odot$)\dotfill & 1.022$\pm$0.083 & 6 \\
$M_\star$\dotfill & Stellar mass ($R_\odot$)\dotfill & 1.118$\pm$0.059 & 6 \\
$A_{\rm V}$\dotfill & Interstellar reddening (mag)\dotfill & 0.20$\pm$0.03 & 10 \\
\hline
\end{tabular}
\begin{flushleft}
 \footnotesize{ \textsc{NOTE}---
$\dagger$ Systemic RV uncertainty is the standard deviation of single-transit radial velocities, as quoted in Gaia DR2. 
Provenances are:
$^1$\citet{gaia_collaboration_gaia_2018},
$^2$\citet{stassun_TIC8_2019},
$^3$\citet{skrutskie_tmass_2006},
$^4$\citet{wright_WISE_2010},
$^5$CHIRON spectra,
$^6$Method~2 (cluster isochrone, Section~\ref{subsec:starparams}),
$^7$FEROS spectra,
$^8$TESS light curve,
$^9$IC~2602 ages from isochrone \& lithium depletion analyses (Section~\ref{subsec:clusterchar}),
$^{10}$Method~1 (photometric SED fit, Section~\ref{subsec:starparams}).}
\end{flushleft}
\vspace{-0.5cm}
\end{table*}

\begin{deluxetable*}{lllrrrrr}
  \tablecaption{ Priors and posteriors for the model fitted to the
  TESS and ground-based data.}
\label{tab:posterior}
%
%
%
\tablehead{
  \colhead{Param.} & 
  \colhead{Unit} &
  \colhead{Prior} & 
  \colhead{Median} & 
  \colhead{Mean} & 
  \colhead{Std{.} Dev.} &
  \colhead{3\%} &
  \colhead{97\%}
}

\startdata
{\it Sampled: physical} & & & & & & & \\
\hline
$P$ & d & $\mathcal{N}(8.3249; 0.1000)$ & 8.3248762 & 8.3248762 & 0.0000157 & 8.3248466 & 8.3249057 \\
$t_0^{(1)}$ & d & $\mathcal{N}(1574.273800; 0.1000)$ & 1574.272527 & 1574.2725263 & 0.0005931 & 1574.2713991 & 1574.273626 \\
$\log R_{\rm p}/R_\star$ & -- & $\mathcal{U}(-4.605; 0.000)$ & -2.58156 & -2.56901 & 0.06659 & -2.67426 & -2.44791 \\
$b$ & -- & $\mathcal{U}(0; 1+R_{\mathrm{p}}/R_\star)$ & 0.9358 & 0.9374 & 0.0127 & 0.9164 & 0.9615 \\
$u_1$ & -- & $\mathcal{U}(0.175; 0.475)$$^{(2)}$ & 0.344 & 0.338 & 0.085 & 0.199 & 0.475 \\
$u_2$ & -- & $\mathcal{U}(0.085; 0.385)$$^{(2)}$ & 0.251 & 0.245 & 0.085 & 0.108 & 0.385 \\
$R_\star$ & $R_\odot$ & $\mathcal{T}(1.022; 0.083)$ & 1.042 & 1.042 & 0.076 & 0.902 & 1.189 \\
$\log g$ & cgs & $\mathcal{N}(4.467; 0.049)$ & 4.451 & 4.451 & 0.042 & 4.372 & 4.528 \\
\hline
{\it Sampled: nuisance} & & & & & & & \\
\hline
$a_{00;\mathrm{TESS}}$ & -- & $\mathcal{N}(1.00; 0.01)$ & 0.9986 & 0.9986 & 0.0001 & 0.9984 & 0.9987 \\
$a_{01;\mathrm{TESS}}$ & d$^{-1}$ & $\mathcal{U}(-0.05; 0.05)$ & -0.0004 & -0.0004 & 0.0003 & -0.0010 & 0.0003 \\
$a_{02;\mathrm{TESS}}$ & d$^{-2}$ & $\mathcal{U}(-0.05; 0.05)$ & -0.0183 & -0.0183 & 0.0023 & -0.0226 & -0.0141 \\
$a_{10;\mathrm{TESS}}$ & -- & $\mathcal{N}(1.00; 0.01)$ & 1.0090 & 1.0090 & 0.0001 & 1.0088 & 1.0092 \\
$a_{11;\mathrm{TESS}}$ & d$^{-1}$ & $\mathcal{U}(-0.05; 0.05)$ & -0.0138 & -0.0138 & 0.0003 & -0.0144 & -0.0132 \\
$a_{12;\mathrm{TESS}}$ & d$^{-2}$ & $\mathcal{U}(-0.05; 0.05)$ & -0.0550 & -0.0550 & 0.0022 & -0.0591 & -0.0508 \\
$a_{20;\mathrm{TESS}}$ & -- & $\mathcal{N}(1.00; 0.01)$ & 0.9992 & 0.9992 & 0.0001 & 0.9990 & 0.9993 \\
$a_{21;\mathrm{TESS}}$ & d$^{-1}$ & $\mathcal{U}(-0.05; 0.05)$ & 0.0156 & 0.0156 & 0.0004 & 0.0150 & 0.0163 \\
$a_{22;\mathrm{TESS}}$ & d$^{-2}$ & $\mathcal{U}(-0.05; 0.05)$ & 0.0232 & 0.0232 & 0.0024 & 0.0187 & 0.0276 \\
$a_{30;\mathrm{TESS}}$ & -- & $\mathcal{N}(1.00; 0.01)$ & 1.0013 & 1.0013 & 0.0001 & 1.0011 & 1.0015 \\
$a_{31;\mathrm{TESS}}$ & d$^{-1}$ & $\mathcal{U}(-0.05; 0.05)$ & 0.0021 & 0.0021 & 0.0004 & 0.0014 & 0.0028 \\
$a_{32;\mathrm{TESS}}$ & d$^{-2}$ & $\mathcal{U}(-0.05; 0.05)$ & -0.0097 & -0.0097 & 0.0029 & -0.0150 & -0.0043 \\
$a_{40;\mathrm{TESS}}$ & -- & $\mathcal{N}(1.00; 0.01)$ & 0.9906 & 0.9906 & 0.0001 & 0.9905 & 0.9908 \\
$a_{41;\mathrm{TESS}}$ & d$^{-1}$ & $\mathcal{U}(-0.05; 0.05)$ & 0.0015 & 0.0015 & 0.0003 & 0.0009 & 0.0022 \\
$a_{42;\mathrm{TESS}}$ & d$^{-2}$ & $\mathcal{U}(-0.05; 0.05)$ & 0.0313 & 0.0313 & 0.0023 & 0.0269 & 0.0356 \\
$a_{00;\mathrm{Sauce}}$ & -- & $\mathcal{N}(1.00; 0.01)$ & 0.9996 & 0.9996 & 0.0001 & 0.9993 & 0.9998 \\
$a_{01;\mathrm{Sauce}}$ & d$^{-1}$ & $\mathcal{U}(-0.05; 0.05)$ & -0.0041 & -0.0041 & 0.0023 & -0.0085 & 0.0002 \\
$a_{02;\mathrm{Sauce}}$ & d$^{-2}$ & $\mathcal{U}(-0.05; 0.05)$ & 0.0311 & 0.0256 & 0.0205 & -0.0135 & 0.0500 \\
$a_{10;\mathrm{Sauce}}$ & -- & $\mathcal{N}(1.00; 0.01)$ & 0.9998 & 0.9998 & 0.0001 & 0.9996 & 1.0000 \\
$a_{11;\mathrm{Sauce}}$ & d$^{-1}$ & $\mathcal{U}(-0.05; 0.05)$ & -0.0004 & -0.0004 & 0.0021 & -0.0044 & 0.0035 \\
$a_{12;\mathrm{Sauce}}$ & d$^{-2}$ & $\mathcal{U}(-0.05; 0.05)$ & 0.0360 & 0.0314 & 0.0165 & 0.0005 & 0.05000 \\
$a_{20;\mathrm{Sauce}}$ & -- & $\mathcal{N}(1.00; 0.01)$ & 0.9999 & 0.9999 & 0.0001 & 0.9996 & 1.0001 \\
$a_{21;\mathrm{Sauce}}$ & d$^{-1}$ & $\mathcal{U}(-0.05; 0.05)$ & -0.0009 & -0.0009 & 0.0030 & -0.0066 & 0.0046 \\
$a_{22;\mathrm{Sauce}}$ & d$^{-2}$ & $\mathcal{U}(-0.05; 0.05)$ & 0.0067 & 0.0047 & 0.0278 & -0.0410 & 0.0500 \\
$a_{30;\mathrm{Sauce}}$ & -- & $\mathcal{N}(1.00; 0.01)$ & 0.9996 & 0.9996 & 0.0003 & 0.9991 & 1.0001 \\
$a_{31;\mathrm{Sauce}}$ & d$^{-1}$ & $\mathcal{U}(-0.05; 0.05)$ & 0.0047 & 0.0047 & 0.0069 & -0.0084 & 0.0176 \\
$a_{32;\mathrm{Sauce}}$ & d$^{-2}$ & $\mathcal{U}(-0.05; 0.05)$ & 0.0077 & 0.0052 & 0.0286 & -0.0419 & 0.0500 \\
$a_{00;\mathrm{ASTEP}}$ & -- & $\mathcal{N}(1.00; 0.01)$ & 0.9996 & 0.9996 & 0.0001 & 0.9994 & 0.9998 \\
$a_{01;\mathrm{ASTEP}}$ & d$^{-1}$ & $\mathcal{U}(-0.05; 0.05)$ & 0.0022 & 0.0022 & 0.0007 & 0.0010 & 0.0035 \\
$a_{02;\mathrm{ASTEP}}$ & d$^{-2}$ & $\mathcal{U}(-0.05; 0.05)$ & 0.0073 & 0.0074 & 0.0076 & -0.0067 & 0.0219 \\
$a_{10;\mathrm{ASTEP}}$ & -- & $\mathcal{N}(1.00; 0.01)$ & 1.0000 & 1.0000 & 0.0001 & 0.9998 & 1.0002 \\
$a_{11;\mathrm{ASTEP}}$ & d$^{-1}$ & $\mathcal{U}(-0.05; 0.05)$ & 0.0042 & 0.0042 & 0.0010 & 0.0024 & 0.0061 \\
$a_{12;\mathrm{ASTEP}}$ & d$^{-2}$ & $\mathcal{U}(-0.05; 0.05)$ & 0.0164 & 0.0162 & 0.0145 & -0.0105 & 0.0439 \\
$a_{20;\mathrm{ASTEP}}$ & -- & $\mathcal{N}(1.00; 0.01)$ & 0.9993 & 0.9993 & 0.0001 & 0.9991 & 0.9995 \\
$a_{21;\mathrm{ASTEP}}$ & d$^{-1}$ & $\mathcal{U}(-0.05; 0.05)$ & -0.0074 & -0.0074 & 0.0016 & -0.0103 & -0.0044 \\
$a_{22;\mathrm{ASTEP}}$ & d$^{-2}$ & $\mathcal{U}(-0.05; 0.05)$ & 0.0204 & 0.0173 & 0.0216 & -0.0209 & 0.0500 \\
\hline
{\it Derived} & & & & & & & \\
\hline
$R_{\rm p}/R_\star$ & -- & -- & 0.08 & 0.08 & 0.01 & 0.07 & 0.09 \\
$\rho_\star$ & g$\ $cm$^{-3}$ & -- & 1.40 & 1.40 & 0.15 & 1.13 & 1.68 \\
$R_{\rm p}$ & $R_{\mathrm{Jup}}$ & -- & 0.77 & 0.78 & 0.09 & 0.62 & 0.95 \\
$a/R_\star$ & -- & -- & 17.26 & 17.24 & 0.60 & 16.12 & 18.36 \\
$\cos i$ & -- & -- & 0.054 & 0.054 & 0.003 & 0.050 & 0.059 \\
$T_{14}$ & hr & -- & 1.957 & 1.955 & 0.039 & 1.887 & 2.032 \\
$T_{13}$ & hr & -- & 0.30 & 0.30 & 0.13 & 0.004 & NaN$^{(3)}$ \\
\enddata
\tablecomments{
(1) The most precise ephemeris based on the combination of TESS and
ground-based data is also shown in Equation~\ref{eq:ephem}.
%
%
(2) Assuming an informative quadratic limb-darkening prior with
values about those given for the appropriate $T_{\rm eff}$ and
$\log g$ in TESS-band from \citet{claret_limb_2017}. The precision
achieved in the ground-based data did not appear to necessitate using
bandpass-dependent limb-darkening coefficients.
(3) The second and third contact points do not exist for a grazing transit.
{\it Notation}:
$a_{ij;\mathrm{Instr}}$ denotes the $i^{\rm th}$ transit of a
particular instrument, and the $j^{\rm th}$ polynomial detrending
order.
$\mathcal{U}$ denotes a uniform distribution,
$\mathcal{N}$ a normal distribution, and
$\mathcal{T}$ a truncated normal bounded between zero and an upper limit much larger than the mean.
}
\vspace{-0.3cm}
\end{deluxetable*}

\clearpage
\bibliographystyle{yahapj}                            
\bibliography{bibliography}

\begin{thebibliography}{}
\providecommand\natexlab[1]{#1}
\providecommand\JournalTitle[1]{#1}

\bibitem[{Abe {et~al.}(2013)Abe, Gonçalves, Agabi, Alapini, Guillot,
  Mékarnia, Rivet, Schmider, Crouzet, Fortney, Pont, Barbieri, Daban,
  Fanteï-Caujolle, Gouvret, Bresson, Roussel, Bonhomme, Robini, Dugué,
  Bondoux, Péron, Petit, Szulágyi, Fruth, Erikson, Rauer, Fressin,
  Valbousquet, Blanc, Le~van Suu, \& Aigrain}]{abe_secondary_2013}
Abe, L., Gonçalves, I., Agabi, A., {et~al.} 2013,
  \href{http://dx.doi.org/10.1051/0004-6361/201220351}{\JournalTitle{\aap},
  553, A49}

\bibitem[{Addison {et~al.}(2020)Addison, Horner, Wittenmyer, Plavchan, Wright,
  Nicholson, Marshall, Clark, Kane, Hirano, Kielkopf, Shporer, Tinney, Zhang,
  Ballard, Bedding, Bowler, Mengel, Okumura, \& Gaidos}]{addison_youngest_2020}
Addison, B.~C., Horner, J., Wittenmyer, R.~A., {et~al.} 2020,
  \href{http://arxiv.org/abs/2006.13675}{\JournalTitle{arXiv:2006.13675
  [astro-ph]}}

\bibitem[{{Agol} {et~al.}(2020){Agol}, {Luger}, \&
  {Foreman-Mackey}}]{exoplanet:agol20}
{Agol}, E., {Luger}, R., \& {Foreman-Mackey}, D. 2020,
  \href{http://dx.doi.org/10.3847/1538-3881/ab4fee}{\JournalTitle{\aj}, 159,
  123}

\bibitem[{Aigrain {et~al.}(2007)Aigrain, Hodgkin, Irwin, Hebb, Irwin, Favata,
  Moraux, \& Pont}]{aigrain_monitor_2007}
Aigrain, S., Hodgkin, S., Irwin, J., {et~al.} 2007,
  \href{http://dx.doi.org/10.1111/j.1365-2966.2006.11303.x}{\JournalTitle{\mnras},
  375, 29}

\bibitem[{Aigrain {et~al.}(2012)Aigrain, Pont, \& Zucker}]{aigrain_simple_2012}
Aigrain, S., Pont, F., \& Zucker, S. 2012,
  \href{http://dx.doi.org/10.1111/j.1365-2966.2011.19960.x}{\JournalTitle{\mnras},
  419, 3147}

\bibitem[{Angus {et~al.}(2015)Angus, Aigrain, Foreman-Mackey, \&
  McQuillan}]{angus_calibrating_2015}
Angus, R., Aigrain, S., Foreman-Mackey, D., \& McQuillan, A. 2015,
  \href{http://dx.doi.org/10.1093/mnras/stv423}{\JournalTitle{\mnras}, 450,
  1787}

\bibitem[{{Astropy Collaboration} {et~al.}(2018){Astropy Collaboration},
  {Price-Whelan}, {Sip{\H{o}}cz}, {G{\"u}nther}, {Lim}, {Crawford}, {Conseil},
  {Shupe}, {Craig}, {Dencheva}, {Ginsburg}, {Vand erPlas}, {Bradley},
  {P{\'e}rez-Su{\'a}rez}, {de Val-Borro}, {Aldcroft}, {Cruz}, {Robitaille},
  {Tollerud}, {Ardelean}, {Babej}, {Bach}, {Bachetti}, {Bakanov}, {Bamford},
  {Barentsen}, {Barmby}, {Baumbach}, {Berry}, {Biscani}, {Boquien}, {Bostroem},
  {Bouma}, {Brammer}, {Bray}, {Breytenbach}, {Buddelmeijer}, {Burke},
  {Calderone}, {Cano Rodr{\'\i}guez}, {Cara}, {Cardoso}, {Cheedella}, {Copin},
  {Corrales}, {Crichton}, {D'Avella}, {Deil}, {Depagne}, {Dietrich}, {Donath},
  {Droettboom}, {Earl}, {Erben}, {Fabbro}, {Ferreira}, {Finethy}, {Fox},
  {Garrison}, {Gibbons}, {Goldstein}, {Gommers}, {Greco}, {Greenfield},
  {Groener}, {Grollier}, {Hagen}, {Hirst}, {Homeier}, {Horton}, {Hosseinzadeh},
  {Hu}, {Hunkeler}, {Ivezi{\'c}}, {Jain}, {Jenness}, {Kanarek}, {Kendrew},
  {Kern}, {Kerzendorf}, {Khvalko}, {King}, {Kirkby}, {Kulkarni}, {Kumar},
  {Lee}, {Lenz}, {Littlefair}, {Ma}, {Macleod}, {Mastropietro}, {McCully},
  {Montagnac}, {Morris}, {Mueller}, {Mumford}, {Muna}, {Murphy}, {Nelson},
  {Nguyen}, {Ninan}, {N{\"o}the}, {Ogaz}, {Oh}, {Parejko}, {Parley}, {Pascual},
  {Patil}, {Patil}, {Plunkett}, {Prochaska}, {Rastogi}, {Reddy Janga},
  {Sabater}, {Sakurikar}, {Seifert}, {Sherbert}, {Sherwood-Taylor}, {Shih},
  {Sick}, {Silbiger}, {Singanamalla}, {Singer}, {Sladen}, {Sooley},
  {Sornarajah}, {Streicher}, {Teuben}, {Thomas}, {Tremblay}, {Turner},
  {Terr{\'o}n}, {van Kerkwijk}, {de la Vega}, {Watkins}, {Weaver}, {Whitmore},
  {Woillez}, \& {Zabalza}}]{astropy_2018}
{Astropy Collaboration}, {Price-Whelan}, A.~M., {Sip{\H{o}}cz}, B.~M., {et~al.}
  2018, \href{http://dx.doi.org/10.3847/1538-3881/aabc4f}{\JournalTitle{\aj},
  156, 123}

\bibitem[{Bakos {et~al.}(2006)Bakos, P\'{a}l, Latham, Noyes, \&
  Stefanik}]{bakos_stellar_2006}
Bakos, G.~A., P\'{a}l, A., Latham, D.~W., Noyes, R.~W., \& Stefanik, R.~P.
  2006, \href{http://dx.doi.org/10.1086/503671}{\JournalTitle{\apjl}, 641, L57}

\bibitem[{Baraffe {et~al.}(2003)Baraffe, Chabrier, Barman, Allard, \&
  Hauschildt}]{baraffe_evolutionary_2003}
Baraffe, I., Chabrier, G., Barman, T.~S., Allard, F., \& Hauschildt, P.~H.
  2003,
  \href{http://dx.doi.org/10.1051/0004-6361:20030252}{\JournalTitle{\aap}, 402,
  701}

\bibitem[{Baratella {et~al.}(2020)Baratella, D'Orazi, Carraro, Desidera,
  Randich, Magrini, Adibekyan, Smiljanic, Spina, Tsantaki, Tautvaišienė,
  Sousa, Jofré, Jiménez-Esteban, Delgado-Mena, Martell, Van~der Swaelmen,
  Roccatagliata, Gilmore, Alfaro, Bayo, Bensby, Bragaglia, Franciosini,
  Gonneau, Heiter, Hourihane, Jeffries, Koposov, Morbidelli, Prisinzano, Sacco,
  Sbordone, Worley, Zaggia, \& Lewis}]{baratella_gaia-eso_2020}
Baratella, M., D'Orazi, V., Carraro, G., {et~al.} 2020,
  \href{http://dx.doi.org/10.1051/0004-6361/201937055}{\JournalTitle{\aap},
  634, A34}

\bibitem[{Barnes {et~al.}(2015)Barnes, Weingrill, Granzer, Spada, \&
  Strassmeier}]{barnes_color-period_2015}
Barnes, S.~A., Weingrill, J., Granzer, T., Spada, F., \& Strassmeier, K.~G.
  2015,
  \href{http://dx.doi.org/10.1051/0004-6361/201526129}{\JournalTitle{\aap},
  583, A73}

\bibitem[{Barrag\'{a}n {et~al.}(2019)Barrag\'{a}n, Aigrain, Kubyshkina,
  Gandolfi, Livingston, Fridlund, Fossati, Korth, Parviainen, Malavolta, Palle,
  Deeg, Nowak, Rajpaul, Zicher, Antoniciello, Narita, Albrecht, Bedin, Cabrera,
  Cochran, de~Leon, Eigmüller, Fukui, Granata, Grziwa, Guenther, Hatzes,
  Kusakabe, Latham, Libralato, Luque, Monta~nés Rodríguez, Murgas, Nardiello,
  Pagano, Piotto, Persson, Redfield, \& Tamura}]{barragan_radial_2019}
Barrag\'{a}n, O., Aigrain, S., Kubyshkina, D., {et~al.} 2019,
  \href{http://dx.doi.org/10.1093/mnras/stz2569}{\JournalTitle{\mnras}}

\bibitem[{Berger {et~al.}(2018)Berger, Howard, \&
  Boesgaard}]{berger_identifying_2018}
Berger, T.~A., Howard, A.~W., \& Boesgaard, A.~M. 2018,
  \href{http://dx.doi.org/10.3847/1538-4357/aab154}{\JournalTitle{\apj}, 855,
  115}

\bibitem[{Berger {et~al.}(2020)Berger, Huber, Gaidos, van Saders, \&
  Weiss}]{berger_gaia-kepler_2020}
Berger, T.~A., Huber, D., Gaidos, E., van Saders, J.~L., \& Weiss, L.~M. 2020,
  \href{http://adsabs.harvard.edu/abs/2020arXiv200514671B}{\JournalTitle{arXiv
  e-prints}, 2005}

\bibitem[{Bhatti {et~al.}(2019)Bhatti, Bouma, \&
  Yee}]{bhatti_cdips-pipeline_2019}
Bhatti, W., Bouma, L., \& Yee, S. 2019, \texttt{cdips-pipeline v0.1.0},
  \url{https://doi.org/10.5281/zenodo.3370324}

\bibitem[{Bhatti {et~al.}(2018)Bhatti, Bouma, \&
  Wallace}]{bhatti_astrobase_2018}
Bhatti, W., Bouma, L.~G., \& Wallace, J. 2018, \texttt{astrobase},
  \url{https://doi.org/10.5281/zenodo.1469822}

\bibitem[{Biddle {et~al.}(2018)Biddle, Johns-Krull, Llama, Prato, \&
  Skiff}]{biddle_k2_2018}
Biddle, L.~I., Johns-Krull, C.~M., Llama, J., Prato, L., \& Skiff, B.~A. 2018,
  \href{http://dx.doi.org/10.3847/2041-8213/aaa897}{\JournalTitle{\apjl}, 853,
  L34}

\bibitem[{Bonomo {et~al.}(2017)Bonomo, Desidera, Benatti, Borsa, Crespi,
  Damasso, Lanza, Sozzetti, Lodato, Marzari, Boccato, Claudi, Cosentino,
  Covino, Gratton, Maggio, Micela, Molinari, Pagano, Piotto, Poretti,
  Smareglia, Affer, Biazzo, Bignamini, Esposito, Giacobbe, Hébrard, Malavolta,
  Maldonado, Mancini, Martinez~Fiorenzano, Masiero, Nascimbeni, Pedani, Rainer,
  \& Scandariato}]{bonomo_gaps_2017}
Bonomo, A.~S., Desidera, S., Benatti, S., {et~al.} 2017,
  \href{http://dx.doi.org/10.1051/0004-6361/201629882}{\JournalTitle{\aap},
  602, A107}

\bibitem[{Borucki {et~al.}(2010)Borucki, Koch, Basri, Batalha, Brown, Caldwell,
  Caldwell, Christensen-Dalsgaard, Cochran, DeVore, Dunham, Dupree, Gautier,
  Geary, Gilliland, Gould, Howell, Jenkins, Kondo, Latham, Marcy, Meibom,
  Kjeldsen, Lissauer, Monet, Morrison, Sasselov, Tarter, Boss, Brownlee, Owen,
  Buzasi, Charbonneau, Doyle, Fortney, Ford, Holman, Seager, Steffen, Welsh,
  Rowe, Anderson, Buchhave, Ciardi, Walkowicz, Sherry, Horch, Isaacson,
  Everett, Fischer, Torres, Johnson, Endl, MacQueen, Bryson, Dotson, Haas,
  Kolodziejczak, Van~Cleve, Chandrasekaran, Twicken, Quintana, Clarke, Allen,
  Li, Wu, Tenenbaum, Verner, Bruhweiler, Barnes, \& Prsa}]{borucki_kepler_2010}
Borucki, W.~J., Koch, D., Basri, G., {et~al.} 2010,
  \href{http://dx.doi.org/10.1126/science.1185402}{\JournalTitle{Science}, 327,
  977}

\bibitem[{Bossini {et~al.}(2019)Bossini, Vallenari, Bragaglia, Cantat-Gaudin,
  Sordo, Balaguer-Núñez, Jordi, Moitinho, Soubiran, Casamiquela, Carrera, \&
  Heiter}]{bossini_age_2019}
Bossini, D., Vallenari, A., Bragaglia, A., {et~al.} 2019,
  \href{http://dx.doi.org/10.1051/0004-6361/201834693}{\JournalTitle{\aap},
  623, A108}

\bibitem[{Bouma {et~al.}(2019)Bouma, Hartman, Bhatti, Winn, \&
  Bakos}]{bouma_cluster_2019}
Bouma, L.~G., Hartman, J.~D., Bhatti, W., Winn, J.~N., \& Bakos, G.~{\'A}.
  2019, \href{http://dx.doi.org/10.3847/1538-4365/ab4a7e}{\JournalTitle{ApJS},
  245, 13}

\bibitem[{Bouma {et~al.}(2017)Bouma, Winn, Kosiarek, \&
  McCullough}]{bouma_extend_2017}
Bouma, L.~G., Winn, J.~N., Kosiarek, J., \& McCullough, P.~R. 2017,
  \href{http://arxiv.org/abs/1705.08891}{\JournalTitle{arXiv:1705.08891
  [astro-ph]}}

\bibitem[{{Brahm} {et~al.}(2017){Brahm}, {Jord{\'a}n}, \&
  {Espinoza}}]{brahm_2017_ceres}
{Brahm}, R., {Jord{\'a}n}, A., \& {Espinoza}, N. 2017,
  \href{http://dx.doi.org/10.1088/1538-3873/aa5455}{\JournalTitle{\pasp}, 129,
  034002}

\bibitem[{{Brahm} {et~al.}(2019){Brahm}, {Espinoza}, {Jord{\'a}n}, {Henning},
  {Sarkis}, {Jones}, {D{\'\i}az}, {Jenkins}, {Vanzi}, {Zapata}, {Petrovich},
  {Kossakowski}, {Rabus}, {Rojas}, \& {Torres}}]{brahm:2019}
{Brahm}, R., {Espinoza}, N., {Jord{\'a}n}, A., {et~al.} 2019,
  \href{http://dx.doi.org/10.3847/1538-3881/ab279a}{\JournalTitle{\aj}, 158,
  45}

\bibitem[{Brasseur {et~al.}(2019)Brasseur, Phillip, Fleming, Mullally, \&
  White}]{brasseur_astrocut_2019}
Brasseur, C.~E., Phillip, C., Fleming, S.~W., Mullally, S.~E., \& White, R.~L.
  2019,
  \href{https://ui.adsabs.harvard.edu/abs/2019ascl.soft05007B/abstract}{\JournalTitle{Astrophysics
  Source Code Library}, ascl:1905.007}

\bibitem[{Bravi {et~al.}(2018)Bravi, Zari, Sacco, Randich, Jeffries, Jackson,
  Franciosini, Moraux, López-Santiago, Pancino, Spina, Wright,
  Jiménez-Esteban, Klutsch, Roccatagliata, Gilmore, Bragaglia, Flaccomio,
  Francois, Koposov, Bayo, Carraro, Costado, Damiani, Frasca, Hourihane,
  Jofré, Lardo, Lewis, Magrini, Morbidelli, Prisinzano, Sousa, Worley, \&
  Zaggia}]{bravi_gaia-eso_2018}
Bravi, L., Zari, E., Sacco, G.~G., {et~al.} 2018,
  \href{http://dx.doi.org/10.1051/0004-6361/201832645}{\JournalTitle{\aap},
  615, A37}

\bibitem[{Bressan {et~al.}(2012)Bressan, Marigo, Girardi, Salasnich, Dal~Cero,
  Rubele, \& Nanni}]{bressan_parsec_2012}
Bressan, A., Marigo, P., Girardi, L., {et~al.} 2012,
  \href{http://dx.doi.org/10.1111/j.1365-2966.2012.21948.x}{\JournalTitle{\mnras},
  427, 127}

\bibitem[{Brucalassi {et~al.}(2017)Brucalassi, Koppenhoefer, Saglia, Pasquini,
  Ruiz, Bonifacio, Bedin, Libralato, Biazzo, Melo, Lovis, \&
  Randich}]{brucalassi_search_2017}
Brucalassi, A., Koppenhoefer, J., Saglia, R., {et~al.} 2017,
  \href{http://dx.doi.org/10.1051/0004-6361/201527562}{\JournalTitle{\aap},
  603, A85}

\bibitem[{Buchhave {et~al.}(2010)Buchhave, Bakos, Hartman, Torres, Kovács,
  Latham, Noyes, Esquerdo, Everett, Howard, Marcy, Fischer, Johnson, Andersen,
  Fűrész, Perumpilly, Sasselov, Stefanik, Béky, Lázár, Papp, \&
  Sári}]{buchhave_hatp16b_class_2010}
Buchhave, L.~A., Bakos, G.~A., Hartman, J.~D., {et~al.} 2010,
  \href{http://dx.doi.org/10.1088/0004-637X/720/2/1118}{\JournalTitle{\apj},
  720, 1118}

\bibitem[{Burke {et~al.}(2006)Burke, Gaudi, DePoy, \&
  Pogge}]{burke_survey_2006}
Burke, C.~J., Gaudi, B.~S., DePoy, D.~L., \& Pogge, R.~W. 2006,
  \href{http://dx.doi.org/10.1086/504468}{\JournalTitle{\aj}, 132, 210}

\bibitem[{Cameron \& Ward(1976)}]{cameron_origin_1976}
Cameron, A. G.~W., \& Ward, W.~R.
  \href{http://adsabs.harvard.edu/abs/1976LPI.....7..120C}{1976, 7},
  {C}onference Name: Lunar and Planetary Science Conference

\bibitem[{Cantat-Gaudin {et~al.}(2018)Cantat-Gaudin, Jordi, Vallenari,
  Bragaglia, Balaguer-Núñez, Soubiran, Bossini, Moitinho, Castro-Ginard,
  Krone-Martins, Casamiquela, Sordo, \& Carrera}]{cantatgaudin_gaia_2018}
Cantat-Gaudin, T., Jordi, C., Vallenari, A., {et~al.} 2018,
  \href{http://dx.doi.org/10.1051/0004-6361/201833476}{\JournalTitle{\aap},
  618, A93}

\bibitem[{Canup \& Asphaug(2001)}]{canup_origin_2001}
Canup, R.~M., \& Asphaug, E. 2001,
  \href{http://adsabs.harvard.edu/abs/2001Natur.412..708C}{\JournalTitle{\nat},
  412, 708}

\bibitem[{{Castelli} \& {Kurucz}(2004)}]{Castelli:2004}
{Castelli}, F., \& {Kurucz}, R.~L. 2004, \JournalTitle{ArXiv Astrophysics
  e-prints}, \href{http://arxiv.org/abs/astro-ph/0405087}{{\sffamily
  astro-ph/0405087}}

\bibitem[{Chatterjee {et~al.}(2008)Chatterjee, Ford, Matsumura, \&
  Rasio}]{chatterjee_dynamical_2008}
Chatterjee, S., Ford, E.~B., Matsumura, S., \& Rasio, F.~A. 2008,
  \href{http://dx.doi.org/10.1086/590227}{\JournalTitle{\apj}, 686, 580}

\bibitem[{Chen {et~al.}(2015)Chen, Bressan, Girardi, Marigo, Kong, \&
  Lanza}]{chen_parsec_2015}
Chen, Y., Bressan, A., Girardi, L., {et~al.} 2015,
  \href{http://dx.doi.org/10.1093/mnras/stv1281}{\JournalTitle{\mnras}, 452,
  1068}

\bibitem[{Chen {et~al.}(2014)Chen, Girardi, Bressan, Marigo, Barbieri, \&
  Kong}]{chen_improving_2014}
Chen, Y., Girardi, L., Bressan, A., {et~al.} 2014,
  \href{http://dx.doi.org/10.1093/mnras/stu1605}{\JournalTitle{\mnras}, 444,
  2525}

\bibitem[{Choi {et~al.}(2016)Choi, Dotter, Conroy, Cantiello, Paxton, \&
  Johnson}]{choi_mesa_2016}
Choi, J., Dotter, A., Conroy, C., {et~al.} 2016,
  \href{http://dx.doi.org/10.3847/0004-637X/823/2/102}{\JournalTitle{\apj},
  823, 102}

\bibitem[{Ciardi {et~al.}(2018)Ciardi, Crossfield, Feinstein, Schlieder,
  Petigura, David, Bristow, Patel, Arnold, Benneke, Christiansen, Dressing,
  Fulton, Howard, Isaacson, Sinukoff, \& Thackeray}]{ciardi_k2-136_2018}
Ciardi, D.~R., Crossfield, I. J.~M., Feinstein, A.~D., {et~al.} 2018,
  \href{http://dx.doi.org/10.3847/1538-3881/aa9921}{\JournalTitle{\aj}, 155,
  10}

\bibitem[{Claret(2017)}]{claret_limb_2017}
Claret, A. 2017,
  \href{http://dx.doi.org/10.1051/0004-6361/201629705}{\JournalTitle{\aap},
  600, A30}

\bibitem[{Collins {et~al.}(2017)Collins, Kielkopf, Stassun, \&
  Hessman}]{collins_astroimagej_2017}
Collins, K.~A., Kielkopf, J.~F., Stassun, K.~G., \& Hessman, F.~V. 2017,
  \href{http://dx.doi.org/10.3847/1538-3881/153/2/77}{\JournalTitle{\aj}, 153,
  77}

\bibitem[{Cropper {et~al.}(2018)Cropper, Katz, Sartoretti, Prusti, de~Bruijne,
  Chassat, Charvet, Boyadjian, Perryman, Sarri, Gare, Erdmann, Munari, Zwitter,
  Wilkinson, Arenou, Vallenari, Gómez, Panuzzo, Seabroke, Allende~Prieto,
  Benson, Marchal, Huckle, Smith, Dolding, Janßen, Viala, Blomme, Baker,
  Boudreault, Crifo, Soubiran, Frémat, Jasniewicz, Guerrier, Guy, Turon,
  Jean-Antoine-Piccolo, Thévenin, David, Gosset, \&
  Damerdji}]{cropper_gaia_2018}
Cropper, M., Katz, D., Sartoretti, P., {et~al.} 2018,
  \href{http://dx.doi.org/10.1051/0004-6361/201832763}{\JournalTitle{\aap},
  616, A5}

\bibitem[{Crouzet {et~al.}(2018)Crouzet, Chapellier, Guillot, Mékarnia, Agabi,
  Fanteï-Caujolle, Abe, Rivet, Schmider, Fressin, Bondoux, Challita, Pouzenc,
  Valbousquet, Bayliss, Bonhomme, Daban, Gouvret, \&
  Blazit}]{crouzet_four_2018}
Crouzet, N., Chapellier, E., Guillot, T., {et~al.} 2018,
  \href{http://dx.doi.org/10.1051/0004-6361/201732565}{\JournalTitle{\aap},
  619, A116}

\bibitem[{Curtis {et~al.}(2019{\natexlab{a}})Curtis, Ag\"{u}eros, Douglas, \&
  Meibom}]{curtis_temporary_2019}
Curtis, J.~L., Ag\"{u}eros, M.~A., Douglas, S.~T., \& Meibom, S.
  2019{\natexlab{a}},
  \href{http://dx.doi.org/10.3847/1538-4357/ab2393}{\JournalTitle{\apj}, 879,
  49}

\bibitem[{Curtis {et~al.}(2019{\natexlab{b}})Curtis, Ag\"{u}ueros, Mamajek,
  Wright, \& Cummings}]{curtis_tess_2019}
Curtis, J.~L., Ag\"{u}ueros, M.~A., Mamajek, E.~E., Wright, J.~T., \& Cummings,
  J.~D. 2019{\natexlab{b}},
  \href{http://dx.doi.org/10.3847/1538-3881/ab2899}{\JournalTitle{\aj}, 158,
  77}

\bibitem[{Daban {et~al.}(2010)Daban, Gouvret, Guillot, Agabi, Crouzet, Rivet,
  Mekarnia, Abe, Bondoux, Fanteï-Caujolle, Fressin, Schmider, Valbousquet,
  Blanc, Le~van Suu, Rauer, Erikson, Pont, \& Aigrain}]{daban_astep_2010}
Daban, J.-B., Gouvret, C., Guillot, T., {et~al.}
  \href{http://dx.doi.org/10.1117/12.854946}{2010, 7733, 77334T}

\bibitem[{Damasso {et~al.}(2020)Damasso, Lanza, Benatti, Rajpaul, Mallonn,
  Desidera, Biazzo, D'Orazi, Malavolta, Nardiello, Rainer, Borsa, Affer,
  Bignamini, Bonomo, Carleo, Claudi, Cosentino, Covino, Giacobbe, Gratton,
  Harutyunyan, Knapic, Leto, Maggio, Maldonado, Mancini, Micela, Molinari,
  Nascimbeni, Pagano, Piotto, Poretti, Scandariato, Sozzetti, Capuzzo~Dolcetta,
  Di~Mauro, Carosati, Fiorenzano, Frustagli, Pedani, Pinamonti, Stoev, \&
  Turrini}]{damasso_gaps_2020}
Damasso, M., Lanza, A.~F., Benatti, S., {et~al.} 2020,
  \href{http://adsabs.harvard.edu/abs/2020arXiv200809445D}{\JournalTitle{arXiv
  e-prints}, 2008, arXiv:2008.09445}

\bibitem[{Damiani {et~al.}(2019)Damiani, Prisinzano, Pillitteri, Micela, \&
  Sciortino}]{damiani_stellar_2019}
Damiani, F., Prisinzano, L., Pillitteri, I., Micela, G., \& Sciortino, S. 2019,
  \href{http://dx.doi.org/10.1051/0004-6361/201833994}{\JournalTitle{\aap},
  623, A112}, publisher: EDP Sciences

\bibitem[{Dauphas \& Pourmand(2011)}]{dauphas_hf-w-th_2011}
Dauphas, N., \& Pourmand, A. 2011,
  \href{http://dx.doi.org/10.1038/nature10077}{\JournalTitle{\nat}, 473, 489}

\bibitem[{David {et~al.}(2016)David, Hillenbrand, \&
  Petigura}]{David_et_al_2017}
David, T., Hillenbrand, L., \& Petigura, E. 2016, \JournalTitle{\nat}, 534, 658

\bibitem[{David \& Hillenbrand(2015)}]{david_ages_2015}
David, T.~J., \& Hillenbrand, L.~A. 2015,
  \href{http://dx.doi.org/10.1088/0004-637X/804/2/146}{\JournalTitle{\apj},
  804, 146}

\bibitem[{David {et~al.}(2019{\natexlab{a}})David, Petigura, Luger,
  Foreman-Mackey, Livingston, Mamajek, \& Hillenbrand}]{david_four_2019}
David, T.~J., Petigura, E.~A., Luger, R., {et~al.} 2019{\natexlab{a}},
  \href{http://dx.doi.org/10.3847/2041-8213/ab4c99}{\JournalTitle{\apjl}, 885,
  L12}

\bibitem[{David {et~al.}(2018)David, Mamajek, Vanderburg, Schlieder, Bristow,
  Petigura, Ciardi, Crossfield, Isaacson, Cody, Stauffer, Hillenbrand, Bieryla,
  Latham, Fulton, Rebull, Beichman, Gonzales, Hirsch, Howard, Vasisht, \&
  Ygouf}]{david_discovery_2018}
David, T.~J., Mamajek, E.~E., Vanderburg, A., {et~al.} 2018,
  \href{http://dx.doi.org/10.3847/1538-3881/aaeed7}{\JournalTitle{\aj}, 156,
  302}

\bibitem[{David {et~al.}(2019{\natexlab{b}})David, Cody, Hedges, Mamajek,
  Hillenbrand, Ciardi, Beichman, Petigura, Fulton, Isaacson, Howard, Gagné,
  Saunders, Rebull, Stauffer, Vasisht, \& Hinkley}]{david_warm_2019}
David, T.~J., Cody, A.~M., Hedges, C.~L., {et~al.} 2019{\natexlab{b}},
  \href{http://dx.doi.org/10.3847/1538-3881/ab290f}{\JournalTitle{\aj}, 158,
  79}

\bibitem[{de~Zeeuw {et~al.}(1999)de~Zeeuw, Hoogerwerf, de~Bruijne, Brown, \&
  Blaauw}]{de_zeeuw_hipparcos_1999}
de~Zeeuw, P.~T., Hoogerwerf, R., de~Bruijne, J. H.~J., Brown, A. G.~A., \&
  Blaauw, A. 1999, \href{http://dx.doi.org/10.1086/300682}{\JournalTitle{\aj},
  117, 354}

\bibitem[{D\'{i}az {et~al.}(2014)D\'{i}az, Almenara, Santerne, Moutou,
  Lethuillier, \& Deleuil}]{diaz_pastis_2014}
D\'{i}az, R.~F., Almenara, J.~M., Santerne, A., {et~al.} 2014,
  \href{http://dx.doi.org/10.1093/mnras/stu601}{\JournalTitle{\mnras}, 441,
  983}

\bibitem[{Dobbie {et~al.}(2010)Dobbie, Lodieu, \& Sharp}]{dobbie_ic_2010}
Dobbie, P.~D., Lodieu, N., \& Sharp, R.~G. 2010,
  \href{http://dx.doi.org/10.1111/j.1365-2966.2010.17355.x}{\JournalTitle{\mnras},
  409, 1002}

\bibitem[{{Donati} {et~al.}(1997){Donati}, {Semel}, {Carter}, {Rees}, \&
  {Collier Cameron}}]{donati_1997_spectropolarimetric}
{Donati}, J.-F., {Semel}, M., {Carter}, B.~D., {Rees}, D.~E., \& {Collier
  Cameron}, A. 1997,
  \href{http://dx.doi.org/10.1093/mnras/291.4.658}{\JournalTitle{\mnras}, 291,
  658}

\bibitem[{Donati {et~al.}(2016)Donati, Moutou, Malo, Baruteau, Yu, Hébrard,
  Hussain, Alencar, Ménard, Bouvier, Petit, Takami, Doyon, \&
  Cameron}]{donati_hj_2016}
Donati, J.~F., Moutou, C., Malo, L., {et~al.} 2016,
  \href{http://dx.doi.org/10.1038/nature18305}{\JournalTitle{\nat}, advance
  online publication}

\bibitem[{Donati {et~al.}(2020)Donati, Bouvier, Alencar, Moutou, Malo, Takami,
  Ménard, Dougados, Hussain, \& {The Matysse
  Collaboration}}]{donati_magnetic_2020}
Donati, J.-F., Bouvier, J., Alencar, S.~H., {et~al.} 2020,
  \href{http://dx.doi.org/10.1093/mnras/stz3368}{\JournalTitle{\mnras}, 491,
  5660}

\bibitem[{Dotter(2016)}]{dotter_mesa_2016}
Dotter, A. 2016,
  \href{http://dx.doi.org/10.3847/0067-0049/222/1/8}{\JournalTitle{\apjs}, 222,
  8}

\bibitem[{Douglas {et~al.}(2017)Douglas, Agüeros, Covey, \&
  Kraus}]{douglas_poking_2017}
Douglas, S.~T., Agüeros, M.~A., Covey, K.~R., \& Kraus, A. 2017,
  \href{http://dx.doi.org/10.3847/1538-4357/aa6e52}{\JournalTitle{\apj}, 842,
  83}

\bibitem[{Douglas {et~al.}(2019)Douglas, Curtis, Agüeros, Cargile, Brewer,
  Meibom, \& Jansen}]{douglas_k2_2019}
Douglas, S.~T., Curtis, J.~L., Agüeros, M.~A., {et~al.} 2019,
  \href{http://dx.doi.org/10.3847/1538-4357/ab2468}{\JournalTitle{\apj}, 879,
  100}

\bibitem[{Dullemond \& Monnier(2010)}]{dullemond_inner_2010}
Dullemond, C.~P., \& Monnier, J.~D. 2010,
  \href{http://dx.doi.org/10.1146/annurev-astro-081309-130932}{\JournalTitle{\araa},
  48, 205}

\bibitem[{Eastman {et~al.}(2010)Eastman, Siverd, \&
  Gaudi}]{eastman_achieving_2010}
Eastman, J., Siverd, R., \& Gaudi, B.~S. 2010,
  \href{http://dx.doi.org/10.1086/655938}{\JournalTitle{\pasp}, 122, 935}

\bibitem[{Ekstr\"om {et~al.}(2012)Ekstr\"om, Georgy, Eggenberger, Meynet,
  Mowlavi, Wyttenbach, Granada, Decressin, Hirschi, Frischknecht, Charbonnel,
  \& Maeder}]{ekstrom_grids_2012}
Ekstr\"om, S., Georgy, C., Eggenberger, P., {et~al.} 2012,
  \href{http://dx.doi.org/10.1051/0004-6361/201117751}{\JournalTitle{\aap},
  537, A146}

\bibitem[{Evans {et~al.}(2018)Evans, Riello, De~Angeli, Carrasco, Montegriffo,
  Fabricius, Jordi, Palaversa, Diener, Busso, Cacciari, van Leeuwen, Burgess,
  Davidson, Harrison, Hodgkin, Pancino, Richards, Altavilla, Balaguer-Núñez,
  Barstow, Bellazzini, Brown, Castellani, Cocozza, De~Luise, Delgado,
  Ducourant, Galleti, Gilmore, Giuffrida, Holl, Kewley, Koposov, Marinoni,
  Marrese, Osborne, Piersimoni, Portell, Pulone, Ragaini, Sanna, Terrett,
  Walton, Wevers, \& Wyrzykowski}]{evans_gaia_2018}
Evans, D.~W., Riello, M., De~Angeli, F., {et~al.} 2018,
  \href{http://dx.doi.org/10.1051/0004-6361/201832756}{\JournalTitle{\aap},
  616, A4}

\bibitem[{Fabrycky \& Tremaine(2007)}]{fabrycky_shrinking_2007}
Fabrycky, D., \& Tremaine, S. 2007,
  \href{http://dx.doi.org/10.1086/521702}{\JournalTitle{\apj}, 669, 1298}

\bibitem[{Fedele {et~al.}(2010)Fedele, van~den Ancker, Henning, Jayawardhana,
  \& Oliveira}]{fedele_timescale_2010}
Fedele, D., van~den Ancker, M.~E., Henning, T., Jayawardhana, R., \& Oliveira,
  J.~M. 2010,
  \href{http://dx.doi.org/10.1051/0004-6361/200912810}{\JournalTitle{\aap},
  510, A72}

\bibitem[{Feiden \& Chaboyer(2013)}]{feiden_magnetic_2013}
Feiden, G.~A., \& Chaboyer, B. 2013,
  \href{http://dx.doi.org/10.1088/0004-637X/779/2/183}{\JournalTitle{\apj},
  779, 183}

\bibitem[{F\H{u}r\'{e}sz {et~al.}(2008)F\H{u}r\'{e}sz, Szentgyorgyi, \&
  Meibom}]{furesz_tres_2008}
F\H{u}r\'{e}sz, G., Szentgyorgyi, A.~H., \& Meibom, S.
  \href{http://dx.doi.org/10.1007/978-3-540-75485-5_68}{2008, 287}

\bibitem[{Flagg {et~al.}(2019)Flagg, Johns-Krull, Nofi, Llama, Prato, Sullivan,
  Jaffe, \& Mace}]{flagg_co_2019}
Flagg, L., Johns-Krull, C.~M., Nofi, L., {et~al.} 2019,
  \href{http://dx.doi.org/10.3847/2041-8213/ab276d}{\JournalTitle{\apj}, 878,
  L37}

\bibitem[{Foreman-Mackey(2016)}]{corner_2016}
Foreman-Mackey, D. 2016,
  \href{http://dx.doi.org/10.21105/joss.00024}{\JournalTitle{Journal of Open
  Source Software}, 1, 24}

\bibitem[{Foreman-Mackey {et~al.}(2020)Foreman-Mackey, Czekala, Luger, Agol,
  Barentsen, \& Barclay}]{exoplanet:exoplanet}
Foreman-Mackey, D., Czekala, I., Luger, R., {et~al.} 2020,
  exoplanet-dev/exoplanet v0.2.6

\bibitem[{Fortney {et~al.}(2007)Fortney, Marley, \&
  Barnes}]{Fortney_et_al_2007}
Fortney, J.~J., Marley, M.~S., \& Barnes, J.~W. 2007, \JournalTitle{ApJ}, 659,
  1661

\bibitem[{Fulton {et~al.}(2018)Fulton, Petigura, Blunt, \&
  Sinukoff}]{fulton_radvel_2018}
Fulton, B.~J., Petigura, E.~A., Blunt, S., \& Sinukoff, E. 2018,
  \href{http://dx.doi.org/10.1088/1538-3873/aaaaa8}{\JournalTitle{\pasp}, 130,
  044504}

\bibitem[{Fulton {et~al.}(2017)Fulton, Petigura, Howard, Isaacson, Marcy,
  Cargile, Hebb, Weiss, Johnson, Morton, Sinukoff, Crossfield, \&
  Hirsch}]{Fulton_et_al_2017}
Fulton, B.~J., Petigura, E.~A., Howard, A.~W., {et~al.} 2017,
  \JournalTitle{AJ}, 154, 109

\bibitem[{{Gaia Collaboration} {et~al.}(2016){Gaia Collaboration}, Prusti,
  de~Bruijne, Brown, Vallenari, Babusiaux, Bailer-Jones, Bastian, Biermann,
  Evans, Eyer, Jansen, Jordi, Klioner, Lammers, Lindegren, Luri, Mignard,
  Milligan, Panem, Poinsignon, Pourbaix, Randich, Sarri, Sartoretti, Siddiqui,
  Soubiran, Valette, van Leeuwen, Walton, Aerts, Arenou, Cropper, Drimmel,
  Høg, Katz, Lattanzi, O'Mullane, Grebel, Holland, Huc, Passot, Bramante,
  Cacciari, Castañeda, Chaoul, Cheek, De~Angeli, Fabricius, Guerra,
  Hernández, Jean-Antoine-Piccolo, Masana, Messineo, Mowlavi, Nienartowicz,
  Ordóñez-Blanco, Panuzzo, Portell, Richards, Riello, Seabroke, Tanga,
  Thévenin, Torra, Els, Gracia-Abril, Comoretto, Garcia-Reinaldos, Lock,
  Mercier, Altmann, Andrae, Astraatmadja, Bellas-Velidis, Benson, Berthier,
  Blomme, Busso, Carry, Cellino, Clementini, Cowell, Creevey, Cuypers,
  Davidson, De~Ridder, de~Torres, Delchambre, Dell'Oro, Ducourant, Frémat,
  García-Torres, Gosset, Halbwachs, Hambly, Harrison, Hauser, Hestroffer,
  Hodgkin, Huckle, Hutton, Jasniewicz, Jordan, Kontizas, Korn, Lanzafame,
  Manteiga, Moitinho, Muinonen, Osinde, Pancino, Pauwels, Petit, Recio-Blanco,
  Robin, Sarro, Siopis, Smith, Smith, Sozzetti, Thuillot, van Reeven, Viala,
  Abbas, Abreu~Aramburu, Accart, Aguado, Allan, Allasia, Altavilla, Álvarez,
  Alves, Anderson, Andrei, Anglada~Varela, Antiche, Antoja, Antón, Arcay,
  Atzei, Ayache, Bach, Baker, Balaguer-Núñez, Barache, Barata, Barbier,
  Barblan, Baroni, Barrado~y Navascués, Barros, Barstow, Becciani, Bellazzini,
  Bellei, Bello~García, Belokurov, Bendjoya, Berihuete, Bianchi, Bienaymé,
  Billebaud, Blagorodnova, Blanco-Cuaresma, Boch, Bombrun, Borrachero,
  Bouquillon, Bourda, Bouy, Bragaglia, Breddels, Brouillet, Brüsemeister,
  Bucciarelli, Budnik, Burgess, Burgon, Burlacu, Busonero, Buzzi, Caffau,
  Cambras, Campbell, Cancelliere, Cantat-Gaudin, Carlucci, Carrasco,
  Castellani, Charlot, Charnas, Charvet, Chassat, Chiavassa, Clotet, Cocozza,
  Collins, Collins, Costigan, Crifo, Cross, Crosta, Crowley, Dafonte, Damerdji,
  Dapergolas, David, David, De~Cat, de~Felice, de~Laverny, De~Luise, De~March,
  de~Martino, de~Souza, Debosscher, del Pozo, Delbo, Delgado, Delgado,
  di~Marco, Di~Matteo, Diakite, Distefano, Dolding, Dos~Anjos, Drazinos,
  Durán, Dzigan, Ecale, Edvardsson, Enke, Erdmann, Escolar, Espina, Evans,
  Eynard~Bontemps, Fabre, Fabrizio, Faigler, Falcão, Farràs~Casas, Faye,
  Federici, Fedorets, Fernández-Hernández, Fernique, Fienga, Figueras,
  Filippi, Findeisen, Fonti, Fouesneau, Fraile, Fraser, Fuchs, Furnell, Gai,
  Galleti, Galluccio, Garabato, García-Sedano, Garé, Garofalo, Garralda,
  Gavras, Gerssen, Geyer, Gilmore, Girona, Giuffrida, Gomes, González-Marcos,
  González-Núñez, González-Vidal, Granvik, Guerrier, Guillout, Guiraud,
  Gúrpide, Gutiérrez-Sánchez, Guy, Haigron, Hatzidimitriou, Haywood, Heiter,
  Helmi, Hobbs, Hofmann, Holl, Holland, Hunt, Hypki, Icardi, Irwin, Jevardat~de
  Fombelle, Jofré, Jonker, Jorissen, Julbe, Karampelas, Kochoska, Kohley,
  Kolenberg, Kontizas, Koposov, Kordopatis, Koubsky, Kowalczyk, Krone-Martins,
  Kudryashova, Kull, Bachchan, Lacoste-Seris, Lanza, Lavigne,
  Le~Poncin-Lafitte, Lebreton, Lebzelter, Leccia, Leclerc, Lecoeur-Taibi,
  Lemaitre, Lenhardt, Leroux, Liao, Licata, Lindstrøm, Lister, Livanou, Lobel,
  Löffler, López, Lopez-Lozano, Lorenz, Loureiro, MacDonald,
  Magalhães~Fernandes, Managau, Mann, Mantelet, Marchal, Marchant, Marconi,
  Marie, Marinoni, Marrese, Marschalkó, Marshall, Martín-Fleitas, Martino,
  Mary, Matijevič, Mazeh, McMillan, Messina, Mestre, Michalik, Millar,
  Miranda, Molina, Molinaro, Molinaro, Molnár, Moniez, Montegriffo, Monteiro,
  Mor, Mora, Morbidelli, Morel, Morgenthaler, Morley, Morris, Mulone, Muraveva,
  Musella, Narbonne, Nelemans, Nicastro, Noval, Ordénovic, Ordieres-Meré,
  Osborne, Pagani, Pagano, Pailler, Palacin, Palaversa, Parsons, Paulsen,
  Pecoraro, Pedrosa, Pentikäinen, Pereira, Pichon, Piersimoni, Pineau, Plachy,
  Plum, Poujoulet, Prša, Pulone, Ragaini, Rago, Rambaux, Ramos-Lerate,
  Ranalli, Rauw, Read, Regibo, Renk, Reylé, Ribeiro, Rimoldini, Ripepi, Riva,
  Rixon, Roelens, Romero-Gómez, Rowell, Royer, Rudolph, Ruiz-Dern, Sadowski,
  Sagristà~Sellés, Sahlmann, Salgado, Salguero, Sarasso, Savietto, Schnorhk,
  Schultheis, Sciacca, Segol, Segovia, Segransan, Serpell, Shih, Smareglia,
  Smart, Smith, Solano, Solitro, Sordo, Soria~Nieto, Souchay, Spagna, Spoto,
  Stampa, Steele, Steidelmüller, Stephenson, Stoev, Suess, Süveges, Surdej,
  Szabados, Szegedi-Elek, Tapiador, Taris, Tauran, Taylor, Teixeira, Terrett,
  Tingley, Trager, Turon, Ulla, Utrilla, Valentini, van Elteren, Van~Hemelryck,
  van Leeuwen, Varadi, Vecchiato, Veljanoski, Via, Vicente, Vogt, Voss,
  Votruba, Voutsinas, Walmsley, Weiler, Weingrill, Werner, Wevers, Whitehead,
  Wyrzykowski, Yoldas, Žerjal, Zucker, Zurbach, Zwitter, Alecu, Allen,
  Allende~Prieto, Amorim, Anglada-Escudé, Arsenijevic, Azaz, Balm, Beck,
  Bernstein, Bigot, Bijaoui, Blasco, Bonfigli, Bono, Boudreault, Bressan,
  Brown, Brunet, Bunclark, Buonanno, Butkevich, Carret, Carrion, Chemin,
  Chéreau, Corcione, Darmigny, de~Boer, de~Teodoro, de~Zeeuw, Delle~Luche,
  Domingues, Dubath, Fodor, Frézouls, Fries, Fustes, Fyfe, Gallardo, Gallegos,
  Gardiol, Gebran, Gomboc, Gómez, Grux, Gueguen, Heyrovsky, Hoar, Iannicola,
  Isasi~Parache, Janotto, Joliet, Jonckheere, Keil, Kim, Klagyivik, Klar,
  Knude, Kochukhov, Kolka, Kos, Kutka, Lainey, LeBouquin, Liu, Loreggia,
  Makarov, Marseille, Martayan, Martinez-Rubi, Massart, Meynadier, Mignot,
  Munari, Nguyen, Nordlander, Ocvirk, O'Flaherty, Olias~Sanz, Ortiz, Osorio,
  Oszkiewicz, Ouzounis, Palmer, Park, Pasquato, Peltzer, Peralta, Péturaud,
  Pieniluoma, Pigozzi, Poels, Prat, Prod'homme, Raison, Rebordao, Risquez,
  Rocca-Volmerange, Rosen, Ruiz-Fuertes, Russo, Sembay, Serraller~Vizcaino,
  Short, Siebert, Silva, Sinachopoulos, Slezak, Soffel, Sosnowska, Straižys,
  ter Linden, Terrell, Theil, Tiede, Troisi, Tsalmantza, Tur, Vaccari, Vachier,
  Valles, Van~Hamme, Veltz, Virtanen, Wallut, Wichmann, Wilkinson, Ziaeepour,
  \& Zschocke}]{gaia_collaboration_gaia_2016}
{Gaia Collaboration}, Prusti, T., de~Bruijne, J. H.~J., {et~al.} 2016,
  \href{http://dx.doi.org/10.1051/0004-6361/201629272}{\JournalTitle{\aap},
  595, A1}

\bibitem[{{Gaia Collaboration} {et~al.}(2018){Gaia Collaboration}, Brown,
  Vallenari, Prusti, de~Bruijne, Babusiaux, Bailer-Jones, Biermann, Evans,
  Eyer, Jansen, Jordi, Klioner, Lammers, Lindegren, Luri, Mignard, Panem,
  Pourbaix, Randich, Sartoretti, Siddiqui, Soubiran, van Leeuwen, Walton,
  Arenou, Bastian, Cropper, Drimmel, Katz, Lattanzi, Bakker, Cacciari,
  Castañeda, Chaoul, Cheek, De~Angeli, Fabricius, Guerra, Holl, Masana,
  Messineo, Mowlavi, Nienartowicz, Panuzzo, Portell, Riello, Seabroke, Tanga,
  Thévenin, Gracia-Abril, Comoretto, Garcia-Reinaldos, Teyssier, Altmann,
  Andrae, Audard, Bellas-Velidis, Benson, Berthier, Blomme, Burgess, Busso,
  Carry, Cellino, Clementini, Clotet, Creevey, Davidson, De~Ridder, Delchambre,
  Dell'Oro, Ducourant, Fernández-Hernández, Fouesneau, Frémat, Galluccio,
  García-Torres, González-Núñez, González-Vidal, Gosset, Guy, Halbwachs,
  Hambly, Harrison, Hernández, Hestroffer, Hodgkin, Hutton, Jasniewicz,
  Jean-Antoine-Piccolo, Jordan, Korn, Krone-Martins, Lanzafame, Lebzelter,
  Löffler, Manteiga, Marrese, Martín-Fleitas, Moitinho, Mora, Muinonen,
  Osinde, Pancino, Pauwels, Petit, Recio-Blanco, Richards, Rimoldini, Robin,
  Sarro, Siopis, Smith, Sozzetti, Süveges, Torra, van Reeven, Abbas,
  Abreu~Aramburu, Accart, Aerts, Altavilla, Álvarez, Alvarez, Alves, Anderson,
  Andrei, Anglada~Varela, Antiche, Antoja, Arcay, Astraatmadja, Bach, Baker,
  Balaguer-Núñez, Balm, Barache, Barata, Barbato, Barblan, Barklem, Barrado,
  Barros, Barstow, Bartholomé~Muñoz, Bassilana, Becciani, Bellazzini,
  Berihuete, Bertone, Bianchi, Bienaymé, Blanco-Cuaresma, Boch, Boeche,
  Bombrun, Borrachero, Bossini, Bouquillon, Bourda, Bragaglia, Bramante,
  Breddels, Bressan, Brouillet, Brüsemeister, Brugaletta, Bucciarelli,
  Burlacu, Busonero, Butkevich, Buzzi, Caffau, Cancelliere, Cannizzaro,
  Cantat-Gaudin, Carballo, Carlucci, Carrasco, Casamiquela, Castellani,
  Castro-Ginard, Charlot, Chemin, Chiavassa, Cocozza, Costigan, Cowell, Crifo,
  Crosta, Crowley, Cuypers, Dafonte, Damerdji, Dapergolas, David, David,
  de~Laverny, De~Luise, De~March, de~Martino, de~Souza, de~Torres, Debosscher,
  del Pozo, Delbo, Delgado, Delgado, Di~Matteo, Diakite, Diener, Distefano,
  Dolding, Drazinos, Durán, Edvardsson, Enke, Eriksson, Esquej,
  Eynard~Bontemps, Fabre, Fabrizio, Faigler, Falcão, Farràs~Casas, Federici,
  Fedorets, Fernique, Figueras, Filippi, Findeisen, Fonti, Fraile, Fraser,
  Frézouls, Gai, Galleti, Garabato, García-Sedano, Garofalo, Garralda, Gavel,
  Gavras, Gerssen, Geyer, Giacobbe, Gilmore, Girona, Giuffrida, Glass, Gomes,
  Granvik, Gueguen, Guerrier, Guiraud, Gutiérrez-Sánchez, Haigron,
  Hatzidimitriou, Hauser, Haywood, Heiter, Helmi, Heu, Hilger, Hobbs, Hofmann,
  Holland, Huckle, Hypki, Icardi, Janßen, Jevardat~de Fombelle, Jonker,
  Juhász, Julbe, Karampelas, Kewley, Klar, Kochoska, Kohley, Kolenberg,
  Kontizas, Kontizas, Koposov, Kordopatis, Kostrzewa-Rutkowska, Koubsky,
  Lambert, Lanza, Lasne, Lavigne, Le~Fustec, Le~Poncin-Lafitte, Lebreton,
  Leccia, Leclerc, Lecoeur-Taibi, Lenhardt, Leroux, Liao, Licata, Lindstrøm,
  Lister, Livanou, Lobel, López, Managau, Mann, Mantelet, Marchal, Marchant,
  Marconi, Marinoni, Marschalkó, Marshall, Martino, Marton, Mary, Massari,
  Matijevič, Mazeh, McMillan, Messina, Michalik, Millar, Molina, Molinaro,
  Molnár, Montegriffo, Mor, Morbidelli, Morel, Morris, Mulone, Muraveva,
  Musella, Nelemans, Nicastro, Noval, O'Mullane, Ordénovic, Ordóñez-Blanco,
  Osborne, Pagani, Pagano, Pailler, Palacin, Palaversa, Panahi, Pawlak,
  Piersimoni, Pineau, Plachy, Plum, Poggio, Poujoulet, Prša, Pulone, Racero,
  Ragaini, Rambaux, Ramos-Lerate, Regibo, Reylé, Riclet, Ripepi, Riva, Rivard,
  Rixon, Roegiers, Roelens, Romero-Gómez, Rowell, Royer, Ruiz-Dern, Sadowski,
  Sagristà~Sellés, Sahlmann, Salgado, Salguero, Sanna, Santana-Ros, Sarasso,
  Savietto, Schultheis, Sciacca, Segol, Segovia, Ségransan, Shih, Siltala,
  Silva, Smart, Smith, Solano, Solitro, Sordo, Soria~Nieto, Souchay, Spagna,
  Spoto, Stampa, Steele, Steidelmüller, Stephenson, Stoev, Suess, Surdej,
  Szabados, Szegedi-Elek, Tapiador, Taris, Tauran, Taylor, Teixeira, Terrett,
  Teyssandier, Thuillot, Titarenko, Torra~Clotet, Turon, Ulla, Utrilla, Uzzi,
  Vaillant, Valentini, Valette, van Elteren, Van~Hemelryck, van Leeuwen,
  Vaschetto, Vecchiato, Veljanoski, Viala, Vicente, Vogt, von Essen, Voss,
  Votruba, Voutsinas, Walmsley, Weiler, Wertz, Wevers, Wyrzykowski, Yoldas,
  Žerjal, Ziaeepour, Zorec, Zschocke, Zucker, Zurbach, \&
  Zwitter}]{gaia_collaboration_gaia_2018}
{Gaia Collaboration}, Brown, A. G.~A., Vallenari, A., {et~al.} 2018,
  \href{http://dx.doi.org/10.1051/0004-6361/201833051}{\JournalTitle{\aap},
  616, A1}

\bibitem[{Gaudi \& Winn(2007)}]{gaudi_prospects_2007}
Gaudi, B.~S., \& Winn, J.~N. 2007,
  \href{http://dx.doi.org/10.1086/509910}{\JournalTitle{\apj}, 655, 550}

\bibitem[{Gelman \& Rubin(1992)}]{gelman_inference_1992}
Gelman, A., \& Rubin, D.~B. 1992,
  \href{https://www.jstor.org/stable/2246093}{\JournalTitle{Statistical
  Science}, 7, 457}, publisher: Institute of Mathematical Statistics

\bibitem[{Giacalone \& Dressing(2020)}]{giacalone_triceratops_2020}
Giacalone, S., \& Dressing, C.~D. 2020,
  \href{https://ui.adsabs.harvard.edu/abs/2020arXiv200200691G/abstract}{\JournalTitle{arXiv
  e-prints}, arXiv:2002.00691}

\bibitem[{Gilbert {et~al.}(2018)Gilbert, Bergmann, Bloxham, Boz, Brookfield,
  Carkic, Carter, Case, Churilov, Ellis, Gausachs, Gers, Gray, Herrald,
  Ireland, Jones, Kripak, Lawrence, O'Brien, Price, Robertson, Schwab, Tinney,
  Vaccarella, Vest, Wright, \& Zhelem}]{gilbert_veloce_2018}
Gilbert, J., Bergmann, C., Bloxham, G., {et~al.}
  \href{http://dx.doi.org/10.1117/12.2312399}{2018, 0702, 107020Y},
  {C}onference Name: Ground-based and Airborne Instrumentation for Astronomy
  VII ISBN: 9781510619579 arXiv:1807.01938

\bibitem[{Ginsburg {et~al.}(2018)Ginsburg, Sipocz, {Madhura Parikh}, Woillez,
  Groener, Liedtke, Robitaille, Deil, {Jcsegovia}, Norman, Svoboda, {C. E.
  Brasseur}, Tollerud, Persson, {Adamginsburg}, S\'eguin-Charbonneau,
  Armstrong, Val-Borro, Morris, Mirocha, {Ayush Yadav}, Seifert, Droettboom,
  Moolekamp, {James-Allen}, {Azalee Bostroem}, Egeland, Singer, Rol, \&
  Grollier}]{astroquery_2018}
Ginsburg, A., Sipocz, B., {Madhura Parikh}, {et~al.} 2018,
  Astropy/{Astroquery}: {V}0.3.7 {Release}

\bibitem[{Ginzburg {et~al.}(2016)Ginzburg, Schlichting, \&
  Sari}]{ginzburg_superearth_2016}
Ginzburg, S., Schlichting, H.~E., \& Sari, R. 2016,
  \href{http://dx.doi.org/10.3847/0004-637X/825/1/29}{\JournalTitle{\apj}, 825,
  29}

\bibitem[{Girardi {et~al.}(2005)Girardi, Groenewegen, Hatziminaoglou, \&
  da~Costa}]{girardi_star_2005}
Girardi, L., Groenewegen, M. A.~T., Hatziminaoglou, E., \& da~Costa, L. 2005,
  \href{http://dx.doi.org/10.1051/0004-6361:20042352}{\JournalTitle{\aap}, 436,
  895}

\bibitem[{{Gray}(2005)}]{gray_2005_book}
{Gray}, D.~F. 2005, {The Observation and Analysis of Stellar Photospheres}

\bibitem[{{Gray} \& {Corbally}(1994)}]{gray_1994_spectrum}
{Gray}, R.~O., \& {Corbally}, C.~J. 1994,
  \href{http://dx.doi.org/10.1086/116893}{\JournalTitle{\aj}, 107, 742}

\bibitem[{Gupta \& Schlichting(2019)}]{gupta_sculpting_2019}
Gupta, A., \& Schlichting, H.~E. 2019,
  \href{http://dx.doi.org/10.1093/mnras/stz1230}{\JournalTitle{\mnras}, 487,
  24}

\bibitem[{Gupta \& Schlichting(2020)}]{gupta_signatures_2020}
---. 2020,
  \href{http://dx.doi.org/10.1093/mnras/staa315}{\JournalTitle{\mnras}, 493,
  792}

\bibitem[{Hartman {et~al.}(2009)Hartman, Gaudi, Holman, McLeod, Stanek,
  Barranco, Pinsonneault, Meibom, \& Kalirai}]{hartman_MMT_IV_2009}
Hartman, J.~D., Gaudi, B.~S., Holman, M.~J., {et~al.} 2009,
  \href{http://dx.doi.org/10.1088/0004-637X/695/1/336}{\JournalTitle{\apj},
  695, 336}

\bibitem[{Hippke {et~al.}(2019)Hippke, David, Mulders, \&
  Heller}]{hippke_wotan_2019}
Hippke, M., David, T.~J., Mulders, G.~D., \& Heller, R. 2019,
  \href{http://arxiv.org/abs/1906.00966}{\JournalTitle{arXiv:1906.00966
  [astro-ph]}}

\bibitem[{Hirano {et~al.}(2020)Hirano, Krishnamurthy, Gaidos, Flewelling, Mann,
  Narita, Plavchan, Kotani, Tamura, Harakawa, Hodapp, Ishizuka, Jacobson,
  Konishi, Kudo, Kurokawa, Kuzuhara, Nishikawa, Omiya, Serizawa, Ueda, \&
  Vievard}]{hirano_limits_2020}
Hirano, T., Krishnamurthy, V., Gaidos, E., {et~al.} 2020,
  \href{http://arxiv.org/abs/2006.13243}{\JournalTitle{arXiv:2006.13243
  [astro-ph]}}

\bibitem[{Hoffman \& Gelman(2014)}]{hoffman_no-u-turn_2014}
Hoffman, M.~D., \& Gelman, A. 2014,
  \href{http://jmlr.org/papers/v15/hoffman14a.html}{\JournalTitle{Journal of
  Machine Learning Research}, 15, 1593}

\bibitem[{Howard {et~al.}(2012)Howard, Marcy, Bryson, Jenkins, Rowe, Batalha,
  Borucki, Koch, Dunham, Gautier, Van~Cleve, Cochran, Latham, Lissauer, Torres,
  Brown, Gilliland, Buchhave, Caldwell, Christensen-Dalsgaard, Ciardi, Fressin,
  Haas, Howell, Kjeldsen, Seager, Rogers, Sasselov, Steffen, Basri,
  Charbonneau, Christiansen, Clarke, Dupree, Fabrycky, Fischer, Ford, Fortney,
  Tarter, Girouard, Holman, Johnson, Klaus, Machalek, Moorhead, Morehead,
  Ragozzine, Tenenbaum, Twicken, Quinn, Isaacson, Shporer, Lucas, Walkowicz,
  Welsh, Boss, Devore, Gould, Smith, Morris, Prsa, Morton, Still, Thompson,
  Mullally, Endl, \& MacQueen}]{howard_planet_2012}
Howard, A.~W., Marcy, G.~W., Bryson, S.~T., {et~al.} 2012,
  \href{http://dx.doi.org/10.1088/0067-0049/201/2/15}{\JournalTitle{\apjs},
  201, 15}

\bibitem[{Howell {et~al.}(2014)Howell, Sobeck, Haas, Still, Barclay, Mullally,
  {John Troeltzsch}, Aigrain, Bryson, Caldwell, Chaplin, Cochran, Huber, Marcy,
  Miglio, Najita, Smith, Twicken, \& Fortney}]{howell_k2_2014}
Howell, S.~B., Sobeck, C., Haas, M., {et~al.} 2014,
  \href{http://dx.doi.org/10.1086/676406}{\JournalTitle{\pasp}, 126, 398}

\bibitem[{Huang {et~al.}(2018{\natexlab{a}})Huang, Shporer, Dragomir,
  Fausnaugh, Levine, Morgan, Nguyen, Ricker, Wall, Woods, \&
  Vanderspek}]{huang_expected_2018}
Huang, C.~X., Shporer, A., Dragomir, D., {et~al.} 2018{\natexlab{a}},
  \href{http://arxiv.org/abs/1807.11129}{\JournalTitle{arXiv:1807.11129
  [astro-ph]}}

\bibitem[{Huang {et~al.}(2018{\natexlab{b}})Huang, Burt, Vanderburg, Günther,
  Shporer, Dittmann, Winn, Wittenmyer, Sha, Kane, Ricker, Vanderspek, Latham,
  Seager, Jenkins, Caldwell, Collins, Guerrero, Smith, Quinn, Udry, Pepe,
  Bouchy, Ségransan, Lovis, Ehrenreich, Marmier, Mayor, Wohler, Haworth,
  Morgan, Fausnaugh, Ciardi, Christiansen, Charbonneau, Dragomir, Deming,
  Glidden, Levine, McCullough, Yu, Narita, Nguyen, Morton, Pepper, Pál,
  Rodriguez, Stassun, Torres, Sozzetti, Doty, Christensen-Dalsgaard, Laughlin,
  Clampin, Bean, Buchhave, Bakos, Sato, Ida, Kaltenegger, Palle, Sasselov,
  Butler, Lissauer, Ge, \& Rinehart}]{huang_pimen_2018}
Huang, C.~X., Burt, J., Vanderburg, A., {et~al.} 2018{\natexlab{b}},
  \href{http://dx.doi.org/10.3847/2041-8213/aaef91}{\JournalTitle{\apj}, 868,
  L39}

\bibitem[{Hunter(2007)}]{hunter_matplotlib_2007}
Hunter, J.~D. 2007, \JournalTitle{Computing in Science \& Engineering}, 9, 90

\bibitem[{Irwin \& Bouvier(2009)}]{irwin_rotational_2009}
Irwin, J., \& Bouvier, J. 2009,
  \href{http://dx.doi.org/10.1017/S1743921309032025}{in }, eprint:
  arXiv:0901.3342, 363

\bibitem[{Irwin {et~al.}(2007)Irwin, Irwin, Aigrain, Hodgkin, Hebb, \&
  Moraux}]{irwin_monitordata_2007}
Irwin, J., Irwin, M., Aigrain, S., {et~al.} 2007,
  \href{http://dx.doi.org/10.1111/j.1365-2966.2006.11408.x}{\JournalTitle{\mnras},
  375, 1449}

\bibitem[{Jenkins {et~al.}(2016)Jenkins, Twicken, McCauliff, Campbell,
  Sanderfer, Lung, Mansouri-Samani, Girouard, Tenenbaum, Klaus, Smith,
  Caldwell, Chacon, Henze, Heiges, Latham, Morgan, Swade, Rinehart, \&
  Vanderspek}]{jenkins_tess_2016}
Jenkins, J.~M., Twicken, J.~D., McCauliff, S., {et~al.} 2016,
  \href{http://dx.doi.org/10.1117/12.2233418}{\JournalTitle{Software and
  Cyberinfrastructure for Astronomy IV}, 9913, 99133E}

\bibitem[{{Jensen}(2013)}]{Jensen:2013}
{Jensen}, E. 2013, {Tapir: A web interface for transit/eclipse observability},
  Astrophysics Source Code Library,
  \href{http://arxiv.org/abs/1306.007}{{\sffamily ascl:1306.007}}

\bibitem[{Johns-Krull {et~al.}(2016)Johns-Krull, McLane, Prato, Crockett,
  Jaffe, Hartigan, Beichman, Mahmud, Chen, Skiff, Cauley, Jones, \&
  Mace}]{johns-krull_candidate_2016}
Johns-Krull, C.~M., McLane, J.~N., Prato, L., {et~al.} 2016,
  \href{http://dx.doi.org/10.3847/0004-637X/826/2/206}{\JournalTitle{\apj},
  826, 206}

\bibitem[{Jones {et~al.}(2001)Jones, Oliphant, Peterson,
  {et~al.}}]{jones_scipy_2001}
Jones, E., Oliphant, T., Peterson, P., {et~al.} 2001, Open {source}
  {scientific} {tools} for {Python}

\bibitem[{{Jord{\'a}n} {et~al.}(2020){Jord{\'a}n}, {Brahm}, {Espinoza},
  {Henning}, {Jones}, {Kossakowski}, {Sarkis}, {Trifonov}, {Rojas}, {Torres},
  {Drass}, {Nandakumar}, \& {Barbieri}}]{jordan:2020}
{Jord{\'a}n}, A., {Brahm}, R., {Espinoza}, N., {et~al.} 2020,
  \href{http://dx.doi.org/10.3847/1538-3881/ab6f67}{\JournalTitle{\aj}, 159,
  145}

\bibitem[{Kaufer {et~al.}(1999)Kaufer, Stahl, Tubbesing, Nørregaard, Avila,
  Francois, Pasquini, \& Pizzella}]{kaufer_commissioning_1999}
Kaufer, A., Stahl, O., Tubbesing, S., {et~al.} 1999,
  \href{http://adsabs.harvard.edu/abs/1999Msngr..95....8K}{\JournalTitle{The
  Messenger}, 95, 8}

\bibitem[{Kharchenko {et~al.}(2005)Kharchenko, Piskunov, Röser, Schilbach, \&
  Scholz}]{kharchenko_astrophysical_2005}
Kharchenko, N.~V., Piskunov, A.~E., Röser, S., Schilbach, E., \& Scholz, R.-D.
  2005,
  \href{http://dx.doi.org/10.1051/0004-6361:20042523}{\JournalTitle{\aap}, 438,
  1163}

\bibitem[{Kharchenko {et~al.}(2013)Kharchenko, Piskunov, Schilbach, R\"oser, \&
  Scholz}]{Kharchenko_et_al_2013}
Kharchenko, N.~V., Piskunov, A.~E., Schilbach, E., R\"oser, S., \& Scholz,
  R.-D. 2013, \JournalTitle{\aap}, 558, A53

\bibitem[{King \& Wheatley(2020)}]{king_euv_2020}
King, G.~W., \& Wheatley, P.~J. 2020,
  \href{http://arxiv.org/abs/2007.13731}{\JournalTitle{arXiv:2007.13731
  [astro-ph]}}

\bibitem[{{Kipping}(2013)}]{exoplanet:kipping13}
{Kipping}, D.~M. 2013,
  \href{http://dx.doi.org/10.1093/mnras/stt1435}{\JournalTitle{\mnras}, 435,
  2152}

\bibitem[{Kleine {et~al.}(2009)Kleine, Touboul, Bourdon, Nimmo, Mezger, Palme,
  Jacobsen, Yin, \& Halliday}]{kleine_hf-w_2009}
Kleine, T., Touboul, M., Bourdon, B., {et~al.} 2009,
  \href{http://dx.doi.org/10.1016/j.gca.2008.11.047}{\JournalTitle{Geochimica
  et Cosmochimica Acta}, 73, 5150}

\bibitem[{K\"onig {et~al.}(2011)K\"onig, M\"unker, Hohl, Paulick, Barth, Lagos,
  Pf\"ander, \& B\"uchl}]{konig_earths_2011}
K\"onig, S., M\"unker, C., Hohl, S., {et~al.} 2011,
  \href{http://dx.doi.org/10.1016/j.gca.2011.01.031}{\JournalTitle{Geochimica
  et Cosmochimica Acta}, 75, 2119}

\bibitem[{Kounkel \& Covey(2019)}]{kounkel_untangling_2019}
Kounkel, M., \& Covey, K. 2019,
  \href{http://dx.doi.org/10.3847/1538-3881/ab339a}{\JournalTitle{\aj}, 158,
  122}

\bibitem[{Krumholz {et~al.}(2019)Krumholz, McKee, \&
  Bland-Hawthorn}]{krumholz_star_2019}
Krumholz, M.~R., McKee, C.~F., \& Bland-Hawthorn, J. 2019,
  \href{http://dx.doi.org/10.1146/annurev-astro-091918-104430}{\JournalTitle{\araa},
  57, 227}

\bibitem[{Kumar {et~al.}(2019)Kumar, Carroll, Hartikainen, \&
  Martin}]{arviz_2019}
Kumar, R., Carroll, C., Hartikainen, A., \& Martin, O.~A. 2019,
  \href{http://dx.doi.org/10.21105/joss.01143}{\JournalTitle{The Journal of
  Open Source Software}}

\bibitem[{Kurucz(2013)}]{kurucz_atlas12_2013}
Kurucz, R.~L. 2013,
  \href{http://adsabs.harvard.edu/abs/2013ascl.soft03024K}{\JournalTitle{Astrophysics
  Source Code Library}, ascl:1303.024}

\bibitem[{Lada \& Lada(2003)}]{lada_embedded_2003}
Lada, C.~J., \& Lada, E.~A. 2003,
  \href{http://dx.doi.org/10.1146/annurev.astro.41.011802.094844}{\JournalTitle{\araa},
  41, 57}

\bibitem[{{Landsman}(1995)}]{landsman_1995}
{Landsman}, W.~B. 1995, in Astronomical Society of the Pacific Conference
  Series, Vol.~77, Astronomical Data Analysis Software and Systems IV, ed.
  R.~A. {Shaw}, H.~E. {Payne}, \& J.~J.~E. {Hayes}, 437

\bibitem[{Li {et~al.}(2019)Li, Tenenbaum, Twicken, Burke, Jenkins, Quintana,
  Rowe, \& Seader}]{li_kepler_2019}
Li, J., Tenenbaum, P., Twicken, J.~D., {et~al.} 2019,
  \href{http://dx.doi.org/10.1088/1538-3873/aaf44d}{\JournalTitle{\pasp}, 131,
  024506}

\bibitem[{{Lightkurve Collaboration} {et~al.}(2018){Lightkurve Collaboration},
  {Cardoso}, {Hedges}, {Gully-Santiago}, {Saunders}, {Cody}, {Barclay}, {Hall},
  {Sagear}, {Turtelboom}, {Zhang}, {Tzanidakis}, {Mighell}, {Coughlin}, {Bell},
  {Berta-Thompson}, {Williams}, {Dotson}, \& {Barentsen}}]{lightkurve_2018}
{Lightkurve Collaboration}, {Cardoso}, J.~V.~d.~M., {Hedges}, C., {et~al.}
  2018, {Lightkurve: Kepler and TESS time series analysis in Python},
  Astrophysics Source Code Library,
  \href{http://arxiv.org/abs/1812.013}{{\sffamily ascl:1812.013}}

\bibitem[{Lin {et~al.}(1996)Lin, Bodenheimer, \& Richardson}]{lin_orbital_1996}
Lin, D. N.~C., Bodenheimer, P., \& Richardson, D.~C. 1996,
  \href{http://dx.doi.org/10.1038/380606a0}{\JournalTitle{\nat}, 380, 606}

\bibitem[{Lindegren {et~al.}(2018)Lindegren, Hern\'{a}ndez, Bombrun, Klioner,
  Bastian, Ramos-Lerate, de~Torres, Steidelmüller, Stephenson, Hobbs, Lammers,
  Biermann, Geyer, Hilger, Michalik, Stampa, McMillan, Castañeda, Clotet,
  Comoretto, Davidson, Fabricius, Gracia, Hambly, Hutton, Mora, Portell, van
  Leeuwen, Abbas, Abreu, Altmann, Andrei, Anglada, Balaguer-Núñez, Barache,
  Becciani, Bertone, Bianchi, Bouquillon, Bourda, Brüsemeister, Bucciarelli,
  Busonero, Buzzi, Cancelliere, Carlucci, Charlot, Cheek, Crosta, Crowley,
  de~Bruijne, de~Felice, Drimmel, Esquej, Fienga, Fraile, Gai, Garralda,
  González-Vidal, Guerra, Hauser, Hofmann, Holl, Jordan, Lattanzi, Lenhardt,
  Liao, Licata, Lister, Löffler, Marchant, Martin-Fleitas, Messineo, Mignard,
  Morbidelli, Poggio, Riva, Rowell, Salguero, Sarasso, Sciacca, Siddiqui,
  Smart, Spagna, Steele, Taris, Torra, van Elteren, van Reeven, \&
  Vecchiato}]{lindegren_gaiasoln_2018}
Lindegren, L., Hern\'{a}ndez, J., Bombrun, A., {et~al.} 2018,
  \href{http://dx.doi.org/10.1051/0004-6361/201832727}{\JournalTitle{\aap},
  616, A2}

\bibitem[{Lithwick \& Wu(2014)}]{lithwick_secular_2014}
Lithwick, Y., \& Wu, Y. 2014,
  \href{http://dx.doi.org/10.1073/pnas.1308261110}{\JournalTitle{Proceedings of
  the National Academy of Sciences}, 111, 12610}

\bibitem[{Livingston {et~al.}(2018)Livingston, Dai, Hirano, Gandolfi, Nowak,
  Endl, Velasco, Fukui, Narita, Prieto-Arranz, Barragan, Cusano, Albrecht,
  Cabrera, Cochran, Csizmadia, Deeg, Eigmüller, Erikson, Fridlund, Grziwa,
  Guenther, Hatzes, Kawauchi, Korth, Nespral, Palle, Pätzold, Persson, Rauer,
  Smith, Tamura, Tanaka, Eylen, Watanabe, \& Winn}]{livingston_three_2018}
Livingston, J.~H., Dai, F., Hirano, T., {et~al.} 2018,
  \href{http://dx.doi.org/10.3847/1538-3881/aaa841}{\JournalTitle{\aj}, 155,
  115}

\bibitem[{Livingston {et~al.}(2019)Livingston, Dai, Hirano, Gandolfi, Trani,
  Nowak, Cochran, Endl, Albrecht, Barragan, Cabrera, Csizmadia, de~Leon, Deeg,
  Eigmüller, Erikson, Fridlund, Fukui, Grziwa, Guenther, Hatzes, Korth,
  Kuzuhara, Montañes, Narita, Nespral, Palle, Pätzold, Persson,
  Prieto-Arranz, Rauer, Tamura, Van~Eylen, \& Winn}]{livingston_k2-264_2019}
---. 2019,
  \href{http://dx.doi.org/10.1093/mnras/sty3464}{\JournalTitle{\mnras}, 484, 8}

\bibitem[{Lomb(1976)}]{lomb_1976}
Lomb, N.~R. 1976,
  \href{http://dx.doi.org/10.1007/BF00648343}{\JournalTitle{Astrophysics and
  Space Science}, 39, 447}

\bibitem[{Lovis \& Mayor(2007)}]{lovis_mayor_2007}
Lovis, C., \& Mayor, M. 2007, \JournalTitle{\aap}, 472, 657

\bibitem[{{Luger} {et~al.}(2019){Luger}, {Agol}, {Foreman-Mackey}, {Fleming},
  {Lustig-Yaeger}, \& {Deitrick}}]{exoplanet:luger18}
{Luger}, R., {Agol}, E., {Foreman-Mackey}, D., {et~al.} 2019,
  \href{http://dx.doi.org/10.3847/1538-3881/aae8e5}{\JournalTitle{\aj}, 157,
  64}

\bibitem[{Malavolta {et~al.}(2016)Malavolta, Nascimbeni, Piotto, Quinn,
  Borsato, Granata, Bonomo, Marzari, Bedin, Rainer, Desidera, Lanza, Poretti,
  Sozzetti, White, Latham, Cunial, Libralato, Nardiello, Boccato, Claudi,
  Cosentino, Covino, Gratton, Maggio, Micela, Molinari, Pagano, Smareglia,
  Affer, Andreuzzi, Aparicio, Benatti, Bignamini, Borsa, Damasso, Di~Fabrizio,
  Harutyunyan, Esposito, Fiorenzano, Gandolfi, Giacobbe, González~Hernández,
  Maldonado, Masiero, Molinaro, Pedani, \& Scandariato}]{Malavolta_et_al_2016}
Malavolta, L., Nascimbeni, V., Piotto, G., {et~al.} 2016, \JournalTitle{\aap},
  588, A118

\bibitem[{Mamajek(2009)}]{mamajek_initial_2009}
Mamajek, E.~E. \href{http://dx.doi.org/10.1063/1.3215910}{2009, 1158, 3},
  {C}onference Name: Exoplanets and Disks: Their Formation and Diversity Place:
  eprint: arXiv:0906.5011

\bibitem[{Mamajek \& Hillenbrand(2008)}]{mamajek_improved_2008}
Mamajek, E.~E., \& Hillenbrand, L.~A. 2008,
  \href{http://dx.doi.org/10.1086/591785}{\JournalTitle{\apj}, 687, 1264}

\bibitem[{Mann {et~al.}(2013)Mann, Gaidos, \&
  Ansdell}]{mann_spectrothermometry_2013}
Mann, A.~W., Gaidos, E., \& Ansdell, M. 2013,
  \href{http://dx.doi.org/10.1088/0004-637X/779/2/188}{\JournalTitle{\apj},
  779, 188}

\bibitem[{Mann {et~al.}(2016{\natexlab{a}})Mann, Gaidos, Mace, Johnson, Bowler,
  LaCourse, Jacobs, Vanderburg, Kraus, Kaplan, \& Jaffe}]{Mann_K2_25_2016}
Mann, A.~W., Gaidos, E., Mace, G.~N., {et~al.} 2016{\natexlab{a}},
  \JournalTitle{ApJ}, 818

\bibitem[{Mann {et~al.}(2016{\natexlab{b}})Mann, Newton, Rizzuto, Irwin,
  Feiden, Gaidos, Mace, Kraus, James, Ansdell, Charbonneau, Covey, Ireland,
  Jaffe, Johnson, Kidder, \& Vanderburg}]{Mann_K2_33b_2016}
Mann, A.~W., Newton, E.~R., Rizzuto, A.~C., {et~al.} 2016{\natexlab{b}},
  \JournalTitle{AJ}, 152, 61

\bibitem[{Mann {et~al.}(2017)Mann, Gaidos, Vanderburg, Rizzuto, Ansdell,
  Medina, Mace, Kraus, \& Sokal}]{Mann_et_al_2017}
Mann, A.~W., Gaidos, E., Vanderburg, A., {et~al.} 2017, \JournalTitle{AJ}, 153,
  64

\bibitem[{Mann {et~al.}(2018)Mann, Vanderburg, Rizzuto, Kraus, Berlind,
  Bieryla, Calkins, Esquerdo, Latham, Mace, Morris, Quinn, Sokal, \&
  Stefanik}]{mann_ZEITVI_2018}
Mann, A.~W., Vanderburg, A., Rizzuto, A.~C., {et~al.} 2018,
  \href{http://dx.doi.org/10.3847/1538-3881/aa9791}{\JournalTitle{\aj}, 155, 4}

\bibitem[{Mann {et~al.}(2020)Mann, Johnson, Vanderburg, Kraus, Rizzuto, Wood,
  Bush, Rockcliffe, Newton, Latham, Mamajek, Zhou, Quinn, Thao, Benatti,
  Cosentino, Desidera, Harutyunyan, Lovis, Mortier, Pepe, Poretti, Wilson,
  Kristiansen, Gagliano, Jacobs, LaCourse, Omohundro, Schwengeler, Kane, Hill,
  Rabus, Esquerdo, Berlind, Collins, Murawski, Aitken, Sallam, Massey, Ricker,
  Vanderspek, Seager, Winn, Jenkins, Barclay, Caldwell, Dragomir, Doty,
  Glidden, Tenenbaum, Torres, Twicken, \& Villanueva~Jr}]{mann_tess_2020}
Mann, A.~W., Johnson, M.~C., Vanderburg, A., {et~al.} 2020,
  \href{http://arxiv.org/abs/2005.00047}{\JournalTitle{arXiv:2005.00047
  [astro-ph]}}

\bibitem[{Mansfield {et~al.}(2018)Mansfield, Bean, Oklop\v{c}i\'c, Kreidberg,
  Désert, Kempton, Line, Fortney, Henry, Mallonn, Stevenson, Dragomir, Allart,
  \& Bourrier}]{mansfield_detection_2018}
Mansfield, M., Bean, J.~L., Oklop\v{c}i\'c, A., {et~al.} 2018,
  \href{http://dx.doi.org/10.3847/2041-8213/aaf166}{\JournalTitle{\apj}, 868,
  L34}

\bibitem[{Marigo {et~al.}(2017)Marigo, Girardi, Bressan, Rosenfield, Aringer,
  Chen, Dussin, Nanni, Pastorelli, Rodrigues, Trabucchi, Bladh, Dalcanton,
  Groenewegen, Montalbán, \& Wood}]{marigo_new_2017}
Marigo, P., Girardi, L., Bressan, A., {et~al.} 2017,
  \href{http://dx.doi.org/10.3847/1538-4357/835/1/77}{\JournalTitle{\apj}, 835,
  77}

\bibitem[{Martioli {et~al.}(2020)Martioli, Hebrard, Moutou, Donati, Artigau,
  Cale, Cook, Dalal, Delfosse, Forveille, Gaidos, Plavchan, Berberian, Carmona,
  Cloutier, Doyon, Fouque, Klein, Etangs, Manset, Morin, Tanner, Teske, \&
  Wang}]{martioli_magnetism_2020}
Martioli, E., Hebrard, G., Moutou, C., {et~al.} 2020,
  \href{http://arxiv.org/abs/2006.13269}{\JournalTitle{arXiv:2006.13269
  [astro-ph]}}

\bibitem[{Mayor {et~al.}(2011)Mayor, Marmier, Lovis, Udry, Ségransan, Pepe,
  Benz, Bertaux, Bouchy, Dumusque, Lo~Curto, Mordasini, Queloz, \&
  Santos}]{mayor_harps_2011}
Mayor, M., Marmier, M., Lovis, C., {et~al.} 2011,
  \href{http://adsabs.harvard.edu/abs/2011arXiv1109.2497M}{\JournalTitle{ArXiv
  e-prints}, 1109, arXiv:1109.2497}

\bibitem[{McKinney(2010)}]{mckinney-proc-scipy-2010}
McKinney, W. 2010, in Proceedings of the 9th Python in Science Conference, ed.
  S.~van~der Walt \& J.~Millman, 51

\bibitem[{Meibom {et~al.}(2015)Meibom, Barnes, Platais, Gilliland, Latham, \&
  Mathieu}]{meibom_spin-down_2015}
Meibom, S., Barnes, S.~A., Platais, I., {et~al.} 2015,
  \href{http://dx.doi.org/10.1038/nature14118}{\JournalTitle{\nat}, 517, 589}

\bibitem[{Meibom {et~al.}(2013)Meibom, Torres, Fressin, Latham, Rowe, Ciardi,
  Bryson, Rogers, Henze, Janes, Barnes, Marcy, Isaacson, Fischer, Howell,
  Horch, Jenkins, Schuler, \& Crepp}]{Meibom_et_al_2013}
Meibom, S., Torres, G., Fressin, F., {et~al.} 2013, \JournalTitle{\nat}, 499,
  55

\bibitem[{M\'ekarnia {et~al.}(2016)M\'ekarnia, Guillot, Rivet, Schmider, Abe,
  Gonçalves, Agabi, Crouzet, Fruth, Barbieri, Bayliss, Zhou, Aristidi,
  Szulagyi, Daban, Fanteï-Caujolle, Gouvret, Erikson, Rauer, Bouchy, Gerakis,
  \& Bouchez}]{mekarnia_transiting_2016}
M\'ekarnia, D., Guillot, T., Rivet, J.-P., {et~al.} 2016,
  \href{http://dx.doi.org/10.1093/mnras/stw1934}{\JournalTitle{\mnras}, 463,
  45}

\bibitem[{Mermilliod(1981)}]{mermilliod_comparative_1981}
Mermilliod, J.~C. 1981,
  \href{http://adsabs.harvard.edu/abs/1981A%26A....97..235M}{\JournalTitle{\aap},
  97, 235}

\bibitem[{Miller {et~al.}(2008)Miller, Irwin, Aigrain, Hodgkin, \&
  Hebb}]{miller_monitor_2008}
Miller, A.~A., Irwin, J., Aigrain, S., Hodgkin, S., \& Hebb, L. 2008,
  \href{http://dx.doi.org/10.1111/j.1365-2966.2008.13236.x}{\JournalTitle{\mnras},
  387, 349}

\bibitem[{Mochejska {et~al.}(2005)Mochejska, Stanek, Sasselov, Szentgyorgyi,
  Bakos, Hradecky, Marrone, Winn, \& Zaldarriaga}]{mochejska_planets_2005}
Mochejska, B.~J., Stanek, K.~Z., Sasselov, D.~D., {et~al.} 2005,
  \href{http://dx.doi.org/10.1086/430219}{\JournalTitle{\aj}, 129, 2856}

\bibitem[{Mochejska {et~al.}(2006)Mochejska, Stanek, Sasselov, Szentgyorgyi,
  Adams, Cooper, Foster, Hartman, Hickox, Lai, Westover, \&
  Winn}]{mochejska_planets_2006}
---. 2006, \href{http://dx.doi.org/10.1086/499208}{\JournalTitle{\aj}, 131,
  1090}

\bibitem[{Montet {et~al.}(2020)Montet, Feinstein, Luger, Bedell,
  Gully-Santiago, Teske, Wang, Butler, Flowers, Shectman, Crane, \&
  Thompson}]{montet_young_2020}
Montet, B.~T., Feinstein, A.~D., Luger, R., {et~al.} 2020,
  \href{http://dx.doi.org/10.3847/1538-3881/ab6d6d}{\JournalTitle{\aj}, 159,
  112}

\bibitem[{Morbidelli {et~al.}(2012)Morbidelli, Lunine, O'Brien, Raymond, \&
  Walsh}]{morbidelli_building_2012}
Morbidelli, A., Lunine, J.~I., O'Brien, D.~P., Raymond, S.~N., \& Walsh, K.~J.
  2012,
  \href{http://dx.doi.org/10.1146/annurev-earth-042711-105319}{\JournalTitle{Annual
  Review of Earth and Planetary Sciences}, 40, 251}

\bibitem[{Morton(2012)}]{morton_efficient_2012}
Morton, T.~D. 2012,
  \href{http://dx.doi.org/10.1088/0004-637X/761/1/6}{\JournalTitle{\apj}, 761,
  6}

\bibitem[{{Morton}(2015{\natexlab{a}})}]{morton_2015_isochrones}
{Morton}, T.~D. 2015{\natexlab{a}}, {isochrones: Stellar model grid package}

\bibitem[{{Morton}(2015{\natexlab{b}})}]{vespa_2015}
---. 2015{\natexlab{b}}, {VESPA: False positive probabilities calculator},
  Astrophysics Source Code Library,
  \href{http://arxiv.org/abs/1503.011}{{\sffamily ascl:1503.011}}

\bibitem[{Morton {et~al.}(2016)Morton, Bryson, Coughlin, Rowe, Ravichandran,
  Petigura, Haas, \& Batalha}]{morton_false_2016}
Morton, T.~D., Bryson, S.~T., Coughlin, J.~L., {et~al.} 2016,
  \href{http://dx.doi.org/10.3847/0004-637X/822/2/86}{\JournalTitle{\apj}, 822,
  86}

\bibitem[{Nardiello {et~al.}(2020)Nardiello, Piotto, Deleuil, Malavolta,
  Montalto, Bedin, Borsato, Granata, Libralato, \&
  Manthopoulou}]{nardiello_pathosII_2020}
Nardiello, D., Piotto, G., Deleuil, M., {et~al.} 2020,
  \href{http://dx.doi.org/10.1093/mnras/staa1465}{\JournalTitle{\mnras}, 495,
  4924}

\bibitem[{Netopil {et~al.}(2016)Netopil, Paunzen, Heiter, \&
  Soubiran}]{netopil_metallicity_2016}
Netopil, M., Paunzen, E., Heiter, U., \& Soubiran, C. 2016,
  \href{http://dx.doi.org/10.1051/0004-6361/201526370}{\JournalTitle{\aap},
  585, A150}

\bibitem[{Newton {et~al.}(2019)Newton, Mann, Tofflemire, Pearce, Rizzuto,
  Vanderburg, Martinez, Wang, Ruffio, Kraus, Johnson, Thao, Wood, Rampalli,
  Nielsen, Collins, Dragomir, Hellier, Anderson, Barclay, Brown, Feiden, Hart,
  Isopi, Kielkopf, Mallia, Nelson, Rodriguez, Stockdale, Waite, Wright,
  Lissauer, Ricker, Vanderspek, Latham, Seager, Winn, Jenkins, Bouma, Burke,
  Davies, Fausnaugh, Li, Morris, Mukai, Villaseñor, Villeneuva, Rosa,
  Macintosh, Mengel, Okumura, \& Wittenmyer}]{newton_tess_2019}
Newton, E.~R., Mann, A.~W., Tofflemire, B.~M., {et~al.} 2019,
  \href{http://dx.doi.org/10.3847/2041-8213/ab2988}{\JournalTitle{\apj}, 880,
  L17}

\bibitem[{Obermeier {et~al.}(2016)Obermeier, Henning, Schlieder, Crossfield,
  Petigura, Howard, Sinukoff, Isaacson, Ciardi, David, Hillenbrand, Beichman,
  Howell, Horch, Everett, Hirsch, Teske, Christiansen, Lépine, Aller, Liu,
  Saglia, Livingston, \& Kluge}]{obermeier_k2_2016}
Obermeier, C., Henning, T., Schlieder, J.~E., {et~al.} 2016,
  \href{http://dx.doi.org/10.3847/1538-3881/152/6/223}{\JournalTitle{\aj}, 152,
  223}

\bibitem[{Oh {et~al.}(2017)Oh, Price-Whelan, Hogg, Morton, \&
  Spergel}]{oh_comoving_2017}
Oh, S., Price-Whelan, A.~M., Hogg, D.~W., Morton, T.~D., \& Spergel, D.~N.
  2017, \href{http://dx.doi.org/10.3847/1538-3881/aa6ffd}{\JournalTitle{\aj},
  153, 257}

\bibitem[{{Oklop{\v{c}}i{\'c}} \& {Hirata}(2018)}]{oklopcic_new_2018}
{Oklop{\v{c}}i{\'c}}, A., \& {Hirata}, C.~M. 2018,
  \href{http://dx.doi.org/10.3847/2041-8213/aaada9}{\JournalTitle{\apjl}, 855,
  L11}

\bibitem[{Owen \& Wu(2013)}]{Owen_Wu_2013}
Owen, J.~E., \& Wu, Y. 2013, \JournalTitle{ApJ}, 775, 105

\bibitem[{Owen \& Wu(2017)}]{owen_evaporation_2017}
---. 2017,
  \href{http://dx.doi.org/10.3847/1538-4357/aa890a}{\JournalTitle{\apj}, 847,
  29}

\bibitem[{Palle {et~al.}(2020)Palle, Oshagh, Casasayas-Barris, Hirano,
  Stangret, Luque, Strachan, Gaidos, Anglada-Escude, Plavchan, \&
  Addison}]{palle_transmission_2020}
Palle, E., Oshagh, M., Casasayas-Barris, N., {et~al.} 2020,
  \href{http://arxiv.org/abs/2006.13609}{\JournalTitle{arXiv:2006.13609
  [astro-ph]}}

\bibitem[{Paxton {et~al.}(2011)Paxton, Bildsten, Dotter, Herwig, Lesaffre, \&
  Timmes}]{paxton_modules_2011}
Paxton, B., Bildsten, L., Dotter, A., {et~al.} 2011,
  \href{http://dx.doi.org/10.1088/0067-0049/192/1/3}{\JournalTitle{\apjs}, 192,
  3}

\bibitem[{Paxton {et~al.}(2013)Paxton, Cantiello, Arras, Bildsten, Brown,
  Dotter, Mankovich, Montgomery, Stello, Timmes, \&
  Townsend}]{paxton_modules_2013}
Paxton, B., Cantiello, M., Arras, P., {et~al.} 2013,
  \href{http://dx.doi.org/10.1088/0067-0049/208/1/4}{\JournalTitle{\apjs}, 208,
  4}

\bibitem[{Paxton {et~al.}(2015)Paxton, Marchant, Schwab, Bauer, Bildsten,
  Cantiello, Dessart, Farmer, Hu, Langer, Townsend, Townsley, \&
  Timmes}]{paxton_modules_2015}
Paxton, B., Marchant, P., Schwab, J., {et~al.} 2015,
  \href{http://dx.doi.org/10.1088/0067-0049/220/1/15}{\JournalTitle{\apjs},
  220, 15}

\bibitem[{Pepper {et~al.}(2008)Pepper, Stanek, Pogge, Latham, DePoy, Siverd,
  Poindexter, \& Sivakoff}]{pepper_photometric_2008}
Pepper, J., Stanek, K.~Z., Pogge, R.~W., {et~al.} 2008,
  \href{http://dx.doi.org/10.1088/0004-6256/135/3/907}{\JournalTitle{\aj}, 135,
  907}

\bibitem[{P\'erez \& Granger(2007)}]{perez_2007}
P\'erez, F., \& Granger, B.~E. 2007,
  \href{http://dx.doi.org/10.1109/MCSE.2007.53}{\JournalTitle{Computing in
  Science and Engineering}, 9, 21}

\bibitem[{Petigura {et~al.}(2018)Petigura, Marcy, Winn, Weiss, Fulton, Howard,
  Sinukoff, Isaacson, Morton, \& Johnson}]{petigura_metallicity_2018}
Petigura, E.~A., Marcy, G.~W., Winn, J.~N., {et~al.} 2018,
  \href{http://dx.doi.org/10.3847/1538-3881/aaa54c}{\JournalTitle{\aj}, 155,
  89}

\bibitem[{Plavchan {et~al.}(2020)Plavchan, Barclay, Gagné, Gao, Cale, Matzko,
  Dragomir, Quinn, Feliz, Stassun, Crossfield, Berardo, Latham, Tieu,
  Anglada-Escudé, Ricker, Vanderspek, Seager, Winn, Jenkins, Rinehart,
  Krishnamurthy, Dynes, Doty, Adams, Afanasev, Beichman, Bottom, Bowler,
  Brinkwort, Brown, Cancino, Ciardi, Clampin, Clark, Collins, Davison,
  Foreman-Mackey, Furlan, Gaidos, Geneser, Giddens, Gilbert, Hall, Hellier,
  Henry, Horner, Howard, Huang, Huber, Kane, Kenworthy, Kielkopf, Kipping,
  Klenke, Kruse, Latouf, Lowrance, Mennesson, Mengel, Mills, Morton, Narita,
  Newton, Nishimoto, Okumura, Palle, Pepper, Quintana, Roberge, Roccatagliata,
  Schlieder, Tanner, Teske, Tinney, Vanderburg, von Braun, Walp, Wang, Wang,
  Weigand, White, Wittenmyer, Wright, Youngblood, Zhang, \&
  Zilberman}]{plavchan_planet_2020}
Plavchan, P., Barclay, T., Gagné, J., {et~al.} 2020,
  \href{http://dx.doi.org/10.1038/s41586-020-2400-z}{\JournalTitle{\nat}, 582,
  497}

\bibitem[{Pollack {et~al.}(1996)Pollack, Hubickyj, Bodenheimer, Lissauer,
  Podolak, \& Greenzweig}]{pollack_formation_1996}
Pollack, J.~B., Hubickyj, O., Bodenheimer, P., {et~al.} 1996,
  \href{http://dx.doi.org/10.1006/icar.1996.0190}{\JournalTitle{Icarus}, 124,
  62}

\bibitem[{Quinn {et~al.}(2012)Quinn, White, Latham, Buchhave, Cantrell, Dahm,
  Fűr\'esz, Szentgyorgyi, Geary, Torres, Bieryla, Berlind, Calkins, Esquerdo,
  \& Stefanik}]{Quinn_et_al_2012}
Quinn, S.~N., White, R.~J., Latham, D.~W., {et~al.} 2012, \JournalTitle{\apjl},
  756, L33

\bibitem[{Raghavan {et~al.}(2010)Raghavan, McAlister, Henry, Latham, Marcy,
  Mason, Gies, White, \& Brummelaar}]{raghavan_survey_2010}
Raghavan, D., McAlister, H.~A., Henry, T.~J., {et~al.} 2010,
  \href{http://dx.doi.org/10.1088/0067-0049/190/1/1}{\JournalTitle{\apjs}, 190,
  1}

\bibitem[{Rajpaul {et~al.}(2015)Rajpaul, Aigrain, Osborne, Reece, \&
  Roberts}]{rajpaul_gaussian_2015}
Rajpaul, V., Aigrain, S., Osborne, M.~A., Reece, S., \& Roberts, S. 2015,
  \href{http://dx.doi.org/10.1093/mnras/stv1428}{\JournalTitle{\mnras}, 452,
  2269}

\bibitem[{Rajpurohit {et~al.}(2013)Rajpurohit, Reylé, Allard, Homeier,
  Schultheis, Bessell, \& Robin}]{rajpurohit_effective_2013}
Rajpurohit, A.~S., Reylé, C., Allard, F., {et~al.} 2013,
  \href{http://dx.doi.org/10.1051/0004-6361/201321346}{\JournalTitle{\aap},
  556, A15}

\bibitem[{Randich {et~al.}(2018)Randich, Tognelli, Jackson, Jeffries,
  Degl’Innocenti, Pancino, Fiorentin, Spagna, Sacco, Bragaglia, Magrini,
  Moroni, Alfaro, Franciosini, Morbidelli, Roccatagliata, Bouy, Bravi,
  Jiménez-Esteban, Jordi, Zari, Tautvaišiene, Drazdauskas, Mikolaitis,
  Gilmore, Feltzing, Vallenari, Bensby, Koposov, Korn, Lanzafame, Smiljanic,
  Bayo, Carraro, Costado, Heiter, Hourihane, Jofré, Lewis, Monaco, Prisinzano,
  Sbordone, Sousa, Worley, \& Zaggia}]{randich_gaiaeso_2018}
Randich, S., Tognelli, E., Jackson, R., {et~al.} 2018,
  \href{http://dx.doi.org/10.1051/0004-6361/201731738}{\JournalTitle{\aap},
  612, A99}

\bibitem[{Raymond {et~al.}(2014)Raymond, Kokubo, Morbidelli, Morishima, \&
  Walsh}]{raymond_terrestrial_2014}
Raymond, S.~N., Kokubo, E., Morbidelli, A., Morishima, R., \& Walsh, K.~J.
  2014,
  \href{http://dx.doi.org/10.2458/azu_uapress_9780816531240-ch026}{\JournalTitle{Protostars
  and Planets VI}, 595}

\bibitem[{Rebull {et~al.}(2020)Rebull, Stauffer, Cody, Hillenbrand, Bouvier,
  Roggero, \& David}]{rebull_rotation_2020}
Rebull, L.~M., Stauffer, J.~R., Cody, A.~M., {et~al.}
  \href{https://arxiv.org/abs/2004.04236v1}{2020}

\bibitem[{Rebull {et~al.}(2016)Rebull, Stauffer, Bouvier, Cody, Hillenbrand,
  Soderblom, Valenti, Barrado, Bouy, Ciardi, Pinsonneault, Stassun, Micela,
  Aigrain, Vrba, Somers, Christiansen, Gillen, \&
  Collier~Cameron}]{rebull_rotation_2016a}
Rebull, L.~M., Stauffer, J.~R., Bouvier, J., {et~al.} 2016,
  \href{http://dx.doi.org/10.3847/0004-6256/152/5/113}{\JournalTitle{\aj}, 152,
  113}

\bibitem[{Ricker {et~al.}(2015)Ricker, Winn, Vanderspek, Latham, Bakos, Bean,
  Berta-Thompson, Brown, Buchhave, Butler, Butler, Chaplin, Charbonneau,
  Christensen-Dalsgaard, Clampin, Deming, Doty, De~Lee, Dressing, Dunham, Endl,
  Fressin, Ge, Henning, Holman, Howard, Ida, Jenkins, Jernigan, Johnson,
  Kaltenegger, Kawai, Kjeldsen, Laughlin, Levine, Lin, Lissauer, MacQueen,
  Marcy, McCullough, Morton, Narita, Paegert, Palle, Pepe, Pepper, Quirrenbach,
  Rinehart, Sasselov, Sato, Seager, Sozzetti, Stassun, Sullivan, Szentgyorgyi,
  Torres, Udry, \& Villasenor}]{ricker_transiting_2015}
Ricker, G.~R., Winn, J.~N., Vanderspek, R., {et~al.} 2015,
  \href{http://dx.doi.org/10.1117/1.JATIS.1.1.014003}{\JournalTitle{Journal of
  Astronomical Telescopes, Instruments, and Systems}, 1, 014003}

\bibitem[{Rizzuto {et~al.}(2018)Rizzuto, Vanderburg, Mann, Kraus, Dressing,
  Agüeros, Douglas, \& Krolikowski}]{rizzuto_zeitVIII_2018}
Rizzuto, A.~C., Vanderburg, A., Mann, A.~W., {et~al.} 2018,
  \href{http://dx.doi.org/10.3847/1538-3881/aadf37}{\JournalTitle{\aj}, 156,
  195}

\bibitem[{Rizzuto {et~al.}(2020)Rizzuto, Newton, Mann, Tofflemire, Vanderburg,
  Kraus, Wood, Quinn, Zhou, Thao, Law, Ziegler, \& Briceno}]{rizzuto_tess_2020}
Rizzuto, A.~C., Newton, E.~R., Mann, A.~W., {et~al.} 2020,
  \href{http://arxiv.org/abs/2005.00013}{\JournalTitle{arXiv:2005.00013
  [astro-ph]}}

\bibitem[{Rohatgi(2019)}]{rohatgi_2019}
Rohatgi, A. 2019, WebPlotDigitizer: v4.2

\bibitem[{{Salvatier} {et~al.}(2016){Salvatier}, {Wiecki{\^a}}, \&
  {Fonnesbeck}}]{salvatier_2016_PyMC3}
{Salvatier}, J., {Wiecki{\^a}}, T.~V., \& {Fonnesbeck}, C. 2016, {PyMC3: Python
  probabilistic programming framework}

\bibitem[{Sanchis-Ojeda {et~al.}(2013)Sanchis-Ojeda, Winn, Marcy, Howard,
  Isaacson, Johnson, Torres, Albrecht, Campante, Chaplin, Davies, Lund, Carter,
  Dawson, Buchhave, Everett, Fischer, Geary, Gilliland, Horch, Howell, \&
  Latham}]{sanchis-ojeda_kepler-63b_2013}
Sanchis-Ojeda, R., Winn, J.~N., Marcy, G.~W., {et~al.} 2013,
  \href{http://dx.doi.org/10.1088/0004-637X/775/1/54}{\JournalTitle{The
  Astrophysical Journal}, 775, 54}

\bibitem[{Santerne {et~al.}(2015)Santerne, D\'{i}az, Almenara, Bouchy, Deleuil,
  Figueira, Hébrard, Moutou, Rodionov, \& Santos}]{santerne_pastis_2015}
Santerne, A., D\'{i}az, R.~F., Almenara, J.-M., {et~al.} 2015,
  \href{http://dx.doi.org/10.1093/mnras/stv1080}{\JournalTitle{MNRAS}, 451,
  2337}

\bibitem[{Sato {et~al.}(2007)Sato, Izumiura, Toyota, Kambe, Takeda, Masuda,
  Omiya, {Daisuke Murata}, Itoh, Ando, Yoshida, Ikoma, Kokubo, \&
  Ida}]{Sato_et_al_2007}
Sato, B., Izumiura, H., Toyota, E., {et~al.} 2007, \JournalTitle{\apj}, 661,
  527

\bibitem[{Scargle(1982)}]{scargle_studies_1982}
Scargle, J.~D. 1982,
  \href{http://dx.doi.org/10.1086/160554}{\JournalTitle{\apj}, 263, 835}

\bibitem[{Seager \& Mall\'en-Ornelas(2003)}]{seager_unique_2003}
Seager, S., \& Mall\'en-Ornelas, G. 2003,
  \href{http://dx.doi.org/10.1086/346105}{\JournalTitle{\apj}, 585, 1038}

\bibitem[{Serv\'en {et~al.}(2018)Serv\'en, Brummitt, \&
  Abedi}]{serven_pygam_2018_1476122}
Serv\'en, D., Brummitt, C., \& Abedi, H. 2018, dswah/pyGAM: v0.8.0

\bibitem[{Skrutskie {et~al.}(2006)Skrutskie, Cutri, Stiening, Weinberg,
  Schneider, Carpenter, Beichman, Capps, Chester, Elias, Huchra, Liebert,
  Lonsdale, Monet, Price, Seitzer, Jarrett, Kirkpatrick, Gizis, Howard, Evans,
  Fowler, Fullmer, Hurt, Light, Kopan, Marsh, McCallon, Tam, Van~Dyk, \&
  Wheelock}]{skrutskie_tmass_2006}
Skrutskie, M.~F., Cutri, R.~M., Stiening, R., {et~al.} 2006,
  \href{http://dx.doi.org/10.1086/498708}{\JournalTitle{\aj}, 131, 1163}

\bibitem[{Skumanich(1972)}]{skumanich_time_1972}
Skumanich, A. 1972,
  \href{http://dx.doi.org/10.1086/151310}{\JournalTitle{\apj}, 171, 565}

\bibitem[{Smith {et~al.}(2016)Smith, Morris, Jenkins, Bryson, Caldwell, \&
  Girouard}]{smith_finding_2016}
Smith, J.~C., Morris, R.~L., Jenkins, J.~M., {et~al.} 2016,
  \href{http://dx.doi.org/10.1088/1538-3873/128/970/124501}{\JournalTitle{\pasp},
  128, 124501}

\bibitem[{Smith {et~al.}(2012)Smith, Stumpe, Cleve, Jenkins, Barclay, Fanelli,
  Girouard, Kolodziejczak, McCauliff, Morris, \& Twicken}]{smith_kepler_2012}
Smith, J.~C., Stumpe, M.~C., Cleve, J. E.~V., {et~al.} 2012,
  \href{http://dx.doi.org/10.1086/667697}{\JournalTitle{\pasp}, 124, 1000}

\bibitem[{Soderblom(2010)}]{soderblom_ages_2010}
Soderblom, D.~R. 2010,
  \href{http://dx.doi.org/10.1146/annurev-astro-081309-130806}{\JournalTitle{\araa},
  48, 581}

\bibitem[{Soderblom {et~al.}(2014)Soderblom, Hillenbrand, Jeffries, Mamajek, \&
  Naylor}]{soderblom_ages_2014}
Soderblom, D.~R., Hillenbrand, L.~A., Jeffries, R.~D., Mamajek, E.~E., \&
  Naylor, T. 2014,
  \href{http://dx.doi.org/10.2458/azu_uapress_9780816531240-ch010}{\JournalTitle{Protostars
  and Planets VI}, 219}

\bibitem[{Spake {et~al.}(2018)Spake, Sing, Evans, Oklopčić, Bourrier,
  Kreidberg, Rackham, Irwin, Ehrenreich, Wyttenbach, Wakeford, Zhou, Chubb,
  Nikolov, Goyal, Henry, Williamson, Blumenthal, Anderson, Hellier,
  Charbonneau, Udry, \& Madhusudhan}]{spake_helium_2018}
Spake, J.~J., Sing, D.~K., Evans, T.~M., {et~al.} 2018,
  \href{http://dx.doi.org/10.1038/s41586-018-0067-5}{\JournalTitle{\nat}, 557,
  68}

\bibitem[{Stassun {et~al.}(2017)Stassun, Collins, \&
  Gaudi}]{stassun_accurate_2017}
Stassun, K.~G., Collins, K.~A., \& Gaudi, B.~S. 2017,
  \href{http://dx.doi.org/10.3847/1538-3881/aa5df3}{\JournalTitle{\aj}, 153,
  136}

\bibitem[{Stassun {et~al.}(2018)Stassun, Oelkers, Pepper, Paegert, Lee, Torres,
  Latham, Charpinet, Dressing, Huber, Kane, {Sébastien Lépine}, Mann,
  Muirhead, Rojas-Ayala, Silvotti, Fleming, {Al Levine}, \&
  Plavchan}]{stassun_TIC_2018}
Stassun, K.~G., Oelkers, R.~J., Pepper, J., {et~al.} 2018,
  \href{http://dx.doi.org/10.3847/1538-3881/aad050}{\JournalTitle{\aj}, 156,
  102}

\bibitem[{Stassun {et~al.}(2019)Stassun, Oelkers, Paegert, Torres, Pepper,
  De~Lee, Collins, Latham, Muirhead, Chittidi, Rojas-Ayala, Fleming, Rose,
  Tenenbaum, Ting, Kane, Barclay, Bean, Brassuer, Charbonneau, Lissauer, Mann,
  McLean, Mulally, Narita, Plavchan, Ricker, Sasselov, Seager, Sharma, Shiao,
  Sozzetti, Stello, Vanderspek, Wallace, \& Winn}]{stassun_TIC8_2019}
Stassun, K.~G., Oelkers, R.~J., Paegert, M., {et~al.} 2019,
  \href{http://arxiv.org/abs/1905.10694}{\JournalTitle{arXiv:1905.10694
  [astro-ph]}}

\bibitem[{Stauffer {et~al.}(1997)Stauffer, Hartmann, Prosser, Randich,
  Balachandran, Patten, Simon, \& Giampapa}]{stauffer_rotational_1997}
Stauffer, J.~R., Hartmann, L.~W., Prosser, C.~F., {et~al.} 1997,
  \href{http://dx.doi.org/10.1086/303930}{\JournalTitle{\apj}, 479, 776}

\bibitem[{Stauffer {et~al.}(2003)Stauffer, Jones, Backman, Hartmann, Barrado~y
  Navascués, Pinsonneault, Terndrup, \& Muench}]{stauffer_why_2003}
Stauffer, J.~R., Jones, B.~F., Backman, D., {et~al.} 2003,
  \href{http://dx.doi.org/10.1086/376739}{\JournalTitle{\aj}, 126, 833}

\bibitem[{Stefansson {et~al.}(2020)Stefansson, Mahadevan, Maney, Ninan,
  Robertson, Rajagopal, Haase, Allen, Ford, Winn, Wolfgang, Dawson, Wisniewski,
  Bender, Cañas, Cochran, Diddams, Fredrick, Halverson, Hearty, Hebb, Kanodia,
  Levi, Metcalf, Monson, Ramsey, Roy, Schwab, Terrien, \&
  Wright}]{stefansson_k2-25_2020}
Stefansson, G., Mahadevan, S., Maney, M., {et~al.} 2020,
  \href{http://adsabs.harvard.edu/abs/2020arXiv200712766S}{\JournalTitle{arXiv
  e-prints}, 2007, arXiv:2007.12766}

\bibitem[{Stumpe {et~al.}(2014)Stumpe, Smith, Catanzarite, Cleve, Jenkins,
  Twicken, \& Girouard}]{stumpe_multiscale_2014}
Stumpe, M.~C., Smith, J.~C., Catanzarite, J.~H., {et~al.} 2014,
  \href{http://dx.doi.org/10.1086/674989}{\JournalTitle{\pasp}, 126, 100}

\bibitem[{Sullivan {et~al.}(2015)}]{Sullivan_2015}
Sullivan, P.~W., {et~al.} 2015, \JournalTitle{{ApJ}}, 809, 77

\bibitem[{{Theano Development Team}(2016)}]{exoplanet:theano}
{Theano Development Team}. 2016,
  \href{http://arxiv.org/abs/1605.02688}{\JournalTitle{arXiv e-prints},
  abs/1605.02688}

\bibitem[{Tokovinin(2018)}]{tokovinin_ten_2018}
Tokovinin, A. 2018,
  \href{http://dx.doi.org/10.1088/1538-3873/aaa7d9}{\JournalTitle{\pasp}, 130,
  035002}

\bibitem[{Tokovinin {et~al.}(2013)Tokovinin, Fischer, Bonati, Giguere, Moore,
  Schwab, Spronck, \& Szymkowiak}]{tokovinin_chironfiber_2013}
Tokovinin, A., Fischer, D.~A., Bonati, M., {et~al.} 2013,
  \href{http://dx.doi.org/10.1086/674012}{\JournalTitle{\pasp}, 125, 1336}

\bibitem[{Torres {et~al.}(2011)Torres, Fressin, Batalha, Borucki, Brown,
  Bryson, Buchhave, Charbonneau, Ciardi, Dunham, Fabrycky, Ford, Gautier,
  Gilliland, Holman, Howell, Isaacson, Jenkins, Koch, Latham, Lissauer, Marcy,
  Monet, Prsa, Quinn, Ragozzine, Rowe, Sasselov, Steffen, \&
  Welsh}]{torres_modeling_2011}
Torres, G., Fressin, F., Batalha, N.~M., {et~al.} 2011,
  \href{http://dx.doi.org/10.1088/0004-637X/727/1/24}{\JournalTitle{\apj}, 727,
  24}

\bibitem[{Touboul {et~al.}(2007)Touboul, Kleine, Bourdon, Palme, \&
  Wieler}]{touboul_late_2007}
Touboul, M., Kleine, T., Bourdon, B., Palme, H., \& Wieler, R. 2007,
  \href{http://dx.doi.org/10.1038/nature06428}{\JournalTitle{\nat}, 450, 1206}

\bibitem[{Twicken {et~al.}(2018)Twicken, Catanzarite, Clarke, Girouard,
  Jenkins, Klaus, Li, McCauliff, Seader, Tenenbaum, Wohler, Bryson, Burke,
  Caldwell, Haas, Henze, \& Sanderfer}]{twicken_kepler_2018}
Twicken, J.~D., Catanzarite, J.~H., Clarke, B.~D., {et~al.} 2018,
  \href{http://dx.doi.org/10.1088/1538-3873/aab694}{\JournalTitle{\pasp}, 130,
  064502}

\bibitem[{van Leeuwen(2009)}]{van_leeuwen_parallaxes_2009}
van Leeuwen, F. 2009,
  \href{http://dx.doi.org/10.1051/0004-6361/200811382}{\JournalTitle{\aap},
  497, 209}

\bibitem[{Vanderburg {et~al.}(2018)Vanderburg, Mann, Rizzuto, Bieryla, Kraus,
  Berlind, Calkins, Curtis, Douglas, Esquerdo, Everett, Horch, Howell, Latham,
  Mayo, Quinn, Scott, \& Stefanik}]{vanderburg_zeitVII_2018}
Vanderburg, A., Mann, A.~W., Rizzuto, A., {et~al.} 2018,
  \href{http://dx.doi.org/10.3847/1538-3881/aac894}{\JournalTitle{\aj}, 156,
  46}

\bibitem[{Vanderburg {et~al.}(2019)Vanderburg, Huang, Rodriguez, Becker,
  Ricker, Vanderspek, Latham, Seager, Winn, Jenkins, Addison, Bieryla,
  Briceño, Bowler, Brown, Burke, Burt, Caldwell, Clark, Crossfield, Dittmann,
  Dynes, Fulton, Guerrero, Harbeck, Horner, Kane, Kielkopf, Kraus, Kreidberg,
  Law, Mann, Mengel, Morton, Okumura, Pearce, Plavchan, Quinn, Rabus, Rose,
  Rowden, Shporer, Siverd, Smith, Stassun, Tinney, Wittenmyer, Wright, Zhang,
  Zhou, \& Ziegler}]{vanderburg_hr858_2019}
Vanderburg, A., Huang, C.~X., Rodriguez, J.~E., {et~al.} 2019,
  \href{http://dx.doi.org/10.3847/2041-8213/ab322d}{\JournalTitle{\apj}, 881,
  L19}

\bibitem[{VanderPlas \& Ivezi\'c(2015)}]{vanderplas_periodograms_2015}
VanderPlas, J.~T., \& Ivezi\'c, Z. 2015,
  \href{http://dx.doi.org/10.1088/0004-637X/812/1/18}{\JournalTitle{\apj}, 812,
  18}

\bibitem[{\v{Z}erjal {et~al.}(2017)\v{Z}erjal, Zwitter, Matijevi\v{c}, Grebel,
  Kordopatis, Munari, Seabroke, Steinmetz, Wojno, Bienaymé, Bland-Hawthorn,
  Conrad, Freeman, Gibson, Gilmore, Kunder, Navarro, Parker, Reid, Siviero,
  Watson, \& Wyse}]{zerjal_chromospherically_2017}
\v{Z}erjal, M., Zwitter, T., Matijevi\v{c}, G., {et~al.} 2017,
  \href{http://dx.doi.org/10.3847/1538-4357/835/1/61}{\JournalTitle{\apj}, 835,
  61}

\bibitem[{\v{Z}erjal {et~al.}(2019)\v{Z}erjal, Ireland, Nordlander, Lin,
  Casagrande, Horner, De~Silva, Martell, Čotar, Traven, \&
  Zwitter}]{zerjal_galah_2019}
\v{Z}erjal, M., Ireland, M.~J., Nordlander, T., {et~al.} 2019,
  \href{http://dx.doi.org/10.1093/mnras/stz296}{\JournalTitle{\mnras}, 484,
  4591}

\bibitem[{Walt {et~al.}(2011)Walt, Colbert, \& Varoquaux}]{walt_numpy_2011}
Walt, S. v.~d., Colbert, S.~C., \& Varoquaux, G. 2011, \JournalTitle{Computing
  in Science \& Engineering}, 13, 22

\bibitem[{Weber \& Davis(1967)}]{weber_angular_1967}
Weber, E.~J., \& Davis, Jr., L. 1967,
  \href{http://dx.doi.org/10.1086/149138}{\JournalTitle{\apj}, 148, 217}

\bibitem[{Williams \& Cieza(2011)}]{williams_protoplanetary_2011}
Williams, J.~P., \& Cieza, L.~A. 2011,
  \href{http://dx.doi.org/10.1146/annurev-astro-081710-102548}{\JournalTitle{\araa},
  49, 67}

\bibitem[{Wright {et~al.}(2010)Wright, Eisenhardt, Mainzer, Ressler, Cutri,
  Jarrett, Kirkpatrick, Padgett, McMillan, Skrutskie, Stanford, Cohen, Walker,
  Mather, Leisawitz, Gautier, McLean, Benford, Lonsdale, Blain, Mendez, Irace,
  Duval, Liu, Royer, Heinrichsen, Howard, Shannon, Kendall, Walsh, Larsen,
  Cardon, Schick, Schwalm, Abid, Fabinsky, Naes, \& Tsai}]{wright_WISE_2010}
Wright, E.~L., Eisenhardt, P. R.~M., Mainzer, A.~K., {et~al.} 2010,
  \href{http://dx.doi.org/10.1088/0004-6256/140/6/1868}{\JournalTitle{\aj},
  140, 1868}

\bibitem[{Wright {et~al.}(2012)Wright, Marcy, Howard, Johnson, Morton, \&
  Fischer}]{wright_frequency_2012}
Wright, J.~T., Marcy, G.~W., Howard, A.~W., {et~al.} 2012,
  \href{http://dx.doi.org/10.1088/0004-637X/753/2/160}{\JournalTitle{\apj},
  753, 160}

\bibitem[{Yu {et~al.}(2017)Yu, Donati, Hébrard, Moutou, Malo, Grankin,
  Hussain, Collier~Cameron, Vidotto, Baruteau, Alencar, Bouvier, Petit, Takami,
  Herczeg, Gregory, Jardine, Morin, Ménard, \& {Matysse
  Collaboration}}]{yu_hot_2017}
Yu, L., Donati, J.-F., Hébrard, E.~M., {et~al.} 2017,
  \href{http://dx.doi.org/10.1093/mnras/stx009}{\JournalTitle{\mnras}, 467,
  1342}

\bibitem[{Zahn(1977)}]{zahn_tidal_1977}
Zahn, J.-P. 1977,
  \href{https://ui.adsabs.harvard.edu//#abs/1977A&A....57..383Z/abstract}{\JournalTitle{\aap},
  500, 121}

\bibitem[{{Zhou} {et~al.}(2018){Zhou}, {Rodriguez}, {Vanderburg}, {Quinn},
  {Irwin}, {Huang}, {Latham}, {Bieryla}, {Esquerdo}, {Berlind}, \&
  {Calkins}}]{zhou_2018_hd106315obliq}
{Zhou}, G., {Rodriguez}, J.~E., {Vanderburg}, A., {et~al.} 2018,
  \href{http://dx.doi.org/10.3847/1538-3881/aad085}{\JournalTitle{\aj}, 156,
  93}

\bibitem[{Zhou {et~al.}(2020)Zhou, Winn, Newton, Quinn, Rodriguez, Mann,
  Rizzuto, Vanderburg, Huang, Latham, Teske, Wang, Shectman, Butler, Crane,
  Thompson, Henry, Paredes, Jao, James, \& Hinojosa}]{zhou_well_2020}
Zhou, G., Winn, J.~N., Newton, E.~R., {et~al.} 2020,
  \href{http://dx.doi.org/10.3847/2041-8213/ab7d3c}{\JournalTitle{The
  Astrophysical Journal}, 892, L21}

\bibitem[{Ziegler {et~al.}(2020)Ziegler, Tokovinin, Briceño, Mang, Law, \&
  Mann}]{ziegler_soar_2020}
Ziegler, C., Tokovinin, A., Briceño, C., {et~al.} 2020,
  \href{http://dx.doi.org/10.3847/1538-3881/ab55e9}{\JournalTitle{\aj}, 159,
  19}

\bibitem[{Zuckerman \& Song(2004)}]{zuckerman_young_2004}
Zuckerman, B., \& Song, I. 2004,
  \href{http://dx.doi.org/10.1146/annurev.astro.42.053102.134111}{\JournalTitle{\araa},
  42, 685}

\end{thebibliography}

\listofchanges

\end{document}